%% file: main.tex
\documentclass[10pt,letterpaper]{article}

\usepackage[margin=1in]{geometry}
\usepackage{amsmath}
\usepackage{fontspec}
\usepackage{unicode-math}
\usepackage{microtype}
\usepackage{graphicx}
\usepackage{xcolor}
\usepackage{booktabs}
\usepackage{array}
\usepackage{enumitem}
\usepackage[font=small,labelfont=bf]{caption}
\usepackage{float}
\usepackage{needspace}   %
\usepackage{algorithm}
\usepackage[noend]{algpseudocode}
\usepackage[numbers,sort&compress]{natbib}
\usepackage[hidelinks]{hyperref}
\usepackage[capitalise,noabbrev]{cleveref}
\hypersetup{pdftitle={Cortex: Measurement-Gated Evolution and Pruning of Coding-Agent Skills and Rules},
            pdfauthor={Samuele Giampieri}}
\makeatletter
\providecommand{\theHALG@line}{}
\renewcommand{\theHALG@line}{\thealgorithm.\arabic{ALG@line}}
\makeatother

\setlist{itemsep=2pt,topsep=4pt}

\newcommand{\cmd}[1]{\texttt{#1}}
\newcommand{\CILevel}{95\%}\newcommand{\CI}{\CILevel~CI}   %
\newcommand{\EvalTag}{\cmd{v1.0-eval}}   %
\newcommand{\PaperTag}{\cmd{\ReleaseTag}} %
\newcommand{\RepoURL}{\url{https://github.com/samugit83/cortex}}


\input{generated/numbers}

\title{Cortex: Measurement-Gated Evolution and Pruning\\of Coding-Agent Skills and Rules}
\author{Samuele Giampieri\\\small Independent researcher}
\date{\small October 2026}

\begin{document}
\maketitle

\begin{abstract}
AI coding agents such as Claude Code do not act on their training alone; they also follow text
files that developers write for them. A project file describes how the code should be written,
\emph{skills} are short guides the agent may choose to read when it judges them relevant, and
\emph{rules} are notes tied to particular files, which the agent sees automatically whenever it
opens one of those files. We call these files the agent's \emph{harness}. Developers add to it
each time they have to correct the agent, but checking whether an addition really helps is
harder than it looks: the same agent given the same task can succeed once and fail the next
time, so seeing an instruction work once proves little.

We present Cortex, a tool that tests changes to the harness instead of trusting them. At the
end of a working session it saves any fix the session made as a task it can replay; across
sessions it spots the corrections the developer keeps making, and writes one new skill or rule
at a time to prevent them. It then runs the agent on those tasks several times with the new
text and several times without it, and keeps the text only if it helps overall and makes no
single task noticeably worse. Text already in the harness is tested the same way before it is
deleted. Throughout, Cortex records whether each piece of text reached the agent, and it
neither adds nor deletes text that the agent never used.

We tested Cortex with \RolloutModelName{} on a repository built for the purpose, with a few coding
conventions the model tends to break and one area with nothing to learn, kept as a comparison. A
script played the developer, and Cortex was run several times from scratch. Each harness it built
was then tested on new tasks it had never seen.

The harnesses Cortex built made the agent follow the conventions far more often than a harness
with no skills or rules, left the comparison area unchanged, and added no text that the agent
must read at every step. The clearest lesson is about how advice is delivered rather than what
it says: rules, which the agent sees automatically, did most of the work, while skills, which
the agent must choose to open, helped only when they were offered for the files a task touched;
offered on every turn, they were rarely opened. We also fed Cortex useless and harmful
suggestions on purpose. It kept none of them, but the agent never opened the useless ones or the
harmful skills, and the harmful rules, which did load, were tried on tasks that mostly failed
without them, where Cortex reached no verdict, even when a harmful rule broke a task that passed
without it. This shows that Cortex keeps no text the agent never uses, not yet that it can spot
text that does damage. Finally, in this study Cortex was worth as much as the model needed
correcting: under the stronger \StrongModelName{}, which already followed most of the
conventions, a single run's harness added little, and on a real open-source project where the
agent was rarely corrected, Cortex found nothing to learn. We release the tool, the evaluation
lab and the data behind every number in the paper.
\end{abstract}

\input{sections/01-introduction}
\input{sections/02-background}
\input{sections/03-design}
\input{sections/04-setup}
\input{sections/05-results}
\input{sections/07-related}
\input{sections/08-limitations}
\input{sections/09-conclusion}

\section*{Acknowledgement}
AI participation was substantial and continuous across design, implementation, evidence review,
statistical verification, critical review, and editing.

{\raggedright\input{main.bbl}}

\clearpage
\appendix
\input{sections/appendix-results}

\end{document}

%% file: generated/numbers.tex
\newcommand{\AADiffering}{3}
\newcommand{\AAEst}{\textminus{}3.3}
\newcommand{\AAHi}{0.0}
\newcommand{\AALo}{\textminus{}7.3}
\newcommand{\AAP}{0.13}
\newcommand{\AARuleEst}{\textminus{}4.2}
\newcommand{\AARuleHi}{0.0}
\newcommand{\AARuleLo}{\textminus{}9.2}
\newcommand{\AARun}{R1}
\newcommand{\AblEvolvedRate}{87.5\%}
\newcommand{\AblFlatGap}{12.5}
\newcommand{\AblFlatGapHi}{+28.3}
\newcommand{\AblFlatGapLo}{\textminus{}1.7}
\newcommand{\AblFlatRate}{75.0\%}
\newcommand{\AblIdealGap}{38.9}
\newcommand{\AblIdealGapHi}{+58.1}
\newcommand{\AblIdealGapLo}{+20.0}
\newcommand{\AblIdealRate}{48.6\%}
\newcommand{\AblKitchenGap}{11.1}
\newcommand{\AblKitchenGapHi}{+22.5}
\newcommand{\AblKitchenGapLo}{0.0}
\newcommand{\AblKitchenRate}{76.4\%}
\newcommand{\AblationK}{3}
\newcommand{\AcceptAllEst}{\textminus{}8.9}
\newcommand{\AcceptAllFamA}{\textminus{}13.3}
\newcommand{\AcceptAllFamC}{\textminus{}17.8}
\newcommand{\AcceptAllFamE}{\textminus{}4.4}
\newcommand{\AcceptAllHi}{\textminus{}1.7}
\newcommand{\AcceptAllLo}{\textminus{}16.1}
\newcommand{\AcceptAllP}{0.075}
\newcommand{\AcceptAllRun}{R2}
\newcommand{\AlwaysOnBudget}{3{,}000}
\newcommand{\AlwaysOnInvoked}{0}
\newcommand{\AlwaysOnOffered}{198}
\newcommand{\BootReps}{10{,}000}
\newcommand{\BreadthKeptLowEval}{8\%}
\newcommand{\BreadthKilledHighEval}{58\%}
\newcommand{\BroadMoneyFiredMin}{28}
\newcommand{\BroadMoneyNetRuns}{5}
\newcommand{\BroadMoneyTasks}{3}
\newcommand{\BroadMoneyUnknown}{7}
\newcommand{\BundledEvolved}{5.8\%}
\newcommand{\BundledNone}{5.4\%}
\newcommand{\ClaudeCodeVersion}{2.1.278}
\newcommand{\CompleteTasks}{23}
\newcommand{\ConfirmSweepRollouts}{156}
\newcommand{\ControlRuns}{1}
\newcommand{\CorrControl}{23}
\newcommand{\CorrControlRate}{0.88}
\newcommand{\CorrEval}{60}
\newcommand{\CorrEvalRate}{0.58}
\newcommand{\CostEvolved}{0.149}
\newcommand{\CostNone}{0.121}
\newcommand{\CostRise}{23\%}
\newcommand{\CostRiseC}{38\%}
\newcommand{\CostRiseD}{5\%}
\newcommand{\CostRiseHi}{22\%}
\newcommand{\CostRiseLo}{12\%}
\newcommand{\DecidedItems}{4}
\newcommand{\DescBodySeenC}{6}
\newcommand{\DescItems}{4}
\newcommand{\DescOpenedC}{16}
\newcommand{\DescPassC}{2}
\newcommand{\DescRateC}{11\%}
\newcommand{\DescRecovered}{79\%}
\newcommand{\DescRecoveredHi}{100\%}
\newcommand{\DescRecoveredLo}{58\%}
\newcommand{\DescRewritten}{1}
\newcommand{\DescRolloutsC}{18}
\newcommand{\DescShareC}{16\%}
\newcommand{\DescShareCHi}{56\%}
\newcommand{\DescShareCLo}{0\%}
\newcommand{\DzeroGain}{+42.5}
\newcommand{\EvalRuns}{4}
\newcommand{\EvalSweeps}{51}
\newcommand{\EvolveCostHi}{4.09}
\newcommand{\EvolveCostLo}{2.39}
\newcommand{\EvolvedRunOneC}{70\%}
\newcommand{\ExcludedFrom}{R4}
\newcommand{\ExcludedScenario}{HA6}
\newcommand{\ExtFirstTime}{7}
\newcommand{\ExtLessons}{1}
\newcommand{\ExtTasks}{8}
\newcommand{\FailCRuleEvolved}{43}
\newcommand{\FailEvolved}{108}
\newcommand{\FailNone}{361}
\newcommand{\FailRuleEvolved}{86}
\newcommand{\FailRuleNone}{341}
\newcommand{\FailRuleShareNone}{94\%}
\newcommand{\FamAEst}{+97.0}
\newcommand{\FamAHi}{+100.0}
\newcommand{\FamALo}{+91.0}
\newcommand{\FamBEst}{+53.3}
\newcommand{\FamBHi}{+88.3}
\newcommand{\FamBLo}{+17.5}
\newcommand{\FamCEst}{+50.0}
\newcommand{\FamCHi}{+73.3}
\newcommand{\FamCLo}{+25.0}
\newcommand{\FamEEst}{+26.7}
\newcommand{\FamEHi}{+60.0}
\newcommand{\FamELo}{+1.7}
\newcommand{\FamHolmP}{= 0.0004}
\newcommand{\FisherP}{0.022}
\newcommand{\FlatChars}{1{,}688}
\newcommand{\FlatMinusEst}{\textminus{}12.5}
\newcommand{\FlatMinusHi}{+1.7}
\newcommand{\FlatMinusLo}{\textminus{}28.3}
\newcommand{\FlatRateC}{0\%}
\newcommand{\FloorFlaggedBuriedEval}{3}
\newcommand{\FloorFlaggedKeptEval}{6}
\newcommand{\GateCandidates}{29}
\newcommand{\GateHarmfulKilled}{2}
\newcommand{\GateHarmfulUnscored}{3}
\newcommand{\GateNoiseGain}{10}
\newcommand{\GateNoiseHarmful}{2}
\newcommand{\GateNoisePlacebo}{7}
\newcommand{\GateNoisePositive}{1}
\newcommand{\GateNoiseSingle}{9}
\newcommand{\GatePlaceboKilled}{12}
\newcommand{\GatePlaceboUnscored}{8}
\newcommand{\GatedInvoked}{28}
\newcommand{\GatedOffered}{66}
\newcommand{\GuideReadNone}{0.5\%}
\newcommand{\HaikuGain}{+66.7}
\newcommand{\HaikuHi}{+80.8}
\newcommand{\HaikuLo}{+51.7}
\newcommand{\HaikuNone}{20.8\%}
\newcommand{\Harmful}{5}
\newcommand{\HarmfulPathlessRule}{small-changes-need-no-changelog}
\newcommand{\HarmfulRules}{3}
\newcommand{\HeldEvoA}{97.0\%}
\newcommand{\HeldEvoB}{79.2\%}
\newcommand{\HeldEvoC}{51.7\%}
\newcommand{\HeldEvoE}{81.7\%}
\newcommand{\HeldEvolvedRate}{76.5\%}
\newcommand{\HeldNoneE}{55.0\%}
\newcommand{\HeldNoneRate}{21.5\%}
\newcommand{\HfourMinP}{0.2}
\newcommand{\HoldoutLate}{15}
\newcommand{\HoldoutTasks}{30}
\newcommand{\HoneAllEst}{+56.5}
\newcommand{\HoneAllHi}{+75.0}
\newcommand{\HoneAllLo}{+35.4}
\newcommand{\HoneEst}{+55.0}
\newcommand{\HoneHi}{+74.1}
\newcommand{\HoneLo}{+34.1}
\newcommand{\HoneP}{= 0.0001}
\newcommand{\HtwoEst}{\textminus{}0.8}
\newcommand{\HtwoHi}{0.0}
\newcommand{\HtwoLo}{\textminus{}5.0}
\newcommand{\HzeroBar}{50\%}
\newcommand{\HzeroBarD}{90\%}
\newcommand{\HzeroBelowN}{3}
\newcommand{\IdealAlwaysSkills}{2}
\newcommand{\IdealChars}{728}
\newcommand{\IdealRateA}{0\%}
\newcommand{\IdealRateE}{56\%}
\newcommand{\IdealRollouts}{90}
\newcommand{\ImplCodeFiles}{9}
\newcommand{\ImplCodeLines}{6{,}906}
\newcommand{\ImplJevLines}{1{,}243}
\newcommand{\ImplPromptLines}{1{,}341}
\newcommand{\InitClaudeChars}{559}
\newcommand{\JudgeItems}{4}
\newcommand{\JudgeItemsDev}{2}
\newcommand{\JudgeMissMax}{54}
\newcommand{\JudgeOver}{3}
\newcommand{\JudgeTied}{3}
\newcommand{\KConfirm}{3}
\newcommand{\KScreen}{2}
\newcommand{\KeptItems}{18}
\newcommand{\KeptRules}{16}
\newcommand{\KeptSkills}{2}
\newcommand{\KitchenChars}{2{,}519}
\newcommand{\KitchenGap}{11.1}
\newcommand{\KitchenHi}{+22.5}
\newcommand{\KitchenLo}{0.0}
\newcommand{\KitchenRun}{R1}
\newcommand{\LoopActive}{25}
\newcommand{\LoopBarren}{14}
\newcommand{\LoopConfirmRollouts}{1{,}494}
\newcommand{\LoopCycles}{39}
\newcommand{\LoopFixReads}{2}
\newcommand{\LoopKilled}{5}
\newcommand{\LoopKilledRegression}{2}
\newcommand{\LoopKilledRegressionOnly}{1}
\newcommand{\LoopRollouts}{1{,}818}
\newcommand{\LoopSweepRollouts}{1{,}809}
\newcommand{\LoopSwept}{26}
\newcommand{\LoopUnscored}{3}
\newcommand{\MarginClose}{10}
\newcommand{\MarginDFamily}{\textminus{}10}
\newcommand{\MarginDescShare}{50\%}
\newcommand{\MarginNotBetter}{+5}
\newcommand{\MarginPlaceboUpper}{25\%}
\newcommand{\MarginPositives}{3}
\newcommand{\MaxCorrections}{3}
\newcommand{\MaxInvalidPct}{10\%}
\newcommand{\MaxRunsDefault}{60}
\newcommand{\MaxRunsEval}{200}
\newcommand{\MinClaudeVersion}{2.1.276}
\newcommand{\MinNetRuns}{2}
\newcommand{\MinThemeFailingTasks}{2}
\newcommand{\MinThemeOccurrences}{3}
\newcommand{\ModelChangeK}{3}
\newcommand{\ModelChangeSource}{R1}
\newcommand{\NarrowClockEvolved}{50.0\%}
\newcommand{\NarrowClockFiles}{3}
\newcommand{\NarrowClockFired}{10}
\newcommand{\NarrowClockN}{30}
\newcommand{\NarrowClockNone}{56.7\%}
\newcommand{\NarrowMoneyFiles}{5}
\newcommand{\NarrowMoneyFired}{0}
\newcommand{\NarrowMoneyN}{30}
\newcommand{\NarrowMoneyNone}{26.7\%}
\newcommand{\NarrowPoolEvolved}{52.2\%}
\newcommand{\NarrowPoolNone}{22.6\%}
\newcommand{\NarrowRun}{R3}
\newcommand{\NoneHeldNames}{112}
\newcommand{\NoneHeldRollouts}{595}
\newcommand{\NoneHeldText}{2}
\newcommand{\NoneNamesN}{93}
\newcommand{\NoneNamesPass}{9.7\%}
\newcommand{\NoneNoNamesN}{382}
\newcommand{\NoneNoNamesPass}{23.6\%}
\newcommand{\NoneRateA}{0\%}
\newcommand{\NoneRateB}{26\%}
\newcommand{\NoneRateC}{2\%}
\newcommand{\NoneRateD}{100\%}
\newcommand{\NoneRateE}{55\%}
\newcommand{\NoneRunOneA}{0\%}
\newcommand{\NoneRunOneE}{57\%}
\newcommand{\OthersPoolHi}{87.0\%}
\newcommand{\OthersPoolLo}{80.9\%}
\newcommand{\OverloadFactor}{3}
\newcommand{\PermReps}{10{,}000}
\newcommand{\PlaceboKept}{0}
\newcommand{\PlaceboMoment}{10}
\newcommand{\PlaceboTopic}{10}
\newcommand{\PlaceboUpper}{17\%}
\newcommand{\Placebos}{20}
\newcommand{\Planted}{4}
\newcommand{\PlantedMatched}{2}
\newcommand{\PlantedPlacebos}{2}
\newcommand{\Positives}{4}
\newcommand{\PositivesKept}{2}
\newcommand{\PrimaryK}{5}
\newcommand{\PruneMatched}{6}
\newcommand{\PruneMomentListed}{1{,}263}
\newcommand{\PruneMomentOpened}{46}
\newcommand{\PruneOwnMinLoss}{9}
\newcommand{\PrunePlaceboBase}{18}
\newcommand{\PrunePlaceboLostTasks}{2}
\newcommand{\PruneRun}{R1}
\newcommand{\PruneTested}{8}
\newcommand{\RecheckKept}{2}
\newcommand{\Rechecked}{3}
\newcommand{\RegressionTolerance}{0.34}
\newcommand{\RelKeptLowEval}{17\%}
\newcommand{\RelKilledHighEval}{75\%}
\newcommand{\ReleaseTag}{v1.0.5}
\newcommand{\RelevanceFloorPct}{35\%}
\newcommand{\RemovableItems}{2}
\newcommand{\RolloutModel}{claude-haiku-4-5-20251001}
\newcommand{\RolloutModelName}{Claude Haiku 4.5}
\newcommand{\RolloutTimeoutMin}{10}
\newcommand{\Rounds}{10}
\newcommand{\RubricGood}{15}
\newcommand{\RubricScored}{18}
\newcommand{\RunGainAllHi}{+66.7}
\newcommand{\RunGainAllLo}{+32.5}
\newcommand{\RunOneItems}{4}
\newcommand{\ScreenCandRollouts}{8}
\newcommand{\SecondRepo}{structlog}
\newcommand{\SecondRepoBase}{73393f34b4}
\newcommand{\SecondRepoHoldout}{4}
\newcommand{\SecondRepoK}{5}
\newcommand{\SecondRepoSlug}{hynek/structlog}
\newcommand{\SecsEvolved}{68}
\newcommand{\SecsNone}{57}
\newcommand{\SeenEvoA}{90.3\%}
\newcommand{\SeenEvoB}{88.9\%}
\newcommand{\SeenEvoC}{76.4\%}
\newcommand{\SeenEvoE}{85.4\%}
\newcommand{\SeenEvolvedRate}{85.2\%}
\newcommand{\SeenGainEst}{+78.8}
\newcommand{\SeenGainHi}{+86.7}
\newcommand{\SeenGainLo}{+68.9}
\newcommand{\SeenNoneE}{10.4\%}
\newcommand{\SeenNoneRate}{6.4\%}
\newcommand{\SeenTasks}{22}
\newcommand{\SessControl}{26}
\newcommand{\SessEval}{104}
\newcommand{\SessionCostHi}{11.78}
\newcommand{\SessionCostLo}{10.08}
\newcommand{\SessionsPerRun}{26}
\newcommand{\ShopWideHi}{100\%}
\newcommand{\ShopWideLo}{93\%}
\newcommand{\ShopWideMargin}{1.25}
\newcommand{\ShopWidePassed}{4}
\newcommand{\SonnetAccReadN}{35}
\newcommand{\SonnetAccReadPass}{28}
\newcommand{\SonnetAccUnreadN}{115}
\newcommand{\SonnetAccUnreadPass}{110}
\newcommand{\SonnetChangelogLoss}{5}
\newcommand{\SonnetClockLoss}{1}
\newcommand{\SonnetCostEvolved}{0.205}
\newcommand{\SonnetCostNone}{0.228}
\newcommand{\SonnetEvolved}{94.4\%}
\newcommand{\SonnetFamCEst}{+11.1}
\newcommand{\SonnetFamCHi}{+38.9}
\newcommand{\SonnetFamCLo}{\textminus{}11.1}
\newcommand{\SonnetGain}{+5.6}
\newcommand{\SonnetHi}{+13.9}
\newcommand{\SonnetLo}{\textminus{}2.8}
\newcommand{\SonnetNone}{88.9\%}
\newcommand{\SonnetNoneRollouts}{168}
\newcommand{\SonnetNoneText}{67}
\newcommand{\SonnetRemovalRollouts}{300}
\newcommand{\SonnetRemovalText}{72}
\newcommand{\SplitCycleRun}{R4}
\newcommand{\SplitCycleSwept}{2}
\newcommand{\StopAfterBarren}{2}
\newcommand{\StopAfterBarrenEval}{3}
\newcommand{\StrongModel}{claude-sonnet-5}
\newcommand{\StrongModelName}{Claude Sonnet 5}
\newcommand{\SweepCostHi}{128.78}
\newcommand{\SweepCostLo}{32.76}
\newcommand{\SweepExampleExposed}{6}
\newcommand{\SweepExampleTasks}{26}
\newcommand{\TamperedEvolved}{13}
\newcommand{\TamperedNone}{9}
\newcommand{\TrainK}{3}
\newcommand{\TrainTasks}{26}
\newcommand{\TurnsEvolved}{22.1}
\newcommand{\TurnsNone}{18.2}
\newcommand{\XBenchImpl}{19}
\newcommand{\XBenchRollouts}{40}
\newcommand{\XTrainImpl}{1}
\newcommand{\XTrainSessions}{8}

%% file: sections/01-introduction.tex
\section{Introduction}\label{sec:intro}

Coding agents are increasingly steered not only by their weights but by text their users
edit. A project instruction file (\cmd{CLAUDE.md}, or \cmd{AGENTS.md}~\cite{gloaguen2026agentsmd}),
a folder of \emph{skills} the agent may choose to invoke, and \emph{rules} that load when
the agent reads matching files~\cite{claudecode-memory,claudecode-skills} together make up
what this paper calls the agent's \emph{harness}: the part of its behaviour a developer
can change in a minute, by editing text. The agent works in \emph{turns}: at each step the
model reads its \emph{context}, the text in front of it (the harness text that is loaded,
the request and the work so far), and then acts. Harnesses grow the way documentation
grows. A correction a developer has to make twice becomes a line, the line stays, and
every line that is always in context is paid for on every turn of every session.

Little of it is measured. Whether an instruction helps is often judged the way it was
written: by someone who watched it work once. That is not evidence, because an agent's
success on a fixed task is a noisy quantity: the model samples its output, so the same
agent can pass a task on one attempt and fail it on the next. Results in this paper are
therefore given as a \emph{pass rate}, the share of attempts that succeeded, and
differences between pass rates in percentage \emph{points}. In the evaluation reported
here, one and the same harness, measured twice on the same tasks in separate batches of
rollouts, differed by \AAEst{}~points (\CILevel{} confidence interval, or CI,
[\AALo, \AAHi], resampling tasks; \cref{sec:setup} explains these intervals), a difference
that, seen once, would pass for an effect. Recent studies report that adding skills can
make an agent worse~\cite{assay2026}, that skill gains measured on the tasks one
repository supplies could not be told apart from run-to-run
noise~\cite{skillissue2026}, and that whether an updated harness helps depends on the
model that uses it: weaker models often fail to activate, or to follow, what was
added~\cite{harnessupdating2026}. A harness curated by impression can accumulate
instructions that do nothing, some that hurt, and a context bill for all of them.

A growing body of work optimises or evolves agent text automatically: prompts and
programs~\cite{khattab2024dspy,agrawal2025gepa}, agent designs~\cite{hu2024adas,zhang2025dgm},
and most recently the harness itself~\cite{darwinx2026,metaharness2026,selfharness2026,harnessx2026}.
These systems are typically optimised and evaluated against benchmarks. A developer's
problem is narrower and, in one respect, harder: one repository, its own conventions, a
handful of tasks drawn from their own work, a fixed context budget, and no way to tell
whether a given instruction earns its place.

\begin{figure}[H]
  \centering
  \includegraphics[width=\linewidth]{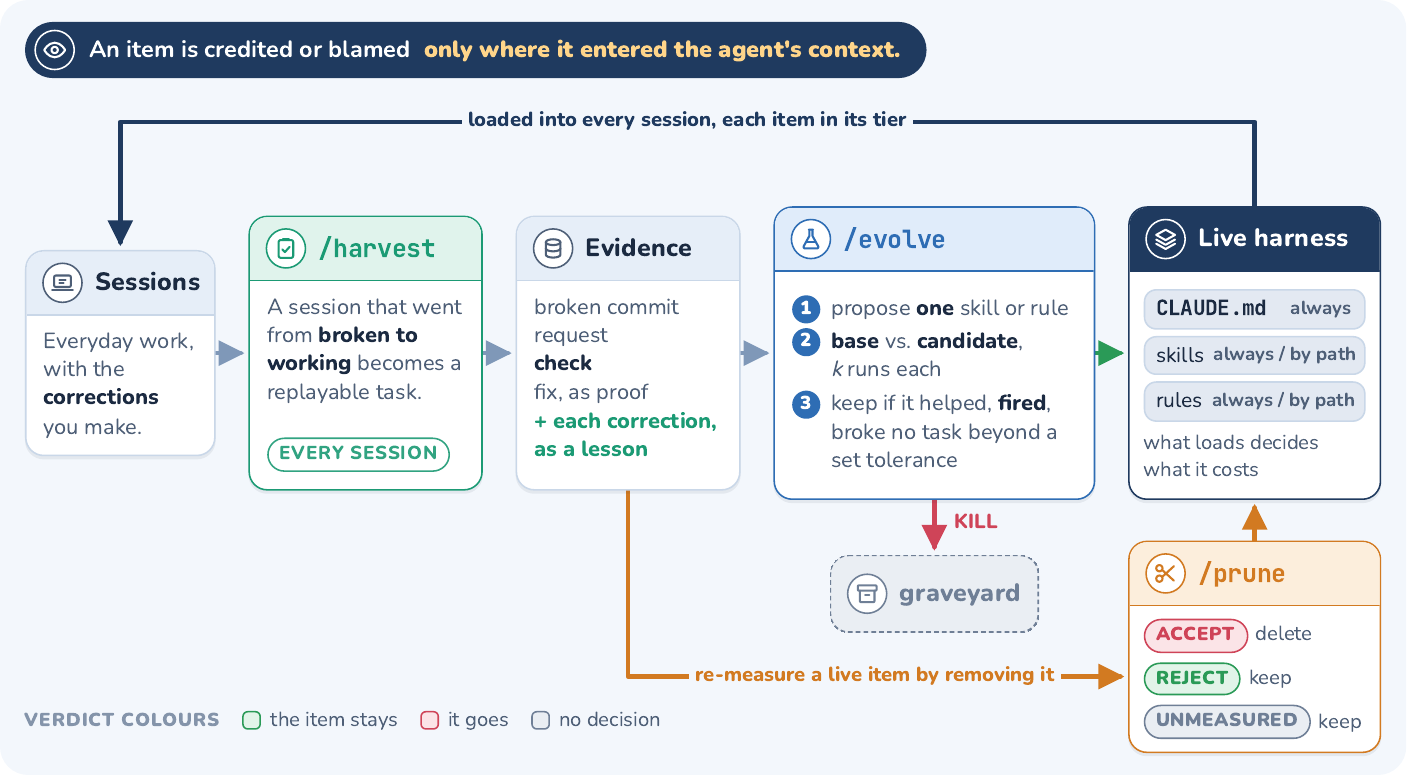}
  \caption{The Cortex loop. \cmd{/harvest} turns a session that went from broken to
  working into a task; \cmd{/evolve} measures one candidate against the live harness and
  keeps it only if it helped, fired, and cost no task more than a small tolerance;
  \cmd{/prune} re-measures live items by removal and keeps any item it never saw fire.}
  \label{fig:loop}
\end{figure}

We present \textbf{Cortex}, a tool for Claude Code that treats the harness as something
to measure rather than to curate (\cref{fig:loop}). Cortex calls each skill or rule an
\emph{item}, and the harness the developer is using now the \emph{live} harness. It
consists of three commands. \cmd{/harvest} runs at the end of a working session. When
something in the session went from broken to working, it saves a replayable
\emph{task}: the broken commit, the developer's request, and a \emph{check}, a script
that decides whether an attempt at the task succeeded. It also logs each correction the
developer made along the way. \cmd{/evolve} reads the corrections that recur and writes
\emph{one} new skill or rule, a \emph{candidate}, for one of them. It also chooses the
candidate's \emph{loading tier}, which decides when its text enters the agent's context
(\cref{sec:background}): if the fixes behind the problem sit in one folder, the
candidate loads only when the agent works there. It then runs the agent on the
harvested tasks several times each, with the live harness and with the live harness plus
the candidate, in disposable clones of the repository. Each attempt is a \emph{rollout},
and a batch of rollouts that compares two harnesses in this way is a \emph{sweep}. The
candidate is kept only if it helped, \emph{fired} in at least one of its rollouts (a rule
loaded into the agent's context, or a skill opened), and made no task worse beyond a set
tolerance, which is zero on a task the live harness never failed; a candidate killed only
for such a drop is swept again first. \cmd{/prune} works the other way round: it
re-measures live items by removing them one at a time, and never deletes one it has not
seen fire. One principle runs through both measurements: \emph{an item is credited or
blamed only where it entered the agent's context.}

We evaluate Cortex in a replicated study (\cref{sec:setup}), with
\RolloutModelName{} as both the agent that is measured and the one that runs Cortex's
commands. The study uses a small synthetic repository with four \emph{house rules}:
conventions of that one project, such as reading the current time only through its own
clock module, which code can break while still working, and which the model often breaks
unaided. Each house rule has a \emph{family} of tasks built around it; a fifth family,
plain bugs with no rule to learn, is the \emph{control}. A script plays the developer.
The \EvalRuns{} independent \emph{runs}, R1 to R4, each start from a fresh copy of the
repository, play the same \Rounds{} rounds of scripted sessions, and let Cortex evolve a
harness from them.
Every harness is then measured on \HoldoutTasks{} \emph{held-out} scenarios: pieces of
work of the same families that no session played. The loop never saw them, so a gain on
them cannot come from remembering the tasks it was shown: what it learned has to carry
over to new work. The main findings are these.

\begin{itemize}
  \item \textbf{It works on held-out scenarios.} On the four rule families, the evolved
    harness beats a harness with no skills or rules by \HoneEst{}~points of pass rate
    (\CI{} [\HoneLo, \HoneHi], permutation $p \HoneP$). The interval allows for chance in
    both which runs and which scenarios were measured. The control family changes by
    \HtwoEst{}~points, and no run kept an item for it.
  \item \textbf{It adds no always-on context.} Every item the loop kept has a \cmd{paths}
    field, which ties it to certain files, so that it loads only when the agent works on
    them. So the evolved harness adds nothing to what every turn carries: beyond the
    skill listing every arm shares, its only text present on every turn is the
    \InitClaudeChars{}-character \cmd{CLAUDE.md} that the harness with no skills or rules
    has too. In the one run that measured it (\KitchenRun), the evolved harness also
    scored \KitchenGap{}~points higher on the rule families than a harness that instead
    appends the project's contribution guide, the file that
    states the house rules, to \cmd{CLAUDE.md} (\CI{} [\KitchenLo, \KitchenHi]); the
    interval reaches zero, and the comparison crosses a batch boundary
    (\cref{sec:results:cost}).
  \item \textbf{It kept nothing known to be useless or harmful.} None of \Placebos{}
    \emph{placebo} candidates (plausible advice about matters the repository does not
    care about) and none of \Harmful{} harmful ones was kept. The test was weaker than
    those counts suggest. Each candidate is first tried in a short, cheap sweep, its
    \emph{screen}, which runs only those of the tasks the candidate was written for that
    the live harness does not always pass. The agent never opened the placebos or the \GateHarmfulKilled{} harmful
    skills on their screens, so none could be kept: Cortex's scoring script refused the
    harmful skills and \GatePlaceboKilled{} placebos for never being used, and reached no
    verdict on the other \GatePlaceboUnscored{} placebos, whose screens could show no
    gain. The \GateHarmfulUnscored{} harmful rules did load, but harm shows only on a task
    that passes without the harmful item, and these screens ran on tasks chosen because
    the live harness had failed them, where no task could show a gain; so Cortex's scoring
    script reached no verdict on them, even on the screen where a rule broke a task that
    passed without it (\cref{sec:results:gates}).
  \item \textbf{Rules are loaded; skills must be chosen.} Of the \KeptItems{} items kept
    across the runs, \KeptRules{} are rules. Claude Code adds a rule to the
    context itself, whenever the agent reads a file that matches the rule's \cmd{paths}.
    Of a skill, only the description is listed to the agent; the body enters the context
    only if the agent decides to open it, and often it did not. A \emph{path-gated}
    candidate, a skill listed only once the agent works with matching files, was listed
    in \GatedOffered{} attempts at a task, in its own sweeps and in the gate test
    (\cref{sec:results:gates}), and opened in \GatedInvoked{}. An \emph{always-on}
    candidate, a skill listed on every turn, was listed in \AlwaysOnOffered{} attempts,
    likewise in its own sweeps and in the gate test, and opened in none of them.
  \item \textbf{Its value shrinks under a stronger model.} The harness of run
    \ModelChangeSource{}, which gained \HaikuGain{}~points under \RolloutModelName{}, the
    model it was evolved with, gained \SonnetGain{}~points under \StrongModelName{}
    (\CI{} [\SonnetLo, \SonnetHi]). One \cmd{/prune} pass after the upgrade, which ran
    each item's tasks \KConfirm{} times with the item and \KConfirm{} times without it,
    flagged \RemovableItems{} of its \DecidedItems{} items as removable.
\end{itemize}

We also report what did not work. Only \PositivesKept{} of \Positives{} known-good
candidates were kept. \cmd{/prune} reached the expected verdict on \PruneMatched{} of the
\PruneTested{} items it was tested on, but on only \PlantedMatched{} of the \Planted{}
\emph{planted} ones, items we added to a harness with a known right answer. It kept a
useless always-on skill, which the agent never opened in its own sweep, because
\PrunePlaceboLostTasks{} tasks passed less often when it was removed. And an always-on
skill that the agent never opens got no verdict where the test asks for
\emph{unmeasured}, the verdict for an item that never fired: \cmd{/prune}'s plan leaves
such an item out, since no sweep could delete it, so the item stayed without being
measured. Both results miss the margins of their
hypotheses (\cref{sec:results:gates,sec:results:model}). On a real open-source
repository the agent was rarely corrected, so nothing recurred and the loop learned
nothing. Finally, Cortex can consult
an optional \emph{judge}, an outside model that advises which candidates to try
(\cref{sec:design:boundary}). It was switched off during the runs; replayed on their
candidates afterwards, it did not anticipate the verdicts (\cref{sec:results:judge}).

\paragraph{Contributions.}
\begin{enumerate}
  \item \textbf{Exposure-aware measurement semantics} for harness items, where an item's
    \emph{exposure} is whether it entered the agent's context in a rollout: the same rule
    governs adding and deleting, so that both verdicts compare the harness with and
    without the item only on the tasks where it entered the context, a candidate that
    never fired is not kept, an item that never fired is not deleted, and unknown
    exposure blocks the verdict (\cref{sec:design}).
  \item \textbf{A known-answer test of the acceptance pipeline}: \Placebos{} placebo,
    \Harmful{} harmful and \Positives{} known-good candidates run through it. It shows that
    a candidate that never fires is never kept; it could not test harm, and the test's
    other weaknesses are reported (\cref{sec:results:gates}).
  \item \textbf{Evidence on loading tiers}: why an item the agent must choose fares
    differently under measurement from one that is simply loaded
    (\cref{sec:results:loop,sec:results:cost}).
  \item \textbf{A replicated evaluation} with a control run in which \cmd{/evolve} is off,
    an A/A arm (the same harness measured twice, to show how far two measurements of one
    harness can differ), ablations (other harnesses, each built to answer one question,
    such as the evolved harness with its \cmd{paths} removed), a model-change study and a
    real-repository attempt. Every number can be regenerated offline from the released
    data: rows exported from the runs' records and from Claude Code's session records
    (the records themselves are not released).
  \item \textbf{The open-source tool and its evaluation lab}, at \RepoURL: tag \EvalTag{}
    is the system as evaluated, and tag \PaperTag{} adds the lab's data and the scripts
    that regenerate every number in the paper.
\end{enumerate}

%% file: sections/02-background.tex
\section{Background}\label{sec:background}

\paragraph{The harness and its loading tiers.} Claude Code reads three kinds of text a
developer writes~\cite{claudecode-memory,claudecode-skills}. A \cmd{CLAUDE.md} at the root
of a repository holds project instructions and is read at the start of every session;
recent versions of Claude Code read an \cmd{AGENTS.md} in its place when a repository has
no \cmd{CLAUDE.md}, and tailoring a coding agent to a repository with such a context file
is widespread practice~\cite{gloaguen2026agentsmd}. A
\emph{rule} is a Markdown file in \cmd{.claude/rules/}. It may carry a \cmd{paths} field,
a list of file patterns such as \cmd{shop/billing/**} (every file under
\cmd{shop/billing/}). A \emph{path-less} rule, one without the field, is loaded at
launch, like \cmd{CLAUDE.md}; a \emph{path-scoped} rule enters the context when the agent
reads a file its patterns match, and, as Cortex's tests found on version
\MinClaudeVersion{}, when the request names such a file with an \cmd{@}-mention. A
\emph{skill} is a folder in \cmd{.claude/skills/} holding a
\cmd{SKILL.md}: a short description, then a body with the instructions. Only the
description is kept in context, in a listing of the available skills; the body is
loaded when the skill is \emph{invoked}: the agent opens it with a call to Claude Code's
Skill tool, or the user types its name as a command. A skill without \cmd{paths} is \emph{always-on}: its
description is listed on every turn. A \emph{path-gated} skill, one with \cmd{paths}, is
listed only once the agent works with matching files. Every description in the listing
costs context on every turn, whether or not its skill is ever
used~\cite{claudecode-skills}. \Cref{tab:tiers} summarises these five \emph{loading
tiers}. The
difference that matters most in this paper is between an item that is \emph{loaded} and
one that must be \emph{chosen}: Claude Code itself puts a rule into the context, but a
skill acts only if the agent, reading its description among the others, decides to open
it.

\begin{table}[!htb]
  \centering\small
  \caption{Loading tiers in Claude Code~\cite{claudecode-memory,claudecode-skills}. The
  rows for items with \cmd{paths} are as Cortex verified them on version
  \MinClaudeVersion, the oldest it accepts; the current documentation also counts writes
  and edits among the triggers of a path-scoped rule. The tier, not the text, decides how
  often an item is paid for, and whether it is \emph{loaded} (rules) or must be
  \emph{chosen} by the agent (skills).}
  \label{tab:tiers}
  \begin{tabular}{@{}lll@{}}
    \toprule
    Tier & What sits in context & When \\
    \midrule
    root \cmd{CLAUDE.md}        & the whole file              & every turn \\
    always-on skill             & its description; the body when invoked & every turn \\
    path-less rule              & the whole rule              & every turn \\
    path-gated skill            & its description; the body when invoked & after a matching file is read or written \\
    path-scoped rule            & the whole rule              & after a matching file is read or @-mentioned \\
    \bottomrule
  \end{tabular}
\end{table}

\paragraph{Tasks with a check.} Coding agents are evaluated on tasks cut from real
repositories: a codebase at a commit where something is broken, a request in natural
language, and tests that decide whether the agent's change fixes it. SWE-bench keeps only
task instances with at least one test that fails before the reference solution, the fix
the project's developers actually merged, and passes after it~\cite{jimenez2024swebench}; SWE-rebench extracts such tasks automatically and
continuously, so that models can be scored on fresh tasks they cannot have
seen~\cite{badertdinov2025swerebench}. Cortex keeps the format but takes its tasks from the
developer's own sessions rather than from public issues (\cref{sec:design:tasks}).

\paragraph{One run is not a measurement.} An agent's attempt at a task is a random draw:
the model samples its output, so two attempts at the same task can take different paths
through the repository, and one may pass where the other fails. A task therefore has a
pass \emph{rate}, and a single pass or fail estimates it poorly. Three studies show how
noisy one run is. On the retail tasks of $\tau$-bench, a GPT-4o agent, then the state of
the art, solved about three tasks in five on a single attempt, but solved the same task
in all eight of eight attempts less than a quarter of the time~\cite{yao2025taubench}.
In the published repeated trials of a coding-agent benchmark, fewer than one task in
five had a one-run result that reliably tracked an agent's score on the whole
benchmark~\cite{deltaselect2026}. When the skill files of three repositories were
optimised automatically, the larger of two optimisers' average gains, about five points,
could not be separated from the agent's run-to-run variance with the tasks one
repository supplies~\cite{skillissue2026}. Treating an evaluation as an experiment,
Miller recommends two remedies~\cite{miller2024errorbars}: sample each question more
than once, so that the luck of single attempts averages out; and compare two systems
through their paired, question-level differences rather than their two overall scores,
so that what makes a question easy or hard for both systems cancels. Cortex does both:
it runs every task $k$ times under the live harness and $k$ times under the candidate,
and compares the two task by task (\cref{sec:design:measure}).

%% file: sections/03-design.tex
\section{Cortex}\label{sec:design}

Cortex changes only text: the files that make up a Claude Code harness. The model is
fixed. This section describes Cortex as it stands at the git tag \EvalTag{}, a named
snapshot of its code, with that snapshot's default settings. Every experiment of
\cref{sec:setup} was made on this version, except the development run D0, which used an
earlier one.
No run changed a gate's threshold, so the rules that decide whether an item is kept are
the same in every experiment. The runs changed a few other settings, none of which
decides a verdict; \cref{sec:setup} lists them.

\Cref{sec:design:workflow} follows the three commands from start to finish. The
subsections after it give the reasons for each step: what an item costs
(\cref{sec:design:tiers}), where the tasks come from (\cref{sec:design:tasks}), how a
change is measured (\cref{sec:design:measure}), when a new item is kept
(\cref{sec:design:gates}), and when an old one is deleted (\cref{sec:design:prune}).
\Cref{tab:terms} collects the terms they use, for reference.

\begin{figure}[!t]
  \captionsetup{type=table}
  \centering\small
  \caption{The terms this paper uses for Cortex, in roughly the order a reader meets
  them, with the subsection that explains each in full.}
  \label{tab:terms}
  \begin{tabular}{@{}>{\raggedright\arraybackslash}p{\dimexpr\linewidth/6\relax}>{\raggedright\arraybackslash}p{\dimexpr\linewidth*2/3+1em\relax}l@{}}
    \toprule
    Term & Meaning & See \\
    \midrule
    harness & the text files that steer the agent: \cmd{CLAUDE.md}, skills and rules & \cref{sec:background} \\
    item & a skill or a rule, the harness files Cortex measures & \cref{sec:design:tiers} \\
    live harness & the harness the developer is using now & \cref{sec:design:measure} \\
    loading tier & when an item's text enters the agent's context (\cref{tab:tiers}) & \cref{sec:background} \\
    task & an episode \cmd{/harvest} saved so it can be replayed: a broken commit, the request, a check and the fix & \cref{sec:design:tasks} \\
    check & a script that exits zero when an attempt solved the task & \cref{sec:design:tasks} \\
    quarantine & set aside a task whose check no longer tells the broken state from the fixed one, or needs the fix's own test code & \cref{sec:design:tasks} \\
    lesson & one logged line for each correction the developer made & \cref{sec:design:workflow} \\
    theme & a problem that recurs in the lessons or the failing tasks & \cref{sec:design:gates} \\
    candidate & a new skill or rule under test, not yet live & \cref{sec:design:gates} \\
    rollout & one attempt by the agent at one task, in a sandbox clone of the repository & \cref{sec:design:measure} \\
    sweep & a batch of rollouts that runs each of a set of tasks $k$ times under each of two harnesses & \cref{sec:design:measure} \\
    run (of a task) & one of the $k$ rollouts of a task under one arm of a sweep; not to be confused with an evaluation run, R1 to R4 (\cref{sec:setup}) & \cref{sec:design:measure} \\
    pinned model & the one model that every rollout uses, when the developer names it in Cortex's configuration & \cref{sec:design:measure} \\
    \emph{base}, \emph{cand} & the two arms of a sweep: the live harness, and the live harness with the change under test & \cref{sec:design:measure} \\
    exposed, fire & a task is \emph{exposed} to an item when the item reached the context in one of its rollouts (a rule loaded, a skill at least listed); an item \emph{fires} when it is used (a rule loaded, a skill opened) & \cref{sec:design:measure} \\
    cycle & one iteration of \cmd{/evolve}: find a theme, write one candidate, measure it & \cref{sec:design:gates} \\
    screen, confirm & the cheap first sweep, on the theme's failing tasks; the stricter second one, on every task & \cref{sec:design:gates} \\
    gate & one condition a candidate must meet to be kept, computed by the scoring script \cmd{score.sh} (\cref{tab:gates}); gates 2 and 3, which stop a candidate that makes a task worse, are the \emph{regression gates} & \cref{sec:design:gates} \\
    recheck (RECHECK) & a repeat sweep of the tasks that dropped, when a candidate failed only the regression gates, before its KILL is final & \cref{sec:design:gates} \\
    KEEP, KILL & the candidate goes live; the candidate is buried & \cref{sec:design:gates} \\
    BARREN & a cycle that found no theme strong enough and ran no rollouts & \cref{sec:design:gates} \\
    RERUN & the sweep cannot decide, because its data cannot be trusted or none of the tasks the candidate reached could show a gain; it is repeated and nothing is decided & \cref{sec:design:gates} \\
    ACCEPT, REJECT, UNMEASURED & \cmd{/prune}'s verdicts: delete the item; keep it, because removing it cost something; keep it, because it never fired, so its removal tested nothing & \cref{sec:design:prune} \\
    graveyard & the folder of buried items, each with the reason it was buried & \cref{sec:design:gates} \\
    journal & the file in which Cortex records every cycle and verdict & \cref{sec:design:workflow} \\
    \bottomrule
  \end{tabular}
\end{figure}

\subsection{The three commands, step by step}\label{sec:design:workflow}

\begin{figure}[tbp]
  \centering
  \includegraphics[width=\linewidth]{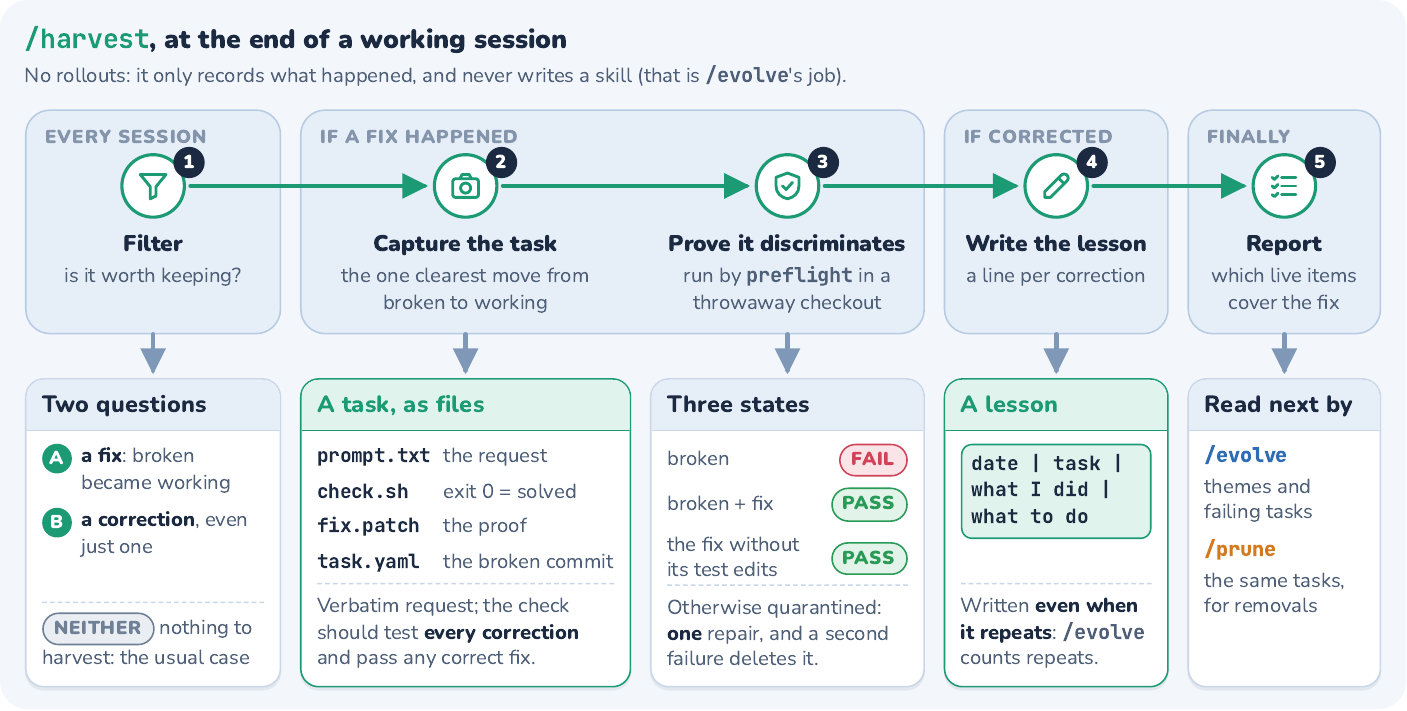}
  \caption{What \cmd{/harvest} does at the end of a working session. It asks two
  questions about the session (step 1). \textbf{A}: did something go from broken to
  working? \textbf{B}: did the developer correct the agent, even once? In most sessions
  the answer to both is no, and nothing is recorded. If the answer to A is yes, the
  command saves that change as a \emph{task} that can be replayed later (step 2): the
  developer's request, word for word; the commit where the code was still broken; the fix
  that made it work; and a check, a short script that says whether an attempt solved the
  problem. Before the task is kept, \cmd{preflight} tests the check itself in a throwaway
  copy of the repository (step 3). The check must fail on the broken code, pass once the
  fix is applied, and still pass when the fix's changes to test code are left out, so
  that an agent is judged on the code it writes, not on reproducing the developer's own
  test. A task that fails any of the three is set aside and gets one repair; if it fails
  again, it is deleted (\cref{sec:design:tasks}). If the answer to B is yes, the command
  writes one line per correction to a log of \emph{lessons}: the date, the task, what the
  agent did, and what it should have done (step 4). A lesson is written even when the same
  one is already there, because \cmd{/evolve} counts the repeats to find the problems
  that recur. Finally, the command reports which items of the current harness already
  cover the files the fix touched (step 5). \cmd{/harvest} only records: it never runs the
  agent on a task and never writes a skill or rule, which is \cmd{/evolve}'s job. The
  tasks it saves are used next by \cmd{/evolve}, to measure new candidates, and by
  \cmd{/prune}, to measure what removing an item costs.}
  \label{fig:harvest}
\end{figure}

\begin{figure}[tbp]
  \centering
  \includegraphics[width=\linewidth]{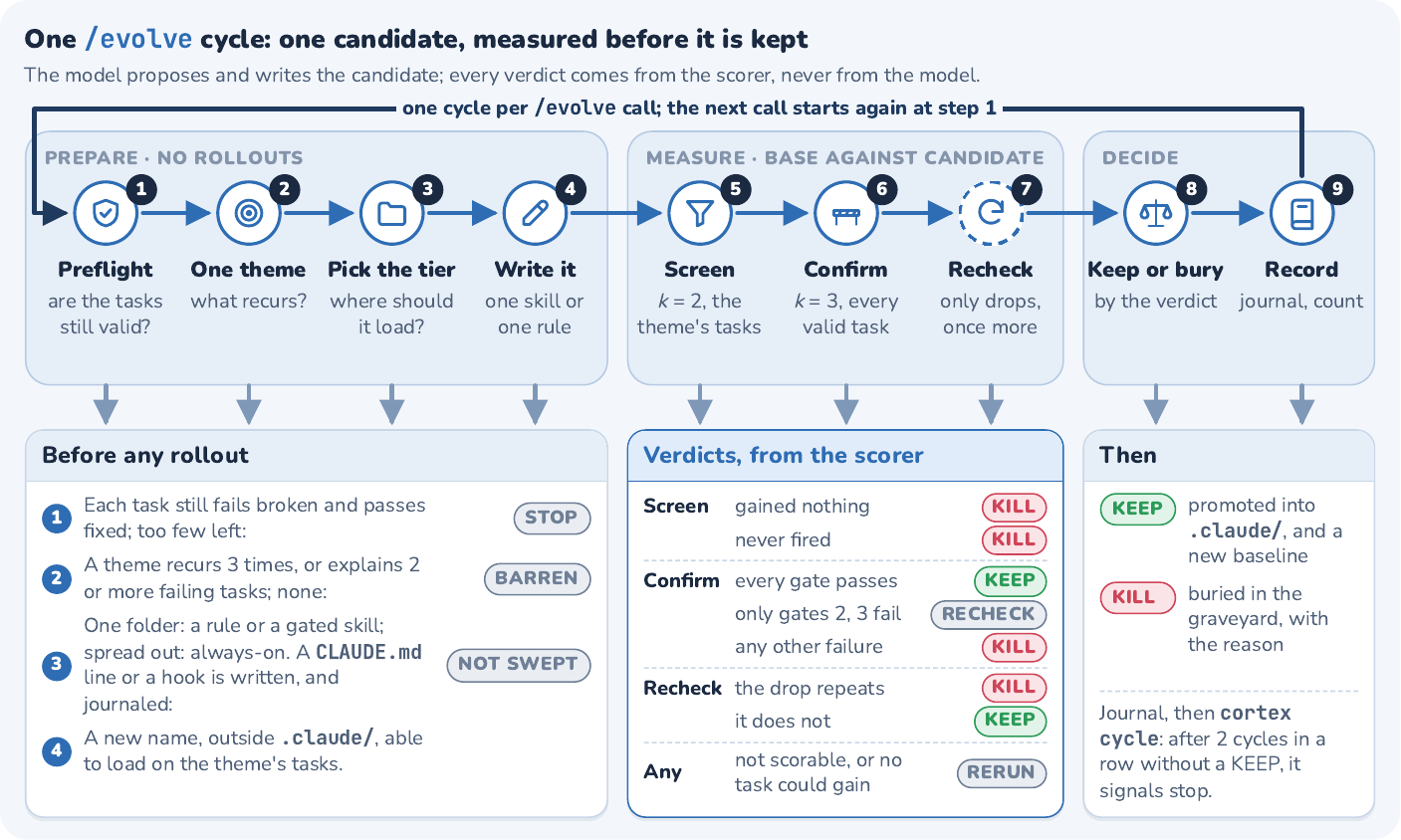}
  \caption{One \cmd{/evolve} cycle, which tests a single new skill or rule before it may
  join the harness. The model proposes and writes the candidate, but never decides its
  fate: every verdict comes from a scoring script (\cref{tab:gates}). Steps 1 to 4 prepare
  the cycle and run the agent on no task; the first three can end the cycle early (grey
  badges). Step 1 re-runs \cmd{preflight} to make sure the saved tasks still tell broken
  code from fixed code; if too few are left, the cycle stops. Step 2 looks for one
  \emph{theme}, a problem that recurs: a lesson logged at least \MinThemeOccurrences{}
  times, or a cause behind \MinThemeFailingTasks{} or more failing tasks. If none clears
  that bar, the cycle ends as BARREN and runs no rollouts. Step 3 chooses where the candidate should load: if the
  fixes behind the theme sit in one folder, it becomes a rule or a skill tied to that
  folder; if they are spread across the repository, it becomes a skill offered on every
  turn. A fact the agent breaks during unrelated work becomes a line in \cmd{CLAUDE.md},
  and a check that must always run becomes a \emph{hook}, a command Claude Code runs by
  itself at a fixed point of a session; such a change is written and recorded in the
  \emph{journal}, Cortex's running log of every cycle and verdict
  (\cmd{.evolve/journal.md}), but not measured. Step 4 writes the candidate under a new name, outside the
  live harness. Steps 5 to 7 measure it: the agent attempts each task with the live
  harness (\emph{base}) and with the harness plus the candidate (\emph{cand}), each attempt
  in its worker's own clone of the repository, reset and verified before every attempt
  (\cref{fig:sandbox}). The \emph{screen} (step 5) is cheap, \KScreen{} runs per task on
  the theme's tasks only; a candidate that gained nothing, or that never fired (a rule
  never loaded, a skill never opened), is killed here. The \emph{confirm} (step 6) is
  stricter, \KConfirm{} runs per task on every valid task, so that it also sees what the
  candidate might break. The candidate is kept only if it wins overall, no task collapses,
  no task that never failed starts failing, and it fired at least once. Because one
  unlucky run can look like a broken task, a candidate that failed only on such a drop
  gets a \emph{recheck} (step 7, drawn dashed): the tasks that dropped are run again, and
  the candidate is killed only if the drop repeats. A sweep whose data cannot be trusted,
  with runs missing or invalid, or in which no task the candidate reached could show a
  gain, is repeated rather than decided (RERUN). Step 8 acts on the verdict: a KEEP moves
  the candidate into the live harness and records the confirm sweep's scores as the new
  baseline, with no new rollouts; a KILL buries it in the \emph{graveyard}, the folder of rejected ideas
  (\cmd{.evolve/graveyard/}), with its reason, so the same idea is not
  proposed again unchanged. Step 9 records the cycle in the journal; after
  \StopAfterBarren{} cycles in a row without a KEEP, \cmd{cortex cycle} signals the loop
  to stop. Badge
  colours follow \cref{fig:loop}: green, the item stays; red, it goes; grey, no decision.}
  \label{fig:evolve}
\end{figure}

\begin{figure}[tbp]
  \centering
  \includegraphics[width=\linewidth]{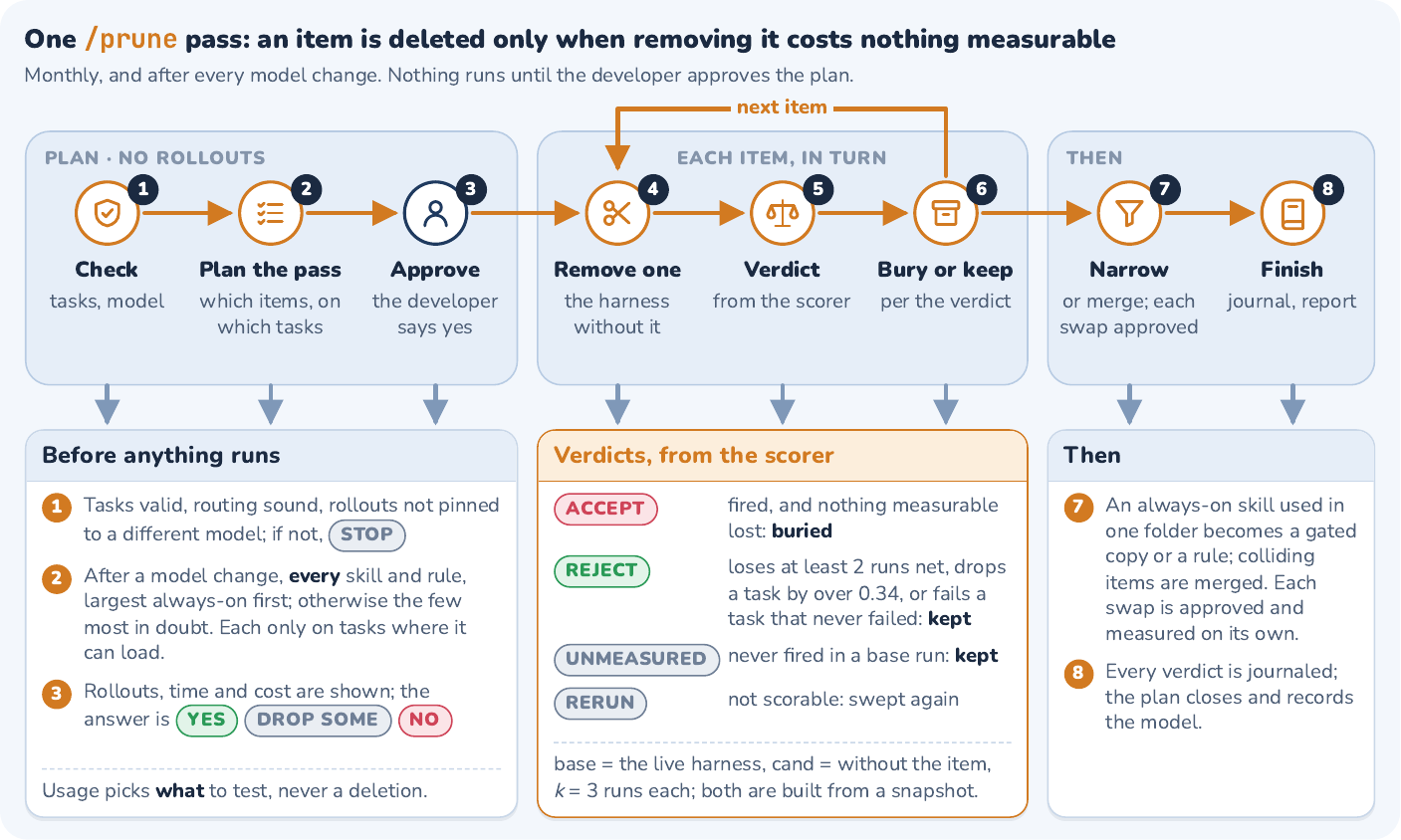}
  \caption{One \cmd{/prune} pass, which tests the items already in the harness and deletes
  one only when removing it costs nothing measurable. It is meant to run monthly and after every
  change of model, since a stronger model may already do what an item teaches. Steps 1 to
  3 plan the pass and run the agent on no task. Step 1 checks that the tasks are still
  valid, that the routing is sound (no name is both a skill and a rule, no symlink leaves
  the repository or dangles), and that the rollouts are not pinned to a model other than
  the one in use; if any check fails, the pass stops. Step 2 chooses which items to test:
  after a model change, every skill and rule, the largest always-on items first;
  otherwise, the few most in doubt. Each item is tested only on the tasks where it can
  load; an always-on skill, only on the tasks where earlier sweeps saw it fire, and not at
  all if they never did. How often an item
  is used helps choose what to test, but never decides a deletion. Step 3 shows the
  developer the plan with its expected rollouts, time and cost, and nothing runs until the
  developer approves it, possibly dropping some items. Steps 4 to 6 then take one item at
  a time: the agent attempts each of its tasks \KConfirm{} times with the live harness
  (\emph{base}) and without the item (\emph{cand}), and the scoring script gives the
  verdict. ACCEPT: the item fired in at least one base run, and removing it lost nothing
  measurable, so it is deleted and buried in the graveyard. REJECT: removing it lost too many runs, made a task
  collapse, or made a task that never failed start failing, so it stays. UNMEASURED: the
  item never fired in any base run, so the test says nothing about it, and it stays; an
  item for a rare situation looks exactly like dead weight until that situation comes.
  RERUN: the data cannot be trusted, and the sweep is repeated. A deleted item is removed
  before the next one is measured, so every test starts from the harness as it really is.
  Step 7 can also move advice to a better place: an always-on skill used in only one
  folder can become a rule or a skill tied to that folder, and overlapping items can be
  merged; each such change is approved and measured on its own. Step 8 records every
  verdict in the journal and closes the plan, noting the model it ran on
  (\cref{sec:design:prune}).}
  \label{fig:prune}
\end{figure}

Cortex consists of three Claude Code commands that the developer runs, each with a
lifecycle of its own, joined into one loop (\cref{fig:loop}). The commands are prompts that the model
follows; for every step that must be exact (running a sweep, scoring it, recording a
cycle), they call a command-line program, \cmd{cortex}, whose subcommands, such as
\cmd{cortex cycle} or \cmd{cortex scope}, appear below. \cmd{/harvest} runs at the end of a
working session and turns what happened into evidence: a replayable task when something
went from broken to working, and a one-line lesson for each correction
(\cref{fig:harvest}). \cmd{/evolve} runs one cycle at a time: it looks in that evidence
for one recurring problem, writes one skill or rule for it, and measures it against the
live harness before it is kept (\cref{fig:evolve}). \cmd{/prune} is meant to run monthly
and after every model change: it measures what removing each item of its plan costs, one
item at a time, and deletes only what costs nothing measurable (\cref{fig:prune}). The three figures
follow each command from start to finish, including every point where it stops early; the
rest of this section gives the reasons for each step.

\paragraph{The three commands as algorithms.} \Cref{alg:harvest,alg:cycle,alg:prune} give
the same three commands as short pseudocode, in the order a developer meets them.
\cmd{/harvest} (\cref{alg:harvest}) only records: it turns a session into a task, lessons,
or nothing, and never runs the agent. \cmd{/evolve} (\cref{alg:cycle}) adds: it picks one
recurring problem, writes one skill or rule for it, and keeps it only if a measurement
against the live harness says so. \cmd{/prune} (\cref{alg:prune}) removes: with the
developer's approval, it measures what taking each item out costs and deletes only the
items whose removal costs nothing measurable. The scores and thresholds they use ($G$, $N$, $D$, $f$,
$P$, $m$, $\tau$) are defined in \cref{sec:design:measure,sec:design:gates}.

\begin{algorithm}[tbp]
  \caption{One \cmd{/harvest}. It runs no rollouts and writes no skill or
  rule: it records a task, lessons, or nothing.}
  \label{alg:harvest}
  \small
  \begin{algorithmic}[1]
    \Require the session that just ended; the repository, with the fix still in place
    \State $A \gets$ something went from broken to working during the session
    \State $B \gets$ the developer corrected the agent at least once
    \If{neither $A$ nor $B$}
      \State report ``nothing to harvest'' and stop \Comment{the usual case}
    \EndIf
    \If{$A$}
      \State $t \gets$ a task from the clearest move from broken to working: the broken
        commit, the fix, the request verbatim, and a check
      \State preflight every task, $t$ included, in a throwaway checkout: fail broken,
        pass fixed, pass fixed minus its test edits \Comment{\cref{sec:design:tasks}}
      \If{$t$ is quarantined}
        \State repair the check once and preflight $t$ again; delete $t$ if it is
          quarantined again
      \EndIf
    \EndIf
    \If{$B$}
      \State append one lesson per correction, even one already logged
        \Comment{\cmd{/evolve} counts repeats}
    \EndIf
    \State report which live items cover the files the fix touched
  \end{algorithmic}
\end{algorithm}

\begin{algorithm}[tbp]
  \caption{One \cmd{/evolve} cycle. The model writes the candidate; every
  verdict comes from \cmd{score.sh}. A sweep that is not scorable, or in which the
  candidate reached tasks but none could show a gain, is repeated, never decided.}
  \label{alg:cycle}
  \small
  \begin{algorithmic}[1]
    \Require live harness $H$; task store; logged corrections; graveyard
    \State quarantine every task whose check no longer separates broken from fixed, or
      needs the fix's own test code \Comment{\cref{sec:design:tasks}}
    \If{too few valid tasks remain, or the routing check finds an error (a name used by
      both a skill and a rule, or a symlink that leaves the repository or dangles)}
      \State stop \Comment{no rollouts}
    \EndIf
    \State $\theta \gets$ a theme logged at least \MinThemeOccurrences{} times, or behind
      \MinThemeFailingTasks{} or more failing tasks, that no live item covers (a buried idea
      may return only revised, under a new name)
    \If{there is no such $\theta$}
      \State record the cycle as BARREN and stop \Comment{no rollouts}
    \EndIf
    \State $x \gets$ one skill or rule for $\theta$, in the tier its fixes point to
      \Comment{\cref{tab:tiers}}
    \State screen: sweep $H$ against $H + x$ on the tasks behind $\theta$, $k = \KScreen$;
      repeat it while it is not scorable, or while $x$ reached some task but none could
      show a gain \Comment{RERUN}
    \If{$G = 0$ or $f = 0$}
      \State bury $x$ and stop \Comment{no confirm is paid for}
    \EndIf
    \State confirm: sweep $H$ against $H + x$ on every valid task, $k = \KConfirm$
    \State $v \gets$ the gates of \cref{tab:gates}
    \If{$v = \text{RECHECK}$}
      \State sweep the regressed tasks again; $v \gets$ KILL if one regresses again, else KEEP
    \EndIf
    \If{$v = \text{KEEP}$}
      \State make $x$ live
    \ElsIf{$v = \text{KILL}$}
      \State bury $x$ with the gates it failed
    \EndIf
  \end{algorithmic}
\end{algorithm}

\subsection{Harness items and what they cost}\label{sec:design:tiers}

Cortex calls each skill or rule a \emph{harness item}; the harness also holds constant
files such as \cmd{CLAUDE.md}, which Cortex never measures. When a file enters the
context decides what it costs (\cref{tab:tiers}): its price is its length multiplied by
how often it is present, so Cortex counts the characters present on every turn
(\cmd{CLAUDE.md}, always-on skill descriptions and path-less rules) and warns above a
budget (\AlwaysOnBudget{} characters by default). Its objective is informally
$\text{pass rate} - \lambda\cdot\text{always-on context}$, where $\lambda$ is the pass
rate a developer would give up to save one character read on every turn. Cortex
never computes this as one number. It enforces it by two rules on the items it measures,
skills and rules: add one only if it raises the pass rate (\cref{sec:design:gates});
remove or narrow one only if that does not lower it (\cref{sec:design:prune}). When
\cmd{/evolve} decides that a theme belongs in \cmd{CLAUDE.md}, a setting, a hook or a
tool rather than in a skill or rule, it writes the change and records it in the journal,
but does not measure it: the gates apply only to skills and rules. The idea of context cost as
an objective is shared with Meta-Harness and SkillReducer~\cite{metaharness2026,skillreducer2026}.
Whether an item is loaded or must be chosen also shaped what survived in our evaluation
(\cref{fig:loaded}).

\subsection{Tasks from the developer's own sessions}\label{sec:design:tasks}

A task is a replayable episode: a broken commit (\cmd{base\_sha}), the developer's
original request, a \cmd{check.sh} that exits zero when the task is solved, and the
\cmd{fix.patch} that proves it can be. \cmd{/harvest} writes one at the end of a session,
and only when something went from broken to working. The request is the developer's
original words, not a cleaned-up version with hints known only afterwards. \cmd{/harvest}
is instructed to write a check that tests not only that the bug is fixed but also each
correction the developer made during the session (for example, that the linter is
clean), so that an item which prevents the corrected mistake can turn a fail into a pass,
and show an effect; preflight confirms that the check separates broken from fixed, not
that it covers every correction. Most sessions produce no task, by design.

As in the benchmarks of \cref{sec:background}, a task is only a ruler if its check
separates the broken state from the fixed one.
\cmd{preflight} runs the check in three states before every cycle and every prune pass: the broken
state must \emph{fail}; the broken state plus the fix must \emph{pass}; and, when the fix
edits test code, the fix \emph{without} those edits must still pass (test data it adds,
such as a fixture, stays). Write $\sigma_t$ for the broken commit, $\delta_t$ for the fix,
$\delta_t^{\mathrm{test}}$ for the fix's edits to test code, $\sigma \oplus \delta$ for
the commit $\sigma$ with the edits $\delta$ applied, and $c_t(\cdot)$ for the result of
the check: \emph{pass} when it exits zero, \emph{fail} when it exits with any other code,
and \emph{hang} when it does not finish within its time limit. A task is valid when
\begin{equation}\label{eq:preflight}
  \operatorname{valid}(t) \iff
  c_t(\sigma_t) = \text{fail} \;\wedge\; c_t(\sigma_t \oplus \delta_t) = \text{pass} \;\wedge\;
  c_t\bigl(\sigma_t \oplus (\delta_t \smallsetminus \delta_t^{\mathrm{test}})\bigr) = \text{pass},
\end{equation}
so a check that hangs fails preflight in any state.
When the fix edits no test code, $\delta_t^{\mathrm{test}}$ is empty and the third
condition repeats the second. The third state exists because a developer often writes a
test while fixing a bug. If the check depended on that test, an agent replaying the task
would have to write the same test again, which a replaying agent has no reason to do,
and the task would fail in both arms forever. A task that fails any state is
\emph{quarantined}: it is set aside and left out of every sweep. One whose environment is
down, as reported by an optional script that checks its services, is skipped instead,
since an outage is not a bad task.

\subsection{Measurement: against the live harness, with exposure}\label{sec:design:measure}

\begin{figure}[tbp]
  \centering
  \includegraphics[width=\linewidth]{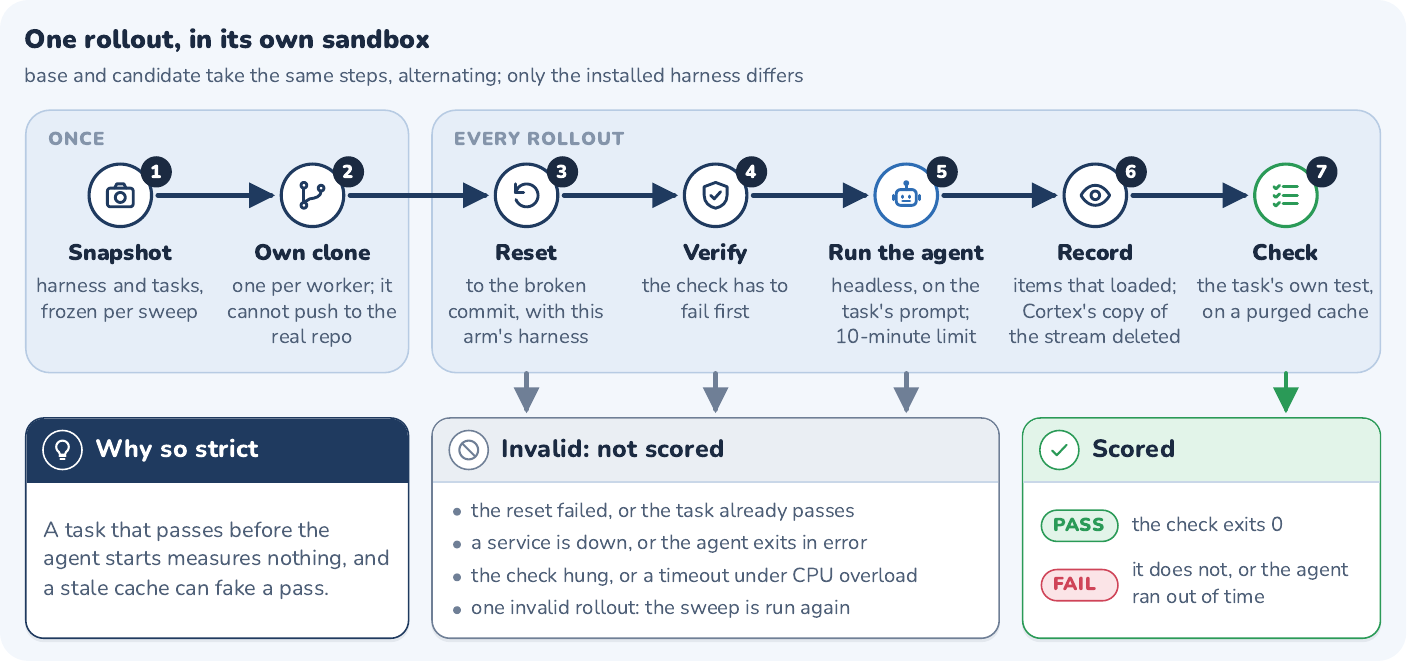}
  \caption{What happens in one rollout, a single attempt by the agent at a task. The
  \emph{base} and \emph{cand} arms take exactly the same steps, in alternation, and differ
  only in the harness that is installed. Once per sweep, Cortex freezes a copy of the
  harness and of the tasks (step 1), so that an edit made while the sweep runs cannot
  reach it, and gives each worker its own clone of the repository, which cannot push to
  the real one (step 2). Then, before every rollout, the clone is reset to the task's
  broken commit and that arm's harness is installed (step 3), and the task's check is run
  once to confirm that it still fails (step 4): a task that already passes before the
  agent starts would measure nothing. The agent then works alone, with no human, on the
  task's request, within a limit of \RolloutTimeoutMin{} minutes (step 5). Cortex records
  which skills and rules entered the agent's context during the attempt (step 6). Finally,
  the task's own check decides the result, after the cached build files are deleted,
  since a stale cache could make old code look as if it passes (step 7). Each attempt ends
  in one of two ways. It is \emph{scored}: a pass if the check succeeds, a fail if it does
  not or the agent ran out of time. Or it is \emph{invalid}, because it could not run
  fairly: the reset failed, the task already passed, a service the task needs was down,
  the agent exited with an error (for example a login failure or a rate limit), the check
  hung, or the agent ran out of time while the machine was oversubscribed. Invalid is not the same as fail, and a sweep with even
  one invalid rollout is run again rather than scored.}
  \label{fig:sandbox}
\end{figure}

Every decision rests on a \emph{sweep}: the agent runs each task $k$ times under two
harnesses. \emph{base} is the harness as it is now, re-measured in every sweep; \emph{cand}
is the same harness plus the one change under test (or minus it, for a removal). Both arms
load every existing item, so what is measured is the change's \emph{marginal} value: a
near-duplicate of an item the developer already has correctly shows no gain. The two
arms' rollouts are interleaved (for each task and each run, \emph{base} and then
\emph{cand}), so a slowdown, or a new model deployment in the middle of a sweep, lands on
both arms alike. The harness and task definitions are snapshotted at the start. Rollouts
run in parallel, one per \emph{worker}, and each worker has its own clone of the
repository (\cref{fig:sandbox}), reset before every rollout to the task's broken commit
with that arm's harness; every reset is verified by running the check, which must fail.
The clone never has Cortex's own folder (\cmd{.evolve/}) on disk: that folder holds the
\emph{task store}, where Cortex keeps its saved tasks with every earlier task's fix, and the lessons
and graveyard, which would give the answer away, so the clone is made without it (a
sparse checkout, which leaves chosen folders out) and every reset deletes it again. It is
still reachable through the clone's git history, which is not pruned;
\cref{sec:limitations} discusses that route.

\enlargethispage{\baselineskip}
Finally, the model that plays the agent can be fixed: by default each rollout uses whatever model Claude Code is set to, but
the developer can name one model in Cortex's configuration (\emph{pin} it), so that every
rollout of every sweep uses the same one.

\begin{algorithm}[H]
  \caption{One \cmd{/prune} pass. Nothing runs until the
  developer approves the plan, and every verdict comes from \cmd{score.sh}. Here
  \emph{base} is the live harness and \emph{cand} the harness without the item, so a
  negative $N$ means the item was earning its place.}
  \label{alg:prune}
  \small
  \begin{algorithmic}[1]
    \Require live harness $H$; task store; each item's use in recent sessions; the model
      of the last pass
    \State quarantine every task whose check no longer separates broken from fixed, or
      needs the fix's own test code \Comment{\cref{sec:design:tasks}}
    \If{too few valid tasks remain, the routing check finds an error (a name used by both
      a skill and a rule, or a symlink that leaves the repository or dangles), or the
      rollouts are pinned to a model other than the one in use}
      \State stop \Comment{no rollouts}
    \EndIf
    \If{the pinned model changed since the last pass}
      \State $X \gets$ every skill and rule, the largest always-on first
    \Else
      \State $X \gets$ the few items most in doubt \Comment{use chooses, never deletes}
    \EndIf
    \State drop from $X$ every item no task can load, and every always-on skill that never
      fired in an earlier sweep \Comment{no sweep could delete it}
    \State show the plan, with its rollouts, time and cost; stop unless the developer
      approves it
    \For{each $x \in X$, in order}
      \State sweep $H$ against $H - x$ on the tasks where $x$ can load (an always-on
        skill: where it fired before), $k = \KConfirm$
      \If{the sweep is not scorable}
        \State sweep again \Comment{RERUN: nothing is decided}
      \ElsIf{$kN \le -m$, or $D > \tau$, or a task in $P$ loses a run}
        \State keep $x$ \Comment{REJECT: removing it costs something}
      \ElsIf{$f = 0$}
        \State keep $x$ \Comment{UNMEASURED: it never fired}
      \Else
        \State bury $x$; $H \gets H - x$ \Comment{ACCEPT, before the next item}
      \EndIf
    \EndFor
    \State narrow always-on items, or merge overlapping ones, each swap approved and swept
      on its own
    \State journal every verdict; close the plan and record the model
  \end{algorithmic}
\end{algorithm}

\needspace{4\baselineskip}
\paragraph{Exposure.} An item can change a result only in a rollout where it entered the
agent's context. Cortex reads each rollout's \emph{event stream}, the record of every
message and tool call that Claude Code emits while the agent works. From it, Cortex
records for each skill whether it was \emph{visible} (listed to the agent) and whether it
was \emph{invoked} (opened with a Skill tool call). A rule leaves no trace of its own, so
Cortex records that it \emph{fired} when the agent read a file its \cmd{paths} match, or
the prompt @-mentioned one. When the stream cannot be read the value is \emph{unknown},
never ``did not fire''. A task is \emph{exposed} to an item if the item reached the
context in any of its rollouts: a rule that loaded, or a skill that was at least listed,
since its description then sits in the context. An always-on skill is listed on every
turn, so every task is exposed to it. On a task that was not exposed, the agent in both
arms read the same text, and Cortex treats any difference between the arms as chance.
Whether the item \emph{fired}, a rule loaded or a skill opened, is a stricter question,
which gate~5 asks (\cref{sec:design:gates}). This one principle drives four decisions:
the gates read only exposed tasks; a candidate that never fired is never kept; a removal
whose item never fired is never accepted; and unknown exposure in any valid rollout
blocks the verdict, so that a tooling failure cannot look like dead weight. Crediting only
what fired is shared with SkillMAS~\cite{skillmas2026}, and it guards against inert edits
such as the never-triggered hook that DemoEvolve's audit found in a harness chosen by
self-rollout search~\cite{demoevolve2026}. It has been applied to deletion as well: SLIM
retires a skill on a comparison with and without it over the validation tasks on which it
was loaded, and only once it has been loaded often enough~\cite{slim2026}, while
Bayesian-Agent and Ratchet retire a skill only on enough evidence from the runs that used
it, scoring the skill's own outcomes rather than the difference its removal
makes~\cite{bayesianagent2026,ratchet2026}. Cortex applies the same rule to adding and to
deleting, and lets unknown exposure block either verdict.

\paragraph{Scores.} Inside a sweep, each of the $k$ rollouts of a task under one arm is
called a \emph{run} of that task; this short sense of ``run'' is unrelated to the
evaluation runs R1 to R4 of \cref{sec:setup}. Let $k$ be the runs per task and arm, and $p_b(t)$ and $p_c(t)$ the
fraction of the valid rollouts of task $t$ that passed under \emph{base} and \emph{cand}.
A task is \emph{measured} when both arms have $k$ valid rollouts on it, and
\emph{exposed} when the item under test was in context in at least one of them: the
candidate in a \emph{cand} rollout or, for a removal, the removed item in a \emph{base}
rollout. Write $n_b(t)$ and $n_c(t)$ for the numbers of valid rollouts of task $t$ under
each arm; $\Omega^{\star}_t$ for its valid rollouts in the arm where the item $x$ under
test exists (\emph{cand} for a candidate, \emph{base} for an item being removed); and,
for a rollout $\omega$, $\operatorname{ctx}(\omega)$ for the items that were in the agent's
context and $\operatorname{fired}(\omega)$ for those that fired, both read from its event
stream. The measured tasks $\mathcal{M}$, the exposed tasks $\mathcal{X}$, the set $S$ of
tasks that are both, and the count $f$ of rollouts in which the item fired, which gate~5
reads (\cref{sec:design:gates}), are
\begin{equation}\label{eq:exposure}
  \begin{gathered}
  \mathcal{M} = \bigl\{t : \min\bigl(n_b(t), n_c(t)\bigr) \ge k\bigr\}, \qquad
  \mathcal{X} = \bigl\{t : \exists\, \omega \in \Omega^{\star}_t,\; x \in \operatorname{ctx}(\omega)\bigr\},\\
  S = \mathcal{M} \cap \mathcal{X}, \qquad
  f = \sum_{t} \sum_{\omega \in \Omega^{\star}_t} \mathbb{1}\bigl[x \in \operatorname{fired}(\omega)\bigr],
  \end{gathered}
\end{equation}
where $\mathbb{1}[\cdot]$ is one when its condition holds and zero otherwise, and $t$ in
the sum runs over every task of the sweep. On $S$, the per-task change and the three
scores are
\begin{equation}\label{eq:scores}
  \Delta_t = p_c(t) - p_b(t), \qquad
  G = \sum_{t \in S} \max(\Delta_t, 0), \qquad
  N = \sum_{t \in S} \Delta_t, \qquad
  D = \max_{t \in S} \max(-\Delta_t, 0),
\end{equation}
the \emph{gain}, the \emph{net} change and the \emph{worst drop}. Each answers a
different question about the change.
\begin{itemize}
  \item $\Delta_t$, the \emph{per-task change}, is how much more often task $t$ passed
    with the change than without it. It is $+1$ when the candidate passed every run and
    the base none, $-1$ in the opposite case, and $0$ when both arms did equally well; in
    between it moves in steps of $1/k$, one run at a time. The other scores are built
    from it.
  \item $G$, the \emph{gain}, adds the tasks that improved and ignores those that got
    worse: it is everything the change won, whatever it lost. A candidate with $G = 0$
    helped on no task, so there is nothing to set against its cost, and the screen
    stops it before the confirm is paid for (or is rerun, if it reached tasks but none
    could show a gain).
  \item $N$, the \emph{net} change, adds every $\Delta_t$, so a loss on one task cancels
    a win on another: it says whether the change is still ahead once what it broke is
    counted. Every $\Delta_t$ is a multiple of $1/k$, so $kN$ is a whole number, the runs
    won net of those lost. The gates ask for $kN \ge m = \MinNetRuns$, because with
    $k = \KConfirm$ runs per arm a net of one run can come from luck alone.
  \item $G - N$, the \emph{regression}, is what the change cost: taking the net away from
    the gain leaves only the losses.
  \item $D$, the \emph{worst drop}, is the largest loss on any single task, and zero when
    $S$ is empty. A net that is comfortably positive can hide one task the change broke;
    $D$ exposes it.
\end{itemize}
For a removal, \emph{cand} is the harness without the item, so the same scores read the
other way: a negative $N$ says the item was earning its place (\cref{sec:design:prune}).

$\Delta_t$, $N$ and $G - N$ are DarwinX's per-task change, net gain and bounded
regression~\cite{darwinx2026}. The gates below use them differently in two ways. They
keep DarwinX's requirement that a change improve some task, and raise its requirement of
a positive net to a net of at least $m$ runs, since luck alone can win a single run. And
they bound the single worst drop rather than the summed regression, because
the sum cannot tell a broken task from scattered noise. Take $k = \KConfirm$. One task
whose pass rate drops from $2/k$ to $0$, and two tasks whose pass rates each drop from
$2/k$ to $1/k$, both give $G - N = 2/k$, but only the first is a task the change broke. The first fails gate 2 ($D = 2/k$); the second passes it
($D = 1/k$), and its two lost runs still count against $kN$. In vector form, with
$\Delta^{+}_t = \max(\Delta_t, 0)$ the gains and $\Delta^{-}_t = \max(-\Delta_t, 0)$ the
losses over $t \in S$, the gain and the regression are $\ell_1$ norms and the worst drop
is an $\ell_\infty$ norm:
\begin{equation}\label{eq:norms}
  G = \bigl\lVert \Delta^{+} \bigr\rVert_{1}, \qquad
  G - N = \bigl\lVert \Delta^{-} \bigr\rVert_{1}, \qquad
  D = \bigl\lVert \Delta^{-} \bigr\rVert_{\infty}, \qquad
  D \;\le\; G - N \;\le\; \bigl|\{t \in S : \Delta_t < 0\}\bigr| \cdot D.
\end{equation}
The same regression can therefore come from one task that lost $G - N$, where it equals
$D$, or be spread over several tasks that each lost at most $D$; $D$ is what tells the two
apart.

Two more quantities complete the gates. The \emph{protected set}
$P = \{t \in S : p_b(t) = 1\}$ holds the tasks the live harness never failed: the ones
the developer already relies on, where a regression costs most. Gate 3 therefore gives
them no tolerance: the candidate must pass every one of their runs too. Finally, $f$
(\cref{eq:exposure}) counts the valid rollouts in which the item \emph{fired}: a skill was invoked, or a rule's
\cmd{paths} matched a file the agent read or the prompt @-mentioned. It is counted where
the item exists, among the \emph{cand} rollouts for a candidate and the \emph{base}
rollouts for an item being removed. The scores measure a difference between the arms,
and $f$ ties that difference to the item: a candidate that never fired cannot have caused
its gain, and an item that never fired was not tested by its removal.

\subsection{Adding an item: proposal and gates}\label{sec:design:gates}

\paragraph{Evidence before proposals.} \cmd{/evolve} improves the harness one
\emph{cycle} at a time (\cref{alg:cycle}): a cycle looks for one recurring problem,
writes one item meant to fix it, and tests that item before the next cycle starts. It
looks for the problem in four places: the corrections the developer made to the agent,
which \cmd{/harvest} logs as one line each; the tasks the live harness currently fails;
the developer's recent transcripts; and the \emph{graveyard}, the items already tested and
rejected, so that a rejected idea is not proposed again unchanged. A problem that recurs
is a \emph{theme}, and a cycle acts on one only when the evidence is strong enough: the
theme recurs at least \MinThemeOccurrences{} times, or it explains
\MinThemeFailingTasks{} or more failing tasks. A single correction may be a one-off, and testing an item costs rollouts, so when
no theme clears this bar the cycle is \emph{BARREN}: it proposes nothing and runs no
rollouts. This is an expected outcome, not a failure. When \StopAfterBarren{} cycles in
a row end without a KEEP, whether barren or killed, \cmd{cortex cycle} signals that the
loop should stop (the tasks may no longer tell a good harness from a bad one, the harness
may already be good, or nothing new is being harvested), and \cmd{/evolve} stops and
tells the developer. The developer may still start another cycle; nothing enforces the
stop.

When a theme clears the bar, the cycle proposes one candidate, and the evidence also
chooses its tier (\cref{sec:design:tiers}) from where the fixes of the failing tasks
landed. If they cluster in one directory, the candidate loads only there: a path-scoped
rule when the fixes edit files that already exist, since a rule is injected whenever the
agent reads a matching file; or a path-gated skill when the fixes only create new files,
since a skill is also offered when a matching file is written. Its \cmd{paths} use the
narrowest glob that covers every fix, because an item costs context each time it loads.
If the fixes are spread across the repository, no directory fits, and the candidate is an
always-on skill tied to a moment rather than a place, such as ``before reporting done''.

\paragraph{Two stages.} A cheap \emph{screen} runs the candidate at $k=\KScreen$ on the
theme's currently failing tasks and stops if it gained nothing or never fired, or is
rerun if it reached tasks but none could show a gain (below). It cannot
show what a change breaks, because it never runs the tasks that pass; only candidates that
survive it pay for the \emph{confirm}, at $k=\KConfirm$ over every valid task. A cheap
screen followed by a stricter confirmation is adapted from DarwinX~\cite{darwinx2026}.
How many rollouts each stage took across the evaluation is shown in \cref{fig:funnel}.

\begin{figure}[tbp]
  \centering\small
  \captionof{table}{The acceptance gates, applied by \cmd{score.sh} to the confirm sweep, in the
  notation of \cref{eq:exposure,eq:scores}, with $\tau = \RegressionTolerance$ and
  $m = \MinNetRuns$ runs. Gates 1--4 read exposed tasks only. A candidate is kept only if
  every gate passes, or if it failed only gates 2 or 3 and the drop did not recur on a
  recheck.}
  \label{tab:gates}
  \begin{tabular}{@{}cll@{}}
    \toprule
    Gate & Condition & Why \\
    \midrule
    0 & scorable and, if a task is exposed, one could gain (otherwise RERUN) & bad data, or no room to gain, is not evidence \\
    1 & $G > 0$                                   & it must win something \\
    2 & $D \le \tau$                             & no task may collapse \\
    3 & $p_c(t) = 1$ for every $t \in P$          & a task that never failed gets no tolerance \\
    4 & $kN \ge m$                                & a one-run win is a coin flip \\
    5 & $f \ge 1$                                 & a candidate that never fired caused nothing \\
    \bottomrule
  \end{tabular}

  \bigskip
  \begin{minipage}[t]{\dimexpr(\linewidth-2em)/2\relax}
    \includegraphics[width=\linewidth]{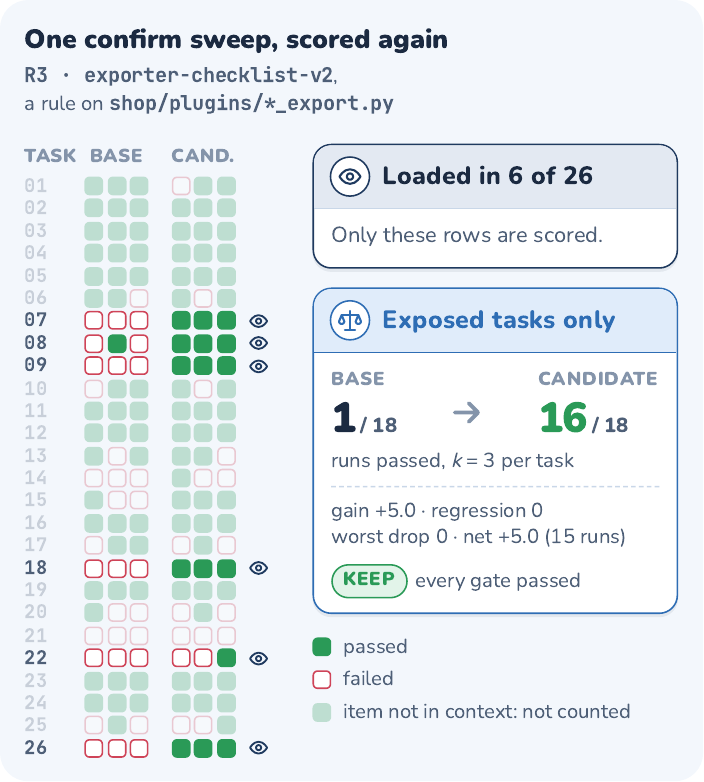}
    \captionof{figure}{One confirm sweep, scored again from its rollouts: evaluation run R3
    (\cref{sec:setup}), the rule \cmd{exporter-checklist-v2}, \KConfirm{} runs per task and
    arm. The rule entered the agent's context on \SweepExampleExposed{} of the sweep's
    \SweepExampleTasks{} tasks, and only those rows are scored; on the others it was
    installed but never loaded, so the two arms differ there only by noise. On those tasks,
    the candidate passed nearly every run and the base nearly none; every gate in \cref{tab:gates} passes, and the rule
    is kept. The summary line gives $G$ (gain), $G - N$ (regression), $D$ (worst drop) and
    the net $N$, in tasks' worth of pass rate, followed in brackets by $kN$, the same net
    counted in runs, which is what gate~4 reads.}
    \label{fig:sweep}
  \end{minipage}\hfill
  \begin{minipage}[t]{\dimexpr(\linewidth-2em)/2\relax}
    \includegraphics[width=\linewidth]{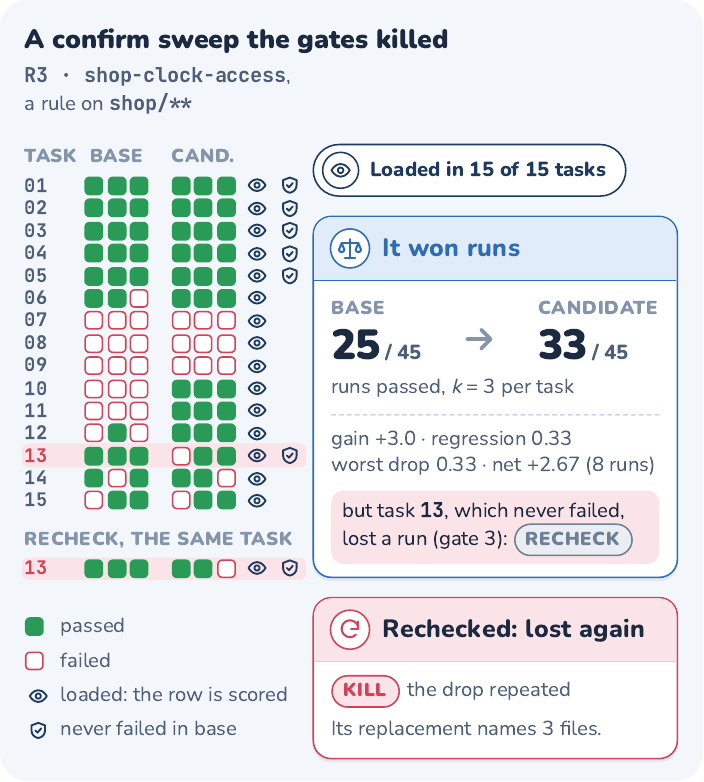}
    \captionof{figure}{A confirm sweep the gates killed, in the same run: the clock rule
    \cmd{shop-clock-access}, on \cmd{shop/**}, which loaded on every task. It won more runs
    than it lost, but a task that the live harness passed in every run lost one (gate~3).
    At \KConfirm{} runs one lost run can be luck, so that task was swept again (RECHECK); it
    lost a run again, and the rule was killed. The rule that replaced it names
    \NarrowClockFiles{} files (\cref{sec:results:loop}).}
    \label{fig:sweepkill}
  \end{minipage}
\end{figure}

\paragraph{The gates.} The confirm verdict is computed by a script, not by a model
(\cref{tab:gates}); \cref{fig:sweep} shows one confirm sweep, scored again from its
rollouts. A change must win something \emph{and} lose nothing that matters: DarwinX's
contract of extending without breaking what is preserved~\cite{darwinx2026}. The net
threshold in runs mirrors SkillGen's~\cite{skillgen2026}, and the no-regression gates
HarnessX's refusal to regress any solved task and Self-Harness's non-regressive
acceptance~\cite{harnessx2026,selfharness2026}. Gates 2 and 3, the \emph{regression
gates}, set two standards for a drop: a task already unstable may lose one run of its
\KConfirm{} ($\tau = \RegressionTolerance$), but a task in $P$ may lose none. Because at $k=\KConfirm$ one unlucky run on a reliable
task looks exactly like a broken task, a KILL that rests \emph{only} on gates 2 and 3 is
not final: the tasks behind it are swept again (RECHECK), and the KILL stands only if one
of them regresses again on the fresh runs; otherwise the candidate is kept.
\Cref{fig:sweepkill} shows a candidate killed this way, although it won more runs than it
lost.

A sweep is not scored at all (RERUN) when any of these holds: it is incomplete; more than
\MaxInvalidPct{} of its rollouts are invalid; some task lacks $k$ valid runs in either
arm, which a single invalid rollout already causes, since a sweep plans exactly $k$;
exposure is unknown in a valid rollout; the Claude Code version changed mid-sweep; or
more than \MaxInvalidPct{} of its rollouts ran on an overloaded machine, with more than
\OverloadFactor{} times as many runnable processes as CPUs. Written out, with $\Omega$ the
sweep's rollouts, $\Omega_{\mathrm{inv}}$ the invalid ones, $\mathcal{T}$ the tasks it
planned, $n_{\mathrm{unk}}$ the number of valid rollouts whose exposure is unknown,
$\operatorname{cli}(\omega)$ the Claude Code version that ran rollout $\omega$,
$L(\omega)$ the mean number of runnable processes sampled during it, $n_{\mathrm{cpu}}$
the number of CPUs of the machine and $\varepsilon = \MaxInvalidPct$, a sweep is scored
only when
\begin{equation}\label{eq:scorable}
  \begin{split}
  \operatorname{scorable} \iff{}
  & \operatorname{complete} \;\wedge\;
    \frac{|\Omega_{\mathrm{inv}}|}{|\Omega|} \le \varepsilon \;\wedge\;
    \mathcal{M} = \mathcal{T} \ne \varnothing \;\wedge\;
    n_{\mathrm{unk}} = 0 \\
  & \wedge\;
    \bigl|\{\operatorname{cli}(\omega)\}_{\omega \in \Omega \smallsetminus \Omega_{\mathrm{inv}}}\bigr| \le 1 \;\wedge\;
    \frac{\bigl|\{\omega \in \Omega : L(\omega) > \OverloadFactor\,n_{\mathrm{cpu}}\}\bigr|}{|\Omega|} \le \varepsilon,
  \end{split}
\end{equation}
where \emph{complete} means that the sweep finished, was not cut short by its cap on
rollouts, and recorded every rollout it planned; the last condition applies when rollouts
run in parallel. A sweep of an addition in which the candidate reached at least one task
is also a RERUN, not a KILL, when none of those tasks could show a gain, because each one
either passed every base run, leaving no room to improve, or never passed in either arm,
so that nothing there tells a useful candidate from a useless one (the scorer suggests
repairing those tasks' checks, or screening tasks the candidate can win). Such a sweep
did not test the candidate, and burying it would record a verdict nobody earned. A
candidate that reached no task at all is killed instead: it gained nothing and never
fired (gates~1, 4 and~5). An agent that timed out is a valid
failure, unless the machine was oversubscribed at the time (more runnable processes than
CPUs, so that every rollout ran slower than it would alone); an agent that could not run
(authentication, rate limits, a failed reset) is invalid, and invalid is not fail.

Gates, recheck and RERUN together decide a confirm sweep of an addition. Let
$\mathcal{I}$ be the exposed tasks that could show a gain: those the base did not pass in
every run, and that passed at least once in either arm. Let $\mathcal{F}$ be the gates of
\cref{tab:gates}, 1 to 5, that the sweep fails, and $\rho(u, v)$ the test that gates 2
and 3 apply to a task whose pass rate went from $u$ under \emph{base} to $v$ under
\emph{cand}. Let $\mathcal{Q}$ be the tasks behind a failure of gate 2 or gate 3, and
$p'_b$, $p'_c$ and $\mathcal{X}'$ the pass rates and the exposed tasks of the recheck,
which sweeps both arms again on the tasks of $\mathcal{Q}$ alone; a recheck that is not
scorable is itself a RERUN. Reading the cases from the top,
\begin{equation}\label{eq:verdict-add}
  \begin{gathered}
  V_{\mathrm{add}} =
  \begin{cases}
    \text{RERUN}   & \text{if } \neg\operatorname{scorable} \;\vee\; \bigl(S \ne \varnothing \wedge \mathcal{I} = \varnothing\bigr),\\
    \text{KEEP}    & \text{if } \mathcal{F} = \varnothing,\\
    \text{RECHECK} & \text{if } \varnothing \ne \mathcal{F} \subseteq \{2, 3\},\\
    \text{KILL}    & \text{otherwise,}
  \end{cases}
  \quad
  \begin{aligned}
    \mathcal{I} &= \bigl\{t \in S : p_b(t) < 1 \,\wedge\, \max\bigl(p_b(t), p_c(t)\bigr) > 0\bigr\},\\
    \mathcal{F} &= \bigl\{g \in \{1,\dots,5\} : \text{gate } g \text{ fails}\bigr\},\\
    \rho(u, v) &= (u - v > \tau) \,\vee\, (u = 1 \wedge v < 1),
  \end{aligned}
  \\[4pt]
  V_{\mathrm{recheck}} =
  \begin{cases}
    \text{KILL} & \text{if } \exists\, t \in \mathcal{Q} \cap \mathcal{X}' : \rho\bigl(p'_b(t), p'_c(t)\bigr),\\
    \text{KEEP} & \text{otherwise,}
  \end{cases}
  \qquad
  \mathcal{Q} = \bigl\{t \in S : \rho\bigl(p_b(t), p_c(t)\bigr)\bigr\}.
  \end{gathered}
\end{equation}

\needspace{9\baselineskip}
\subsection{Removing and moving items}\label{sec:design:prune}

\begin{figure}[tbp]
  \centering
  \includegraphics[width=\linewidth]{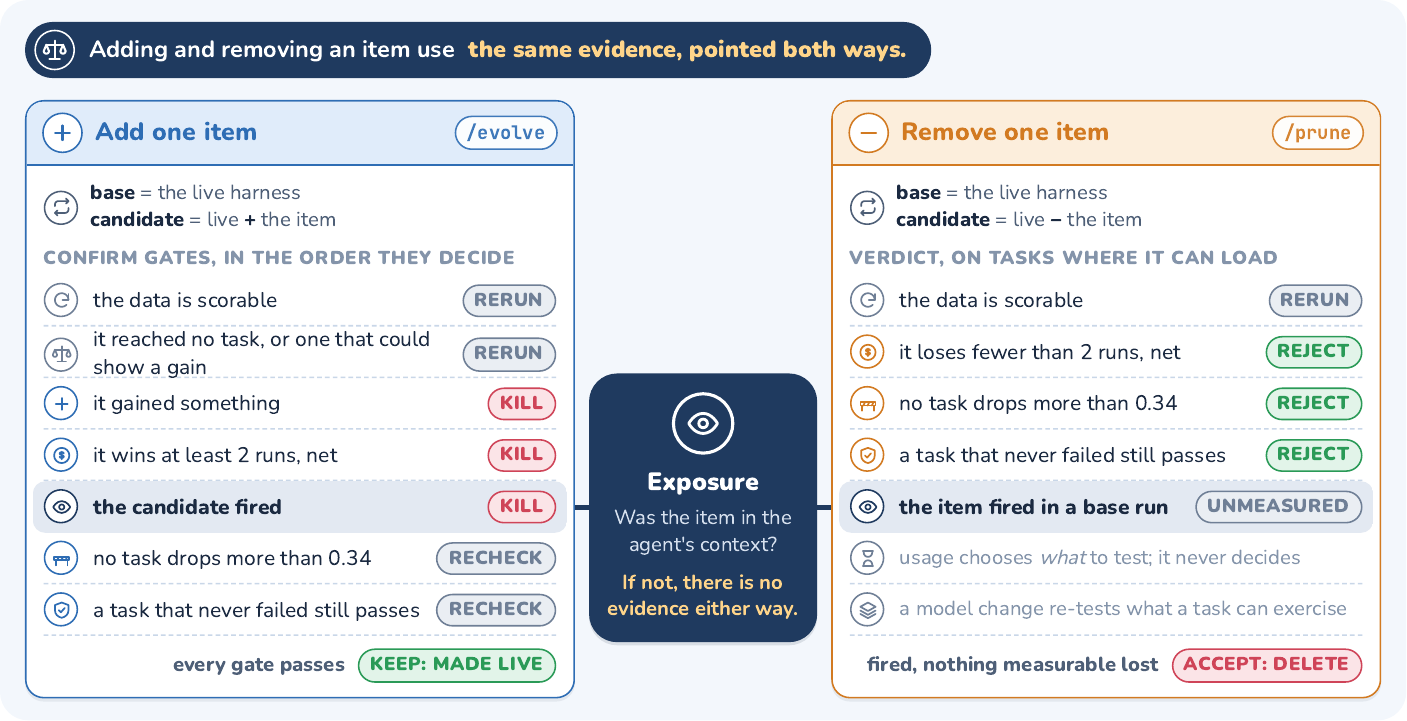}
  \caption{Adding and removing are one test, pointed both ways. \cmd{/evolve} compares
  the live harness with the harness plus a candidate; \cmd{/prune} compares it with the
  harness minus a live item. Both count only the tasks on which the item was in the agent's
  context: a candidate that never fired is never kept, and a live item that never fired is
  never deleted.}
  \label{fig:mirror}
\end{figure}

Deletion needs the same evidence as addition, pointed the other way (\cref{fig:mirror}). \cmd{/prune} removes
one live item and runs a sweep with the item as \emph{base} and without it as \emph{cand},
on the tasks where the item can load. In the notation of
\cref{eq:exposure,eq:scores,eq:scorable}, the verdict is
\begin{equation}\label{eq:verdict-remove}
  V_{\mathrm{remove}} =
  \begin{cases}
    \text{RERUN}      & \text{if } \neg\operatorname{scorable},\\
    \text{REJECT}     & \text{if } kN \le -m \;\vee\; D > \tau \;\vee\; \exists\, t \in P : p_c(t) < 1,\\
    \text{UNMEASURED} & \text{if } f = 0,\\
    \text{ACCEPT}     & \text{otherwise.}
  \end{cases}
\end{equation}
REJECT keeps the item, because removing it cost runs, made a task collapse, or made a task
in $P$ lose a run. UNMEASURED keeps it too: here $f$ counts the valid \emph{base}
rollouts, and $f = 0$ means the item fired in none of them. Only ACCEPT deletes it, to the
graveyard. The cases are read from the top, so an item whose removal cost something gets
REJECT even when it never fired (\cref{sec:results:model}). UNMEASURED is the point: if nothing in the suite used an item,
removing it changed nothing \emph{because nothing needed it}, which is absence of evidence,
not evidence of uselessness: a skill for a twice-a-year situation looks exactly like
dead weight in between. This is the direct counterpoint to deletion rules that retire
items whose measured effect is small~\cite{assay2026}. For the same reason, usage data
chooses \emph{which} items to test and never decides a deletion.

A pass starts as a plan: the items to test, the tasks to test each one on, and the
rollouts, time and cost this will take, estimated from the past rollouts in the same
repository. Nothing runs until the developer approves it. Which items go into the plan
depends on whether the model has changed. If the pinned model differs from the one the last
pass ran on, \emph{every} item is re-tested, because a stronger model may already do what
an item teaches; always-on items come first, the largest first.
Otherwise the pass tests only the few items most in doubt. Either way, an item is left out
when no task can load it, since its removal could only come out UNMEASURED, or when it is
an always-on skill the agent never opened in earlier sweeps, since no sweep could accept
its removal: it could only keep the item, as REJECT or UNMEASURED. A pass can also move
an item to a cheaper tier: an always-on skill whose use is concentrated in one folder can
be rewritten as a skill or a rule tied to that folder. The swap is swept like a removal,
the harness with the new item in place of the old against the live one, and decided by
\cref{eq:verdict-remove}. For a swap, a task is exposed when the old item was in context
in a \emph{base} rollout or the new one in a \emph{cand} rollout, and $f$ counts the
\emph{base} rollouts in which the old item fired. The swap goes ahead only if it loses
nothing measurable on those tasks and the old item fired, and the new item is made live
only if it fired too.

\subsection{The optional judge, Jev}\label{sec:design:boundary}

Cortex can consult an outside model, Jev (TypeSafe's System
One), to help it choose what to try. Jev is sent
some text and a typed question, and it returns a probability for a yes-or-no question,
one option from a list with its confidence, or a score on a fixed scale. It writes no
text and takes no action. It is off by default, and it was off in every run of this
paper, so every theme, tier, candidate and prune order reported in \cref{sec:results} was
chosen without it.

When it is on, Jev acts only in the \emph{proposal} layer, the steps that decide which
theme, candidate or order to try next: it can sort the logged lessons into themes, suggest
a candidate's tier, order a prune plan and tell the developer when a session is worth
harvesting. One command, \cmd{cortex scope}, puts two questions to it about a candidate
before a sweep is paid for. The first gives the candidate's \emph{relevance}: for each
task the candidate would load on, Jev is asked whether the task is actually \emph{about}
the candidate's subject, from the task's title, request and one-line note, never the code
or the fix, and the share of yes answers is the relevance. When it is below
\RelevanceFloorPct{}, Cortex prints a warning, but it never blocks a sweep. Without Jev,
Cortex reports only \emph{breadth}, the share of all tasks the candidate would load on.
The second, for a path-gated skill, is whether an agent doing the task, seeing only the
skill's description, would choose to open it. The tasks a candidate would load on are
predicted from its \cmd{paths}; once the candidate has been swept, they are the tasks
where its sweeps recorded it in context.

\paragraph{What the judge may not do.} Jev may change which theme, candidate or order
Cortex tries; it may never decide what a measurement found. The line is mechanical rather
than a promise. The scorer, the sweep runner
and \cmd{preflight} contain no reference to Jev, and a test in the shipped suite asserts
that they do not. No Jev call runs inside a rollout sandbox, where it would become part of
the harness under test. A wrong suggestion can waste rollouts, but it
cannot turn a KILL into a KEEP, because the gates of \cref{tab:gates} judge every
candidate either way.

\paragraph{How the judge was used in this evaluation.} Only the two \cmd{cortex scope}
questions were exercised: after the runs, they were put to each run's candidates, at no
cost in rollouts (\cref{sec:results:judge}). Each candidate was rebuilt, from the
graveyard or the live harness, in a scratch copy of its run's repository that kept the
run's sweep records. Because every candidate had such a record, \cmd{cortex scope} took
the tasks it would load on from the tasks where its own sweeps had recorded it in
context; on those tasks, from the candidate's text and each task's title, request and
note, never from its verdict, it estimated the candidate's relevance and, for a
path-gated skill, how often the agent would open it.

\subsection{Implementation}\label{sec:design:impl}

Cortex's scripts are \ImplCodeLines{} lines of shell and Python in \ImplCodeFiles{}
files (of which \ImplJevLines{} lines are the optional judge client), plus
\ImplPromptLines{} lines of prompts that define \cmd{/harvest}, \cmd{/evolve} and
\cmd{/prune} as Claude Code commands. The prompts do what needs judgement (writing a
task, proposing a candidate, choosing a tier), and every quantity a verdict depends on
is computed by the scripts. The scripts refuse to make a candidate live without a KEEP
from a confirm sweep or its recheck (or an ACCEPT for a swap); burying a candidate is
done by the \cmd{/evolve} agent, the model session that runs the command. Sweeps run in
parallel, one sandbox
clone per worker, as many workers as the machine's free memory and CPUs allow. Rollouts
are \emph{headless} Claude Code sessions, started by a script with no one at the
keyboard, with a \RolloutTimeoutMin{}-minute limit. Cortex deletes the event stream it
reads from each once exposure is known, because it holds everything the agent saw; Claude
Code keeps its own record of every session, outside the repository.

%% file: sections/04-setup.tex
\section{Evaluation setup}\label{sec:setup}

\begin{figure}[tbp]
  \centering
  \includegraphics[width=\linewidth]{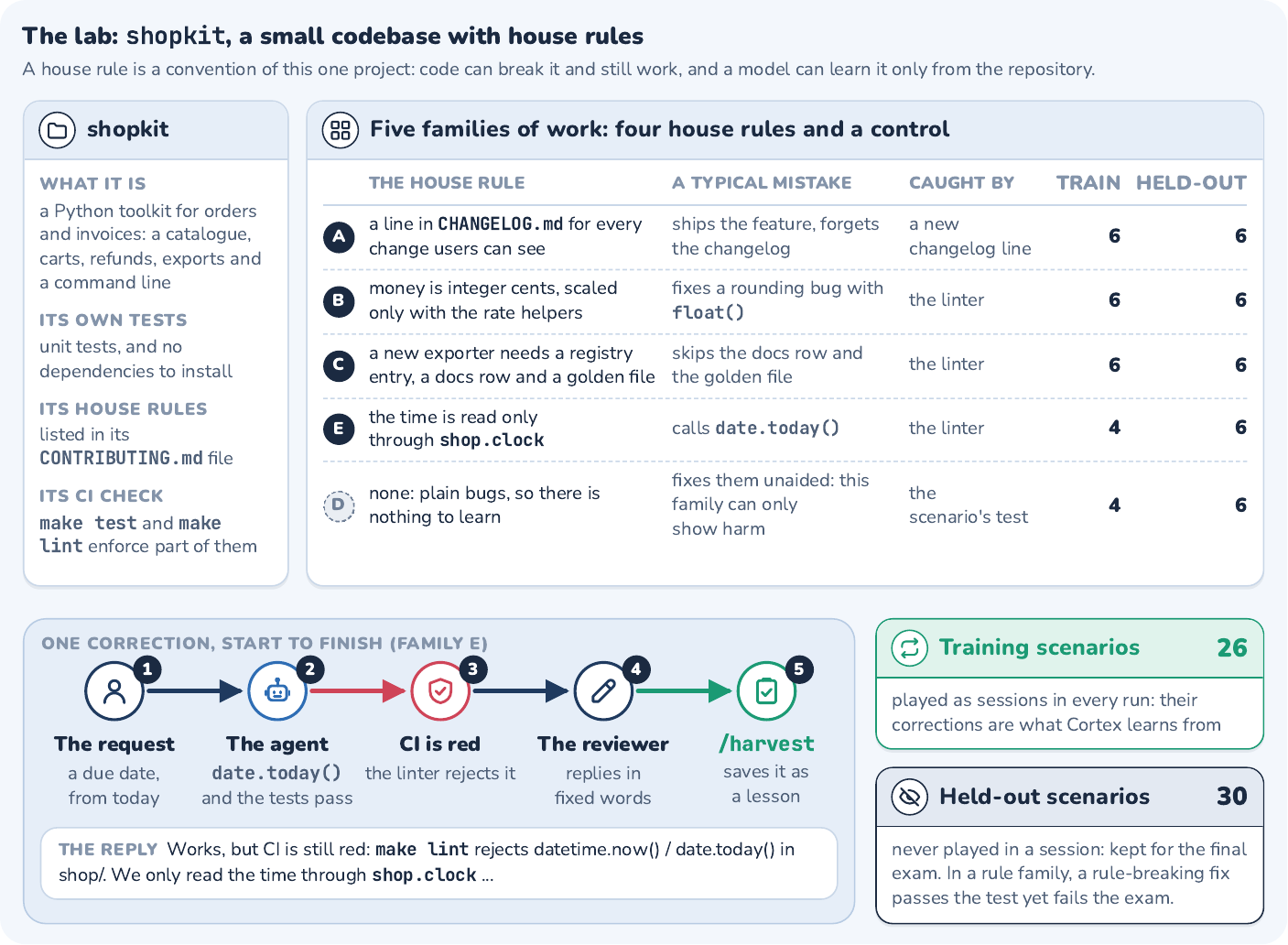}
  \caption{The lab. \emph{Top:} the repository, and its five task families: each rule
  family is built around one house rule written in \cmd{CONTRIBUTING.md}, with the mistake
  an agent typically makes, what catches it, and the counts of training and held-out
  scenarios; family D, plain bugs with no rule, is the control. \emph{Bottom left:} one
  correction from start to finish, for the rule on time. \emph{Bottom right:} the training
  scenarios are played as sessions; the held-out scenarios are kept for the benchmark.}
  \label{fig:lab}
\end{figure}

\begin{figure}[p]
  \centering
  \includegraphics[width=\linewidth]{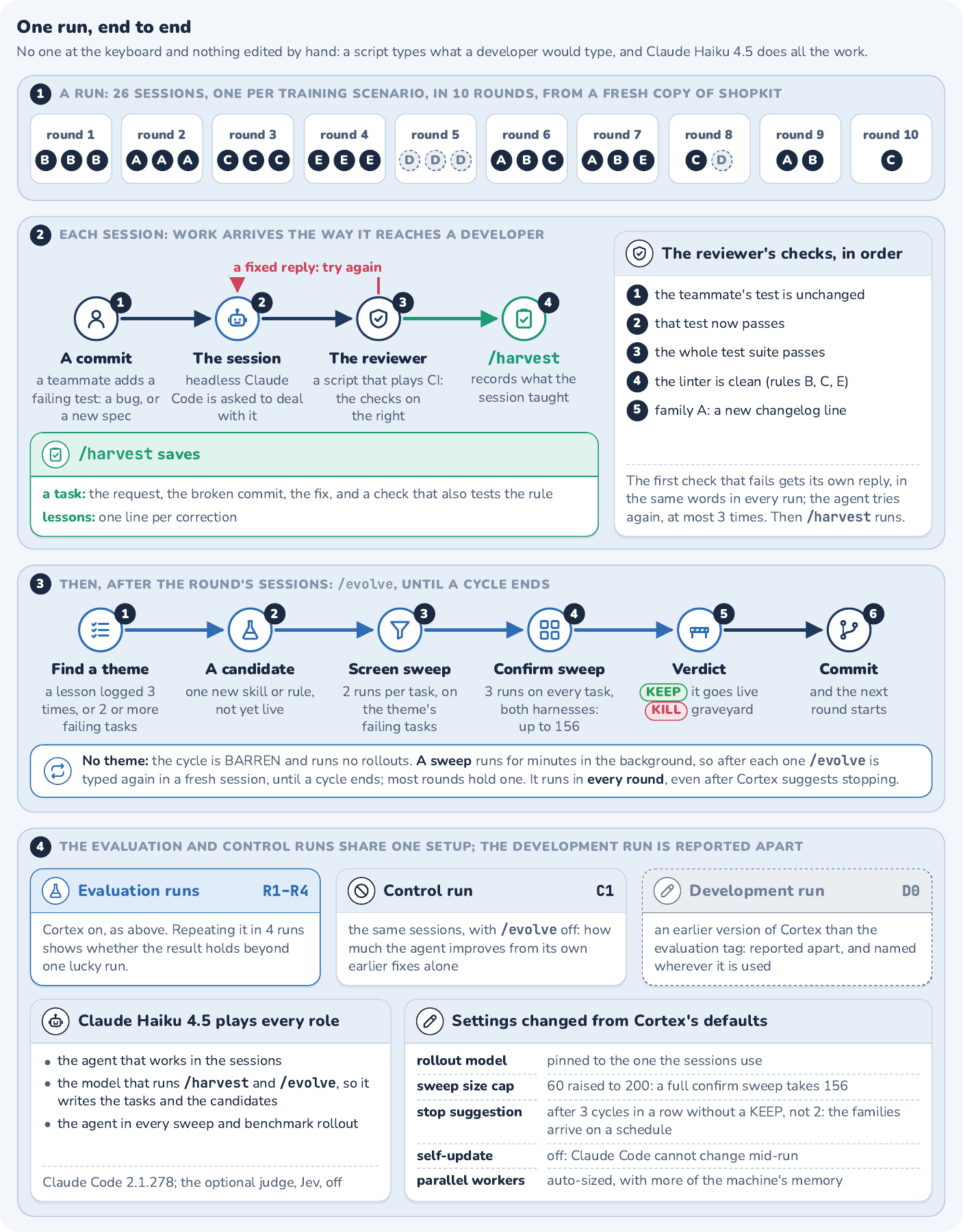}
  \caption{One run, end to end. (1) The rounds, and the family of each session. (2) One
  session: the teammate's commit, the agent, the scripted reviewer, whose fixed reply
  sends the agent back at most \MaxCorrections{} times, and \cmd{/harvest}. (3) The
  \cmd{/evolve} cycle after each round's sessions (\cref{fig:evolve}). (4) The runs:
  evaluation and control share one setup; the development run is reported apart.}
  \label{fig:run}
\end{figure}

\begin{figure}[tbp]
  \centering
  \includegraphics[width=\linewidth]{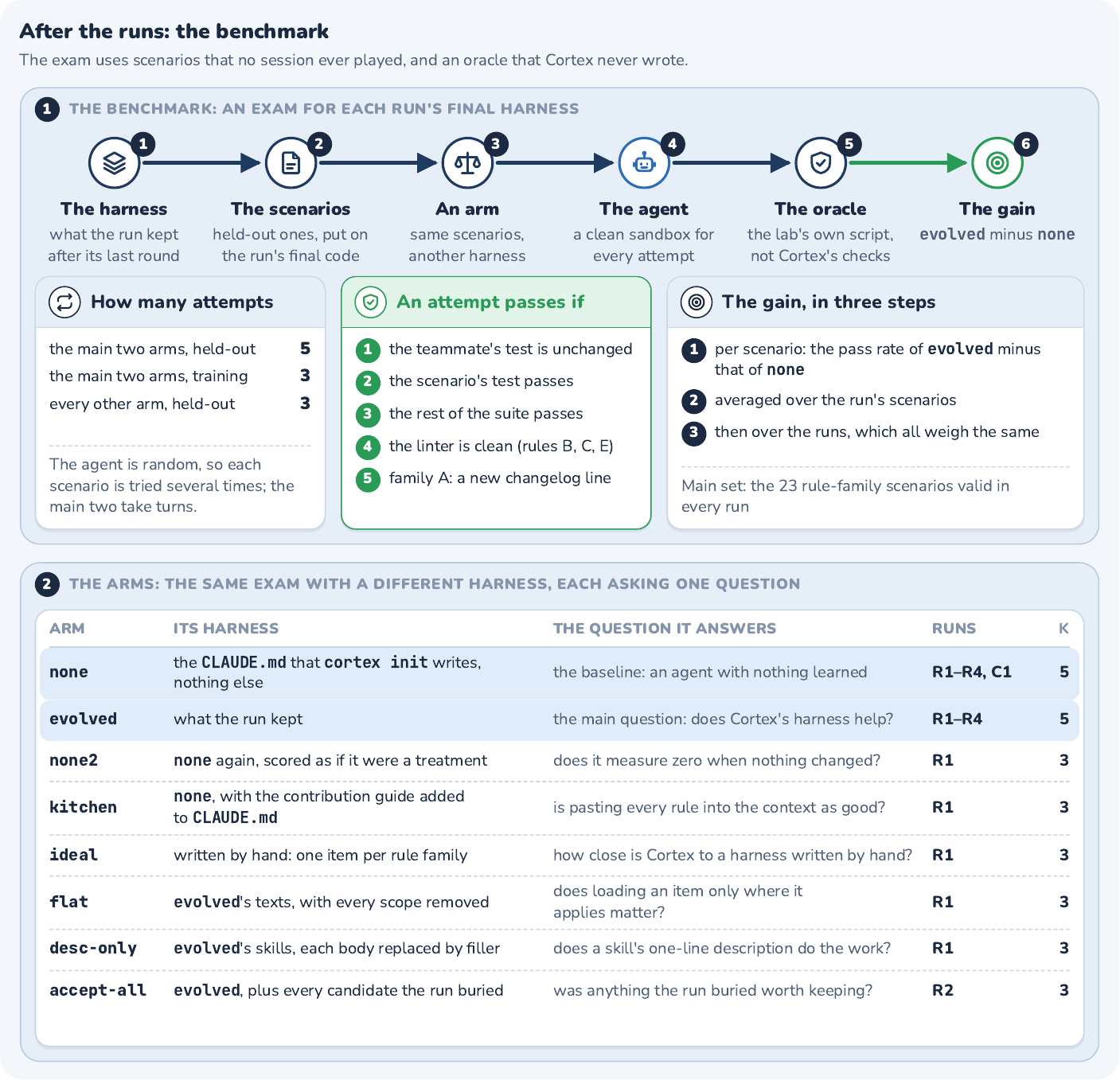}
  \caption{The benchmark. \emph{Top:} how each run's final harness is examined, from the
  scenarios to the gain, which is the difference between the pass rates of \cmd{evolved}
  and \cmd{none}. \emph{Bottom:} the arms, which share the scenarios, the model and the
  oracle and differ only in the harness, with the question each one answers, the runs it
  was measured in, and its number $k$ of attempts per held-out scenario. \cmd{none} and
  \cmd{evolved} also run the training scenarios.}
  \label{fig:bench}
\end{figure}

\begin{figure}[H]
  \centering
  \includegraphics[width=\linewidth]{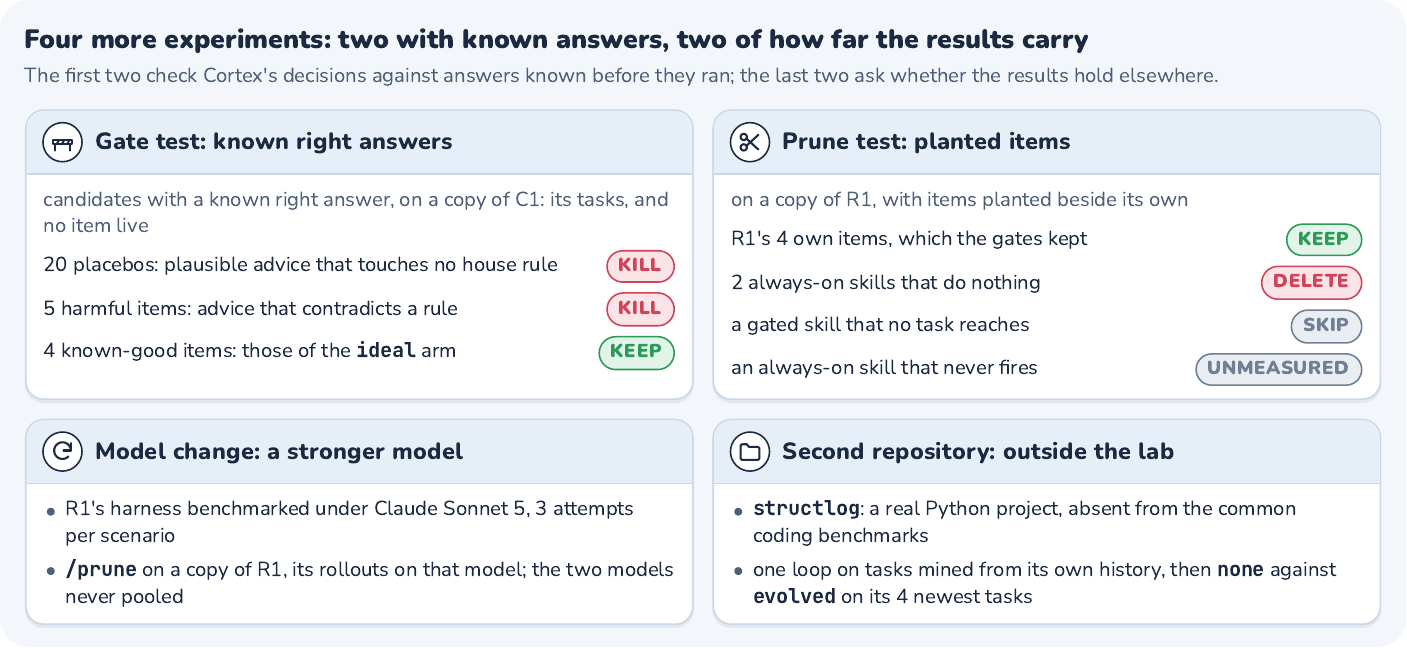}
  \caption{The four further experiments. The gate test and the prune test run the real
  pipeline on copies of finished runs, and each item's right verdict is known before it
  runs; the model change and the second repository ask how far the results carry. The
  badges give each item's right verdict, not the verdict the pipeline reached
  (\cref{sec:results:gates,sec:results:model}).}
  \label{fig:tests}
\end{figure}

The evaluation measures Cortex in a synthetic repository whose house rules a model often
breaks. This section describes the repository (\cref{fig:lab}), what one run does
(\cref{fig:run}), how a run's result is measured (\cref{fig:bench}), two tests with known
answers, two tests of how far the results carry (\cref{fig:tests}), and the statistics.

\paragraph{The lab.} \emph{shopkit} is a small Python toolkit for orders and invoices (a
catalogue, carts, invoices, refunds, exports, reports and a command line) with no
dependencies and its own unit tests. Like many real codebases, it has \emph{house rules}:
conventions this one project has chosen, which code can break while still working. They
are written in its \cmd{CONTRIBUTING.md}, and its continuous-integration check, which runs
the unit tests and a linter, enforces part of them. Take the rule on time: code in
\cmd{shop/} must read the current time only through \cmd{shop.clock}, never with
\cmd{datetime.now()} or \cmd{date.today()}, so that tests can freeze the clock. Asked to
give invoices a due date counted from today, an agent that computes it from
\cmd{date.today()} produces code that works, yet the linter rejects
it, and the reviewer replies: ``Works, but CI is still red: \cmd{make lint} rejects
datetime.now() / date.today() in shop/. We only read the time through shop.clock \ldots''
(CI here is continuous integration). A model cannot learn such a rule from its training, only from the project. It is exactly
what a harness exists to teach, and a model that keeps breaking it keeps being corrected,
which is the evidence Cortex learns from. Four house rules each have a family of tasks
built around them (\cref{fig:lab}), named by a letter: A, the changelog (every change
users can see gets a line in \cmd{CHANGELOG.md}); B, money (amounts are integer cents,
scaled only with the project's rate helpers); C, exporters (a new exporter needs a
registry entry, a row in the documentation and a golden file, a stored sample of its
expected output); and E, time (the clock rule above). A fifth family, D, holds plain bugs
with no house rule, so the loop has nothing to learn there: it is the control, and it can
only show harm.

\paragraph{Training and held-out scenarios.} A \emph{scenario} is one piece of work: a
teammate's commit that adds a failing test, and a request saying what is wanted; the due
date above is one. The scenarios are split in two,
the way a course keeps its exam apart from its exercises. The \TrainTasks{}
\emph{training} scenarios are the exercises: the runs play them as working sessions, the
reviewer corrects the agent's mistakes on them, and Cortex learns from those corrections.
The \HoldoutTasks{} \emph{held-out} scenarios are the exam: no session ever plays them,
so Cortex never sees them, and they serve only to measure harnesses (the benchmark,
below). The split is what lets a gain be read as learning. On a training scenario, a
harness could help by recalling that one task, for instance that the due date is
computed from \cmd{shop.clock}; on a held-out scenario, no task it recalls holds the
answer, and only what carries over to new work, such as the rule, can help. Held-out
scenario HE1 (held-out scenarios are named H, the family's letter and a number), for
one, asks for the orders of the last $N$ days, counted back from today by default: a task no session plays, under the same rule on time. A gain on the held-out
scenarios therefore shows that what Cortex learned carries over to work it has never
seen, which is what a developer needs from a harness. Passing the exam also takes
following the rule, not only making the test pass. Every held-out scenario of the four
rule families comes with a rule-breaking fix, one that makes the scenario's own test pass
but breaks the rule: for HE1, it falls back on \cmd{date.today()}; in families A and C,
it is the correct fix without its changelog line, or without its documentation row and
golden file. Only the rule's part of the check can tell it apart from the correct fix.

\paragraph{A run.} A run is one complete, unattended replay of the project's working life
(\cref{fig:run}). It starts from a fresh copy of the repository and plays the
\TrainTasks{} training scenarios as \SessionsPerRun{} working sessions, one per scenario,
in \Rounds{} rounds; each family is introduced in a round of its own and returns in later
ones. No one sits at the keyboard: a script types what a developer would type, and
nothing in the repository is edited by hand.

\paragraph{One session.} A session begins the way work reaches a developer: a teammate's
commit adds a failing test, for a bug or for a change in what the code should do, and a headless Claude Code session,
one started by the script with no one watching, is asked to deal with it. When the agent
stops, a scripted reviewer plays continuous integration. It checks, in order, that the
teammate's test is unchanged, that this test now passes, that the whole suite passes, that
the linter is clean, and, for family A, that the changelog gained a line. The first check
that fails gets the reply written for that failure, in the same words in every run, like
the reply on time above, and the agent tries again, at most \MaxCorrections{} times.
The session then ends with \cmd{/harvest} (\cref{sec:design:workflow}): if the session went
from broken to working it saves a task, whose check also tests what the reviewer
corrected, and it writes each correction down as a lesson. Because the replies are
scripted, every run receives the same correction for the same mistake.

\paragraph{The loop, after each round.} After the round's sessions, the script runs
\cmd{/evolve} (\cref{fig:evolve}). A cycle looks for a theme, writes one candidate,
screens and confirms it, and keeps or buries it, or ends as BARREN, at no cost in rollouts,
when no theme clears the bar. A sweep takes minutes and runs in the background, so the
session that started it ends; when the sweep has finished, the script types \cmd{/evolve}
again in a fresh session, which reads the result and goes on. It stops once a cycle has
ended with no candidate pending, or when \cmd{/evolve} stops on its own; most rounds hold
one cycle. The script runs \cmd{/evolve} in every round, even after Cortex has signalled
that the loop should stop, so that every run has the same shape. What \cmd{/evolve}
changed, a kept item and the journal, is then committed, as each session's code and task
were at the session's end, and the next round begins.

\paragraph{The runs.} The \EvalRuns{} evaluation runs, R1 to R4, follow this procedure with
Cortex in full; repeating it shows whether a result holds beyond a single run. A control
run, C1, plays the same sessions, receives the same corrections and runs \cmd{/harvest}
as usual, but never runs \cmd{/evolve}, so no skill or rule is ever added. Corrections can
still fall in C1, because the agent can copy a rule from the fixes already in the code,
and C1 measures how much they fall without \cmd{/evolve}. A development run, D0, used an
earlier version of Cortex than \EvalTag{}. It is reported apart and named wherever it is
used (\cref{sec:results:gain,sec:results:judge,sec:limitations}); every other number
comes from the runs made at \EvalTag{}.

\paragraph{One model in every role.} Every evaluation and control run used
\RolloutModelName{} (\cmd{\RolloutModel}) through Claude Code \ClaudeCodeVersion{}, with
Cortex's optional judge, Jev, off. The same model plays every role: it is the agent that
works in the sessions; the model that runs \cmd{/harvest} and \cmd{/evolve}, and so writes
the tasks and proposes the candidates; and the agent in every sweep and benchmark rollout.
No stronger model teaches it. The runs depart from Cortex's defaults in the settings
below, and in where the sandboxes sit on disk and how many run at once; none of them
decides a verdict. The rollout model is pinned to the one the
sessions use, so that every rollout uses that model (by default a rollout uses whatever
model Claude Code is set to). The cap on rollouts per sweep, a guard on cost, is raised
from \MaxRunsDefault{} to \MaxRunsEval{}, because a confirm sweep over every training
task takes \ConfirmSweepRollouts{} rollouts (each of the \TrainTasks{} tasks \KConfirm{}
times under each of the two harnesses). The stop limit, the number of cycles in a row
without a KEEP after which Cortex suggests stopping, is raised from \StopAfterBarren{} to
\StopAfterBarrenEval{}, because the lab introduces its families on a schedule, and quiet
cycles are expected before the next family arrives. And Claude Code never updates itself
mid-run.

\paragraph{The benchmark.} After the last round, each run's final harness sits an exam
(\cref{fig:bench}). Each held-out scenario is committed onto the run's final code, so that
it meets the codebase as that run left it, and the commit adds only the teammate's test:
the agent gets that commit and the scenario's request, as in a session. Each scenario
also has a \emph{reference fix}, a correct solution written with it, and, in the rule
families, its rule-breaking fix; both stay outside the repository and serve only to
validate the scenario, below. The training scenarios are replayed from the commits their
sessions started from. An \emph{arm} is one harness: every arm faces the same scenarios,
with the same model and the same oracle, and the arms differ only in the harness
installed. The two primary arms are \cmd{none}, the minimal \cmd{CLAUDE.md} that Cortex's
installer, \cmd{cortex init}, writes, and nothing else, and \cmd{evolved}, what the run
kept; for short, the text also
calls \cmd{none} ``no harness'', since it has no skills or rules. Each of the others
answers one question (\cref{fig:bench}, bottom). \cmd{none2}, the A/A arm, is \cmd{none}
measured a second time, in a later batch, to show how far two measurements of one harness
can differ.
The rest are the \emph{ablations}: \cmd{kitchen} is \cmd{none} with the whole
contribution guide appended to \cmd{CLAUDE.md}; \cmd{flat} is \cmd{evolved}'s texts with
every \cmd{paths} field removed; \cmd{desc-only} is \cmd{evolved} with each skill's body
replaced by neutral filler; \cmd{accept-all} is \cmd{evolved} plus every candidate the run
buried; and \cmd{ideal} is written by hand: always-on skills for A and E, a rule on
\cmd{shop/billing/**} for B, and a gated skill on \cmd{shop/plugins/**} for C. Every
attempt, or rollout, runs in a clean sandbox, in Claude Code's \cmd{acceptEdits} mode, in
which file edits are accepted without asking and a command runs only if the settings
already allow it, as the repository's own allow list does for \cmd{make}, \cmd{python3}
and the read-only git commands (\cmd{git status}, \cmd{git diff}, \cmd{git log},
\cmd{git show}; \cref{sec:limitations} says what a rollout could read through them).
Success is decided by the lab's own \emph{oracle}, a fixed script that
judges every attempt the way the scripted reviewer does, not by the checks Cortex
harvested, so that Cortex never grades its own work. It makes the reviewer's checks in
the same order and stops at the first that fails: the teammate's test is left unchanged,
that test passes, the rest of the suite passes, the linter is clean (it checks the rules
of B, C and E), and, for family A, the changelog gained a line. The agent does not behave the same way twice, so each scenario
is attempted several times: \cmd{none} and \cmd{evolved} run each held-out scenario
\PrimaryK{} times and each training scenario \TrainK{} times, and the other arms run the
held-out scenarios \AblationK{} times. The primary arms run interleaved, taking turns, so
that a slowdown lands on both; the other arms ran afterwards, in batches of their own, so
every comparison with them crosses a batch boundary. Every arm alike also carried one
shared skill listing on every turn: the descriptions of Claude Code's own bundled skills,
of Cortex's three commands, which \cmd{cortex init} installs in the repository, and of
the skills in the user-level Claude Code configuration; the agent invoked one of the
bundled skills in \BundledNone{} of
\cmd{none} rollouts and \BundledEvolved{} of \cmd{evolved} ones. Because every run's code
ends differently, a held-out scenario counts in a run only if it validates on that run's
final code: its test must fail there while the linter stays clean, its reference fix must
pass the oracle, and, in a rule family, its rule-breaking fix must pass its own test and
be caught only by the rule's part of the check. One did not validate
(\ExcludedScenario{}, in \ExcludedFrom{}), and it is excluded from that run in every arm.
The primary analysis uses the complete-case set, the \CompleteTasks{} rule-family
scenarios valid in every run; the set of all valid scenarios is secondary.

\paragraph{Tests with known answers, and of generality.} Four more experiments complete
the evaluation (\cref{fig:tests}); all ran on \RolloutModelName{} except where the model
change says otherwise. The first two measure Cortex's own decisions against answers known
before they ran; the last two ask whether its results carry beyond the lab and its model.
\begin{itemize}
  \item \textbf{The gate test} asks whether the gates keep only what helps. It runs
    \GateCandidates{} candidates whose right verdict is known through the real pipeline,
    on a copy of C1's final repository, which holds harvested tasks and no live item:
    \Placebos{} placebos, plausible advice that shares no keyword with a house rule, all
    written as always-on skills (\PlaceboTopic{} about a coding topic, such as preferring
    \cmd{pathlib} to \cmd{os.path}, and \PlaceboMoment{} about a moment in the work, such
    as what to do after finishing an edit), which should be killed; \Harmful{} harmful
    items, advice that contradicts a house rule, \HarmfulRules{} of them rules and the rest
    always-on skills, which should be killed; and the \Positives{} items of \cmd{ideal},
    which should be kept. Each is
    screened at $k=\KScreen$ on tasks the live harness had failed (the same tasks for every
    placebo), and confirmed if it survives. The test's procedure stops at the screen whenever the gain is
    not positive or the candidate never fired, and records that stop as a KILL;
    \cmd{score.sh}'s own verdict is recorded beside it, and the two can differ, since
    \cmd{score.sh} returns RERUN when no task on the screen could show a gain
    (\cref{sec:results:gates}).
  \item \textbf{The prune test} asks whether \cmd{/prune} deletes only dead weight. It runs
    \cmd{/prune} on a copy of \PruneRun{}'s final repository, with \Planted{} items planted
    beside \PruneRun{}'s \RunOneItems{} and its limit on items per pass raised to cover them
    all. In its first pass, \cmd{/prune} chose which items to test and planned
    \PruneRun{}'s own, which it should keep. A second pass was given the planted items it
    had left out: \PlantedPlacebos{} placebo always-on skills, which it should delete; a
    gated skill that no task reaches, which it should skip; and an always-on skill that
    never fires, which it should report as unmeasured rather than delete.
  \item \textbf{The model change} asks whether the harness still matters under a stronger
    model. It benchmarks \ModelChangeSource{}'s harness with \StrongModelName{}
    (\cmd{\StrongModel}) as the agent, at $k=\ModelChangeK$, and runs \cmd{/prune} on
    another copy of \ModelChangeSource{} with its
    rollouts pinned to that model, while the session that runs \cmd{/prune} is still
    \RolloutModelName{}. The two models' results are analysed separately and never pooled.
  \item \textbf{The second repository} asks whether the loop works on a project nobody in
    this study built. \SecondRepo{} was chosen as a Python project with a permissive
    licence, a long history, a fast test suite that needs no network or credentials and a
    convention newcomers break, absent from SWE-bench~\cite{jimenez2024swebench},
    SWE-Gym~\cite{swegym2025}, SWE-smith~\cite{swesmith2025} and Aider's coding benchmark.
    The repository is \cmd{\SecondRepoSlug}, taken at commit \cmd{\SecondRepoBase}. It gets
    one loop on tasks mined from its own history, then \cmd{none} against \cmd{evolved} on
    its \SecondRepoHoldout{} newest tasks at $k=\SecondRepoK$.
\end{itemize}

\paragraph{Statistics.} The primary comparison is
\cmd{evolved} against \cmd{none} on the held-out scenarios of the four rule families,
paired by scenario within run. For each run and each scenario we take the difference
between the two arms' pass rates, average it over the run's scenarios and then over the
runs, so that a run that measured more scenarios does not weigh more; this is the gain
reported in \cref{sec:results:gain}. With $\mathcal{R}$ the runs, $\mathcal{V}_r$ the
scenarios of the analysed set that are valid in run $r$, and $y^{a}_{rsi} \in \{0, 1\}$
the outcome of the $i$-th of the $n^{a}_{rs}$ valid rollouts of scenario $s$ under arm $a$
(\emph{evo} for \cmd{evolved}) in run $r$, the gain is
\begin{equation}\label{eq:gain}
  \hat{\Delta} = \frac{1}{|\mathcal{R}|} \sum_{r \in \mathcal{R}} \frac{1}{|\mathcal{V}_r|}
  \sum_{s \in \mathcal{V}_r} d_{rs},
  \qquad
  d_{rs} = \bar{y}^{\,\mathrm{evo}}_{rs} - \bar{y}^{\,\mathrm{none}}_{rs},
  \qquad
  \bar{y}^{\,a}_{rs} = \frac{1}{n^{a}_{rs}} \sum_{i=1}^{n^{a}_{rs}} y^{a}_{rsi},
\end{equation}
and every other comparison between two arms is computed the same way. The study states
its questions as hypotheses, each with a margin that decides it (\cref{sec:results}); the
primary one, H1, is about this gain. Two calculations then say how far that number can be trusted.
\begin{itemize}
  \item \textbf{The confidence interval decides the hypothesis.} A two-way bootstrap
    builds \BootReps{} alternative versions of the study, each drawing the runs and the
    scenarios at random with replacement, independently of each other, and recomputes the
    gain on each. Drawing with replacement means that a run or a scenario can be drawn
    twice and another left out, as if the study had happened to sample differently; the
    spread of the recomputed gains shows how much the result depends on that luck. The
    middle \CILevel{} of those values, with the most extreme cut off equally at both ends,
    is the \CI{}; its two ends are the pair of numbers printed in brackets throughout the
    paper. In replicate $b$, the bootstrap draws $|\mathcal{R}|$ runs $r^{*}_{bi}$ from
    $\mathcal{R}$ and $|\mathcal{V}|$ scenarios $s^{*}_{bj}$ from
    $\mathcal{V} = \bigcup_{r} \mathcal{V}_r$, each uniformly; $J_{bi}$ holds the draws $j$
    whose scenario is valid in run $r^{*}_{bi}$, and $I_b$ the draws $i$ for which
    $J_{bi}$ is not empty. Then
    \begin{equation}\label{eq:bootstrap}
      \hat{\Delta}^{*}_{b} = \frac{1}{|I_b|} \sum_{i \in I_b} \frac{1}{|J_{bi}|}
      \sum_{j \in J_{bi}} d_{\,r^{*}_{bi}\, s^{*}_{bj}},
      \quad
      r^{*}_{bi} \overset{\mathrm{iid}}{\sim} \mathcal{U}(\mathcal{R}),\;
      s^{*}_{bj} \overset{\mathrm{iid}}{\sim} \mathcal{U}(\mathcal{V}),
      \quad
      \mathrm{CI} = \bigl[\,Q_{\alpha/2},\, Q_{1-\alpha/2}\,\bigr]\bigl(\{\hat{\Delta}^{*}_{b}\}_{b=1}^{B}\bigr),
    \end{equation}
    with $B = \BootReps$, $Q_q$ the $q$-quantile of the $B$ replicates, and $1 - \alpha$
    the \CILevel{} level. On the complete-case set every scenario is valid in every run,
    so each $J_{bi}$ holds every draw. The primary hypothesis is
    supported when the lower end of this interval is above zero. Every other hypothesis
    has its own margin, and \cref{sec:results}
    gives each verdict with it.
  \item \textbf{The permutation test gives the $p$-value.} It asks how often a gain at
    least this large, in either direction, would appear if the harness made no
    difference. If it made none, the arm label on each rollout would carry no
    information, so within each scenario of each run the labels are shuffled at random,
    which keeps every scenario's own difficulty fixed, and the gain is recomputed. This is
    repeated $B = \PermReps$ times. Each shuffle $\pi_b$ reorders, uniformly at random and
    independently, the $n^{\mathrm{evo}}_{rs} + n^{\mathrm{none}}_{rs}$ outcomes of each
    scenario $s$ of each run $r$, and gives them back to the two arms in their original
    numbers; $\mathfrak{S}_n$ is the set of the $n!$ orders of $n$ outcomes, and
    $\hat{\Delta}(\pi_b)$ is the gain recomputed after the shuffle. Then
    \begin{equation}\label{eq:perm}
      p = \frac{1 + \sum_{b=1}^{B} \mathbb{1}\bigl[\,\bigl|\hat{\Delta}(\pi_b)\bigr| \ge \bigl|\hat{\Delta}\bigr|\,\bigr]}{B + 1},
      \qquad
      \pi_b \sim \mathcal{U}\Bigl(\,\prod_{(r,s)} \mathfrak{S}_{\,n^{\mathrm{evo}}_{rs} + n^{\mathrm{none}}_{rs}}\Bigr).
    \end{equation}
    The $1$ added above and below counts the observed labels as one of the possible
    arrangements, so $p$ is never zero: it is at least $1/(B + 1)$.
\end{itemize}
Every comparison measured in a single run, the A/A arm, the ablations and the model
change, has an interval that resamples scenarios only. When
the gain is tested in each rule family on its own, Holm's correction adjusts the four
$p$-values for having made four tests, since the more tests are made, the likelier one of
them is to look significant by chance. With the four sorted from smallest to largest,
$p_{(1)} \le \dots \le p_{(4)}$, the adjusted values are
\begin{equation}\label{eq:holm}
  \tilde{p}_{(i)} = \min\Bigl\{1,\; \max_{j \le i}\, (5 - j)\, p_{(j)}\Bigr\},
\end{equation}
so the smallest is multiplied by four, the next by three, and so on, and no adjusted value
is smaller than the one before it. Proportions of candidates get Clopper--Pearson
intervals, an exact interval suited to small counts, which for $x$ successes in $n$
trials is
\begin{equation}\label{eq:cp}
  \Bigl[\,\mathcal{B}^{-1}_{\alpha/2}(x,\; n - x + 1),\;\; \mathcal{B}^{-1}_{1-\alpha/2}(x + 1,\; n - x)\,\Bigr]
  \;\xrightarrow{\;x = 0\;}\;
  \Bigl[\,0,\;\; 1 - \bigl(\tfrac{\alpha}{2}\bigr)^{1/n}\,\Bigr],
\end{equation}
with $\mathcal{B}^{-1}_q(a, b)$ the $q$-quantile of the beta distribution with parameters
$a$ and $b$, the lower end taken as $0$ when $x = 0$ and the upper end as $1$ when
$x = n$; when no candidate of a kind was kept, it reduces to the closed form on the
right, which gives the bound on kept placebos in \cref{sec:results:gates}. Fisher's exact
test, which suits small counts too, asks whether the known-good candidates were kept more
often than the placebos. The A/A arm carries its own rule: if its
interval excludes zero, every other result is suspect until explained. Every random draw
uses a fixed seed, so rebuilding the analysis gives the same intervals and $p$-values.
Invalid rollouts are excluded and counted. Every verdict of the loop reported in
\cref{sec:results} is the gates of \cref{sec:design:gates} applied to the recorded
rollouts, by a re-implementation of \cmd{score.sh} that agrees with it on all
\EvalSweeps{} sweeps of the evaluation runs.

%% file: sections/05-results.tex
\section{Results}\label{sec:results}

Each question of the evaluation is stated as a hypothesis with a \emph{margin}: the result
a measurement must reach for the hypothesis to count as supported. \Cref{tab:hypotheses}
lists the hypotheses tested, each with its margin, its result, its verdict and the
subsection that reports it. Each is reported where its question arises, stated in words,
with its verdict, including those that failed. A hypothesis is supported or not supported
as its margin gives; it is inconclusive when the measurement could not decide the
question, even where the letter of the margin is met, and the text then says why.
Differences are in percentage points of pass rate, \cmd{evolved} minus \cmd{none} unless
stated otherwise. Each interval is a \CI{}, the range of values the data leave plausible
(\cref{sec:setup}).

\begin{table}[p]
  \centering\footnotesize
  \setlength{\tabcolsep}{3pt}
  \caption{The hypotheses tested. The margin is the result a measurement must reach for
  the hypothesis to count as supported, and the verdict is what the margin gives, or
  inconclusive when the measurement could not decide the question. \cmd{none} is the harness with no skills or rules, and
  \cmd{evolved} the one a run built; A, B, C and E are the four rule families, and D is
  the control. Differences are in percentage points of pass rate, \cmd{evolved} minus
  \cmd{none} unless the row says otherwise, and the two numbers in brackets are the lower and upper ends of a \CI{}. The last column is the subsection that reports
  each one.}
  \label{tab:hypotheses}
  \begin{tabular}{@{}l>{\raggedright\arraybackslash}p{\dimexpr\linewidth*2/7\relax}
                  >{\raggedright\arraybackslash}p{\dimexpr\linewidth/5\relax}
                  >{\raggedright\arraybackslash}p{\dimexpr\linewidth*2/7\relax}ll@{}}
    \toprule
    \phantom{H16} & Hypothesis & Margin & Result & Verdict & Sec. \\
    \midrule
    H0 & A check of the lab, not of Cortex: with no skills or rules (\cmd{none}), the
      agent often breaks the house rules, so families A, B, C and E often fail, but it
      fixes the plain bugs of the control family, D. & On the held-out
      scenarios under \cmd{none}, D passes at least \HzeroBarD{} of the time, and at least
      three of A, B, C and E pass less than \HzeroBar{}. & Pass rates under \cmd{none}: D
      \NoneRateD{}; A \NoneRateA{}, B \NoneRateB{}, C \NoneRateC{}, E \NoneRateE{}. D is
      above its bar, and \HzeroBelowN{} of the four rule families are below \HzeroBar{}; E
      is the one above it. & supported & \ref{sec:results:gain} \\
    \addlinespace[3pt]
    H1 & \textbf{Primary.} What Cortex learns carries over to new work. On the held-out
      scenarios of A, B, C and E together, which no session played, the agent passes more
      often with the harness a run evolved (\cmd{evolved}) than with \cmd{none}. & The
      gain, \cmd{evolved} minus \cmd{none}, has an interval whose lower end is above
      zero: even the lowest value the data leave plausible is a gain. & On the \CompleteTasks{} rule-family scenarios valid in every run, the pass rate rose from
      \HeldNoneRate{} to \HeldEvolvedRate{}: a gain of \HoneEst{} [\HoneLo, \HoneHi].
      Permutation $p \HoneP$, the smallest the test can return. & supported & \ref{sec:results:gain} \\
    \addlinespace[3pt]
    H2 & Cortex does no harm where there is nothing to learn. D has no house rule and
      already passes under \cmd{none}, so the evolved harness cannot help there; it should
      not make D worse, and no run should keep an item written for D. & The change in D's
      pass rate has an interval whose lower end is above \MarginDFamily{}~points, so that
      a loss that large is not plausible; and no run keeps an item for D. & D changed by
      \HtwoEst{} [\HtwoLo, \HtwoHi]: the lower end is above \MarginDFamily{}. The upper
      end cannot be above zero, since \cmd{none} passes D \NoneRateD{} of the time and
      leaves nothing to gain. No run kept an item for D. & supported & \ref{sec:results:gain} \\
    \addlinespace[3pt]
    H3 & Cortex puts each kept item in a sensible loading tier, which decides
      when its text enters the agent's context, and the item fires (a rule loads, a skill
      is opened) mainly in the work of its own family. & Every kept item meets three
      criteria: its form (rule or skill) fits how it loads; its scope is limited; it fires more often on
      its own family's rollouts than on the others'. &
      \RubricGood{} of the \RubricScored{} kept items meet all three. The rest are rules
      on \cmd{shop/**}, a scope wide enough to load on almost every task of every family,
      and they fail only the third criterion. & not supported & \ref{sec:results:loop} \\
    \addlinespace[3pt]
    H4 & The harness spares the developer corrections while it is being built. In the
      evaluation runs Cortex's items go live as they are kept, so the scripted developer
      should correct the agent less often there than in the control run, which plays the
      same sessions with \cmd{/evolve} off. & Fewer corrections per session in the
      evaluation runs than in the control run, counting every session, including those
      before any item was kept. & \CorrEval{} corrections in \SessEval{} sessions
      (\CorrEvalRate{} per session) in the evaluation runs, against \CorrControl{} in
      \SessControl{} (\CorrControlRate{}) in the control run. With \ControlRuns{} control
      run, a permutation test cannot return $p$ below \HfourMinP{}, so the difference is
      descriptive, not a tested effect. & supported & \ref{sec:results:gain} \\
    \addlinespace[3pt]
    H5 & The gates, the rules that decide whether a candidate is kept, give the right
      verdict on candidates whose right verdict is known: \Placebos{} placebos (plausible
      advice that touches no house rule), \Harmful{} harmful ones (advice that contradicts
      a house rule) and \Positives{} known-good ones (the items of the hand-written
      harness \cmd{ideal}). They should rarely keep a placebo, kill every harmful
      candidate and keep the known-good ones. & The rate at which placebos are kept
      has an interval whose upper end is below \MarginPlaceboUpper{}; every harmful
      candidate is killed; and at least \MarginPositives{} of the \Positives{} known-good
      ones are kept. & Placebos:
      \PlaceboKept{} of \Placebos{} kept; with so few, the rate could still be as high as
      \PlaceboUpper{},
      below \MarginPlaceboUpper{}: met. Harmful: none kept, but only \GateHarmfulKilled{}
      of \Harmful{} killed, with no verdict on the other \GateHarmfulUnscored{}: not met.
      Known-good: \PositivesKept{} of \Positives{} kept: not met. & not supported &
      \ref{sec:results:gates} \\
    \addlinespace[3pt]
    H6 & The gates add value: what the loop buries would not have helped. \cmd{accept-all}
      is a run's evolved harness plus every candidate that run buried, and it should score
      no better than \cmd{evolved}. & The difference, \cmd{accept-all} minus
      \cmd{evolved}, has an interval whose upper end is below \MarginNotBetter{}~points,
      so that an advantage that large is not plausible. & In \AcceptAllRun{},
      \cmd{accept-all} minus \cmd{evolved} is \AcceptAllEst{} [\AcceptAllLo,
      \AcceptAllHi], so the margin is met. But the arm adds a single candidate, and no
      gate buried it: the scoring script reached no verdict, and the \cmd{/evolve} agent
      buried it as overlapping a live rule. The result rests on that burial, not on the
      gates. & inconclusive &
      \ref{sec:results:gates} \\
    \addlinespace[3pt]
    H7 & The scopes earn their place. Every kept item has a \cmd{paths} field, which loads
      it only when the agent works on matching files. \cmd{flat} is the same texts with
      every \cmd{paths} field removed; it should put more text into every turn's context
      and score no better than \cmd{evolved}. & \cmd{flat} has more always-on characters
      (text present on every turn) than \cmd{evolved}; and the difference, \cmd{flat}
      minus \cmd{evolved}, has an interval whose upper end is below
      \MarginNotBetter{}~points. & In \KitchenRun{}, \cmd{flat} has \FlatChars{} always-on
      characters against \cmd{evolved}'s \InitClaudeChars{}, and \cmd{flat} minus
      \cmd{evolved} is \FlatMinusEst{} [\FlatMinusLo, \FlatMinusHi]: the upper end is
      below \MarginNotBetter{}. Both parts are met. & supported &
      \ref{sec:results:cost} \\
    \bottomrule
  \end{tabular}
\end{table}

\begin{table}[p]
  \ContinuedFloat
  \centering\footnotesize
  \setlength{\tabcolsep}{3pt}
  \caption{The hypotheses tested, continued.}
  \begin{tabular}{@{}l>{\raggedright\arraybackslash}p{\dimexpr\linewidth*2/7\relax}
                  >{\raggedright\arraybackslash}p{\dimexpr\linewidth/5\relax}
                  >{\raggedright\arraybackslash}p{\dimexpr\linewidth*2/7\relax}ll@{}}
    \toprule
    & Hypothesis & Margin & Result & Verdict & Sec. \\
    \midrule
    H8 & Two harnesses state all four house rules: \cmd{kitchen}, which appends the
      contribution guide to \cmd{CLAUDE.md}, and \cmd{ideal}, written by hand with the
      rules known. \cmd{evolved} should come close to them on the rule families, with
      less always-on text. &
      \cmd{evolved} minus the better of the two has an interval whose lower end is above
      \textminus{}\MarginClose{}~points, and \cmd{evolved} has fewer always-on characters
      than either. & In \KitchenRun{}, pass
      rates: \cmd{evolved} \AblEvolvedRate{}, \cmd{kitchen} \AblKitchenRate{}, \cmd{ideal}
      \AblIdealRate{}. \cmd{evolved} leads both: \cmd{kitchen}, the better of the
      two, by \AblKitchenGap{}~points [\AblKitchenGapLo, \AblKitchenGapHi], an
      interval that reaches zero. Always-on characters, in that order:
      \InitClaudeChars{}, \KitchenChars{} and \IdealChars{}. & supported & \ref{sec:results:cost} \\
    \addlinespace[3pt]
    H9 & What the loop learns is repeatable. The \EvalRuns{} evaluation runs are
      independent, each starting from a fresh copy of the repository; they should learn
      the same families, as items of similar form. & A run has learned a family when it
      keeps an item for it. Of the families learned in any run, at least half are learned
      in every run. & All four rule families were learned in every run: each of the
      \EvalRuns{} runs kept an item for A, B, C and E, and none for D. Every run but
      \NarrowRun{} chose the same scopes. & supported & \ref{sec:results:loop} \\
    \addlinespace[3pt]
    H10 & \cmd{/prune}, which re-measures live items by removal, deletes only dead
      weight. Beside \PruneRun{}'s own \RunOneItems{} items, which it should keep, the
      prune test plants \Planted{} items: \PlantedPlacebos{} placebo always-on skills, to
      delete; a gated skill that no task reaches, to skip; and an always-on skill that
      never fires, to report as unmeasured. & Every planted item gets the verdict expected
      of it, and no item that never fired is deleted. & \cmd{/prune} kept \PruneRun{}'s
      own \RunOneItems{} items. Of the \Planted{} planted items, \PlantedMatched{} got the
      expected verdict (\PruneMatched{} of \PruneTested{} overall): one placebo was
      deleted and the unreachable skill skipped. The other placebo was kept:
      \PrunePlaceboLostTasks{} tasks lost runs without it, and the agent never opened it. The skill that never fires got no verdict. No item that
      never fired was deleted. & not supported & \ref{sec:results:model} \\
    \addlinespace[3pt]
    H11 & A skill is offered to the agent by its description, one line in a listing; its
      body enters the context only if the agent opens it. The description should carry much
      of the skill's effect by itself. \cmd{desc-only} is \cmd{evolved} with each skill's
      body replaced by neutral filler. & \cmd{desc-only} recovers at least
      \MarginDescShare{} of \cmd{evolved}'s gain over \cmd{none}, with the lower end of
      that share's interval above zero. & It recovered \DescRecovered{}
      [\DescRecoveredLo, \DescRecoveredHi], which meets the letter of the margin. But only
      \DescRewritten{} of \KitchenRun{}'s \DescItems{} items is a skill and could be
      rewritten; the rest are rules and kept their words. In C, the rewritten skill's
      family, it recovered \DescShareC{}. One skill cannot settle the question. &
      inconclusive & \ref{sec:results:cost} \\
    \addlinespace[3pt]
    H12 & The development run, D0, used an earlier version of Cortex than \EvalTag{}, the
      version evaluated. Its harness, measured
      on the same held-out scenarios with the evaluation's tooling, should gain in line
      with the evaluation runs. & D0's gain lies inside the interval of the evaluation
      runs' gain, taken over every valid scenario. & D0's harness gained
      \DzeroGain{}~points, inside the evaluation runs' interval, [\HoneAllLo, \HoneAllHi],
      and within the range of the single runs' own gains, \RunGainAllLo{} to
      \RunGainAllHi{}. & supported & \ref{sec:results:gain} \\
    \addlinespace[3pt]
    H13 & A harness evolved for one model may be partly redundant under a stronger one.
      \cmd{/prune}, run on \ModelChangeSource{}'s harness with its rollouts on
      \StrongModelName{} instead of \RolloutModelName{}, should find items the weaker
      model needed and the stronger one no longer does. & \cmd{/prune} finds at least one
      such item removable. & \RemovableItems{} of the harness's \DecidedItems{} items are
      removable under \StrongModelName{}: the money rule and the exporter checklist. The
      money rule is the clear case, since the stronger model passed every held-out money
      scenario with no harness; the exporter checklist is less certain. & supported &
      \ref{sec:results:model} \\
    \addlinespace[3pt]
    H14 & The loop also helps outside the lab. \SecondRepo{} is a real open-source project
      that nobody in this study built; the loop runs once on tasks mined from its history,
      and the harness it evolves should raise the pass rate on the project's newest
      tasks. & A positive gain, \cmd{evolved} minus \cmd{none}, on the
      \SecondRepoHoldout{} held-out tasks. & Nothing was learned. In \ExtFirstTime{} of
      the \ExtTasks{} sessions the agent passed the project's tests and linter at the
      first attempt, so nothing recurred and no item was kept. \cmd{evolved} was identical
      to \cmd{none}, which left nothing to compare. & inconclusive &
      \ref{sec:results:external} \\
    \addlinespace[3pt]
    H15 & The optional judge rates a candidate's \emph{relevance}: the share of the
      tasks it would load on that it is actually about. Relevance should separate the
      candidates the gates keep from those they kill on a regression (a task made
      worse); \emph{breadth}, the share of all tasks it would load on, which needs no
      judge, should not. & To separate two groups,
      every kept candidate must score above every killed one. Relevance must separate them,
      and breadth must not. & Relevance did not separate them: the lowest kept candidate
      scored \RelKeptLowEval{} and the highest killed one \RelKilledHighEval{}. Neither
      did breadth (\BreadthKeptLowEval{} against \BreadthKilledHighEval{}). Only
      \LoopKilledRegression{} candidates were killed on a regression, which makes a thin
      comparison. & not supported & \ref{sec:results:judge} \\
    \addlinespace[3pt]
    H16 & For a skill, the judge also predicts its invocation rate, how often the agent
      will open it. The prediction should track the rate measured in the rollouts. & With
      so few skills, only the order is asked for: the predictions rank the skills in the
      same order as the measured rates. & The order is preserved for the
      \JudgeItems{} skills tested, \JudgeItemsDev{} of them kept by D0, which meets the
      letter of the margin. But \JudgeTied{}
      of the predictions are tied at the top, so the order rests on the remaining skill. & inconclusive & \ref{sec:results:judge} \\
    \bottomrule
  \end{tabular}
\end{table}

\subsection{Held-out gain}\label{sec:results:gain}

\paragraph{The baseline.} Without a harness the agent passes the control family and often
breaks the house rules. On the held-out scenarios, \cmd{none}'s pass rate was \NoneRateD{}
on family D, but \NoneRateA{} on family A, \NoneRateB{} on B, \NoneRateC{} on C and
\NoneRateE{} on E. H0, that the model without a harness often breaks the house rules but
passes the control family, is supported. Its margin asks that family D pass at least
\HzeroBarD{} of the time and that at least three of the four rule families pass less than
\HzeroBar{}: D passed \NoneRateD{}, and \HzeroBelowN{} rule families are below
\HzeroBar{}; E is the one above it. H0 checks the lab rather than Cortex: its scenarios
were built so that the model breaks the rules unaided (\cref{sec:limitations}). The rules
are written down, in the repository's
contribution guide (\cmd{CONTRIBUTING.md}), but the agent opened that file in only
\GuideReadNone{} of \cmd{none}'s held-out rollouts.

\begin{figure}[!htbp]
  \centering
  \includegraphics[width=\linewidth]{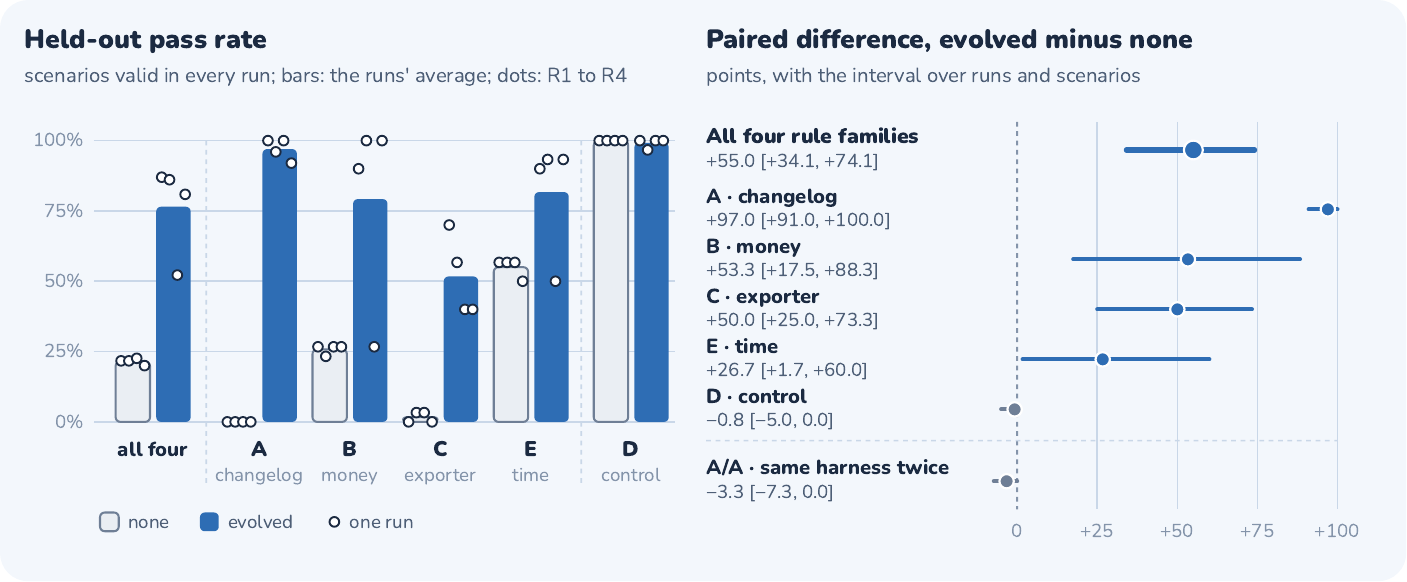}
  \caption{Held-out results on the scenarios valid in every run. Left: pass rate; each bar
  averages the \EvalRuns{} evaluation runs, and each dot is one run, R1 to R4 from left
  to right. Right: \cmd{evolved}
  minus \cmd{none}, with the two-way bootstrap interval over runs and scenarios. The bottom
  row is the A/A check, the \cmd{none} harness measured twice in \AARun{} over all five
  families, whose interval resamples scenarios only. Family D is the control.}
  \label{fig:headline}
\end{figure}

\paragraph{The primary result.} H1, the study's primary hypothesis, is that \cmd{evolved}
beats \cmd{none} on the held-out scenarios of the four rule families taken together; its
margin is an interval whose lower end is above zero. On the \CompleteTasks{} rule-family
scenarios valid in every run, \cmd{evolved} beat \cmd{none} by \HoneEst{}~points (\CI{}
[\HoneLo, \HoneHi], permutation $p \HoneP$), so H1 is supported
(\cref{tab:main,fig:headline}). With \PermReps{} shuffles, that $p$ is the smallest the
test can return: no shuffle of the arm labels produced a gain as large as the one
observed. Every rule family gained: A by \FamAEst{} [\FamALo, \FamAHi], B by \FamBEst{}
[\FamBLo, \FamBHi], C by \FamCEst{} [\FamCLo, \FamCHi] and E by \FamEEst{} [\FamELo,
\FamEHi]. The permutation test gives each family $p \FamHolmP$ after Holm's correction,
which raises each $p$-value to allow for having tested four families instead of one; as
for the four together, no shuffle reached any family's gain. That test holds the measured
scenarios and runs fixed, while the intervals, which are not adjusted for the four
families, also allow for which ones were drawn, and the clock family's lower end lies
close to zero. H2 is that the harness does not
harm the control family (the lower end of its interval stays above
\MarginDFamily{}~points) and that no run keeps an item for it. The control family moved
by \HtwoEst{} [\HtwoLo, \HtwoHi], and no run kept an item for it, so H2 is supported.
Counting every rule-family scenario in each run where it was valid, so also the one
excluded from a single run (\cref{sec:setup}), the gain on the four rule families
together is
\HoneAllEst{} [\HoneAllLo, \HoneAllHi]. Every rollout behind these numbers was valid.

\begin{table}[H]
  \centering\small
  \caption{The held-out gain, family by family: \cmd{evolved} against \cmd{none} on the
  scenarios valid in every evaluation run, R1 to R4. Pass rates are over rollouts, and
  \emph{Rollouts} counts both arms. The gain is the paired difference in pass rate, in
  points, averaged over the scenarios of each run and then over the runs; its \CI{} is the
  two-way bootstrap and $p$ the permutation test (\cref{sec:setup}). Holm's correction
  adjusts the four rule families' $p$-values for having made four tests. In the
  rule-family rows no shuffle reached the observed gain, so each $p$ is the smallest the
  test can return with \PermReps{} shuffles; the test keeps the measured scenarios and
  runs fixed, while the interval also allows for which ones were drawn. Family D is the
  control.}
  \label{tab:main}
  \begin{tabular}{@{}lrrrrrlrr@{}}
    \toprule
    & & & \multicolumn{2}{c}{Pass rate} & & & & \\
    \cmidrule(lr){4-5}
    Family & Scenarios & Rollouts & \cmd{none} & \cmd{evolved} & Gain & \CI{} & $p$ & Holm $p$ \\
    \midrule
    \input{generated/tab-main.tex} 
    \bottomrule
  \end{tabular}
\end{table}

\paragraph{The noise floor.} The A/A check measures how far two measurements of one and
the same harness can differ. \cmd{none2} is the \cmd{none} harness again,
measured as if it were a different one; on all five families of \AARun{}, the two
differed by \AAEst{}~points [\AALo, \AAHi], with an interval that resamples scenarios
(permutation $p = \AAP$). By the rule of \cref{sec:setup}, an A/A interval that excluded
zero would put every other result in doubt. This interval ends at zero, so the rule does not apply,
but only just: on each of the \AADiffering{} scenarios where the two measurements differ,
\cmd{none} passed more often than \cmd{none2}. \cmd{none2} ran in a later batch than
\cmd{none}, together with the ablation arms of \cref{sec:results:cost}, so the difference
may hold a shift between batches (a change in conditions between the two periods in
which the rollouts ran) as well as noise.

\begin{figure}[!htbp]
  \centering
  \includegraphics[width=\linewidth]{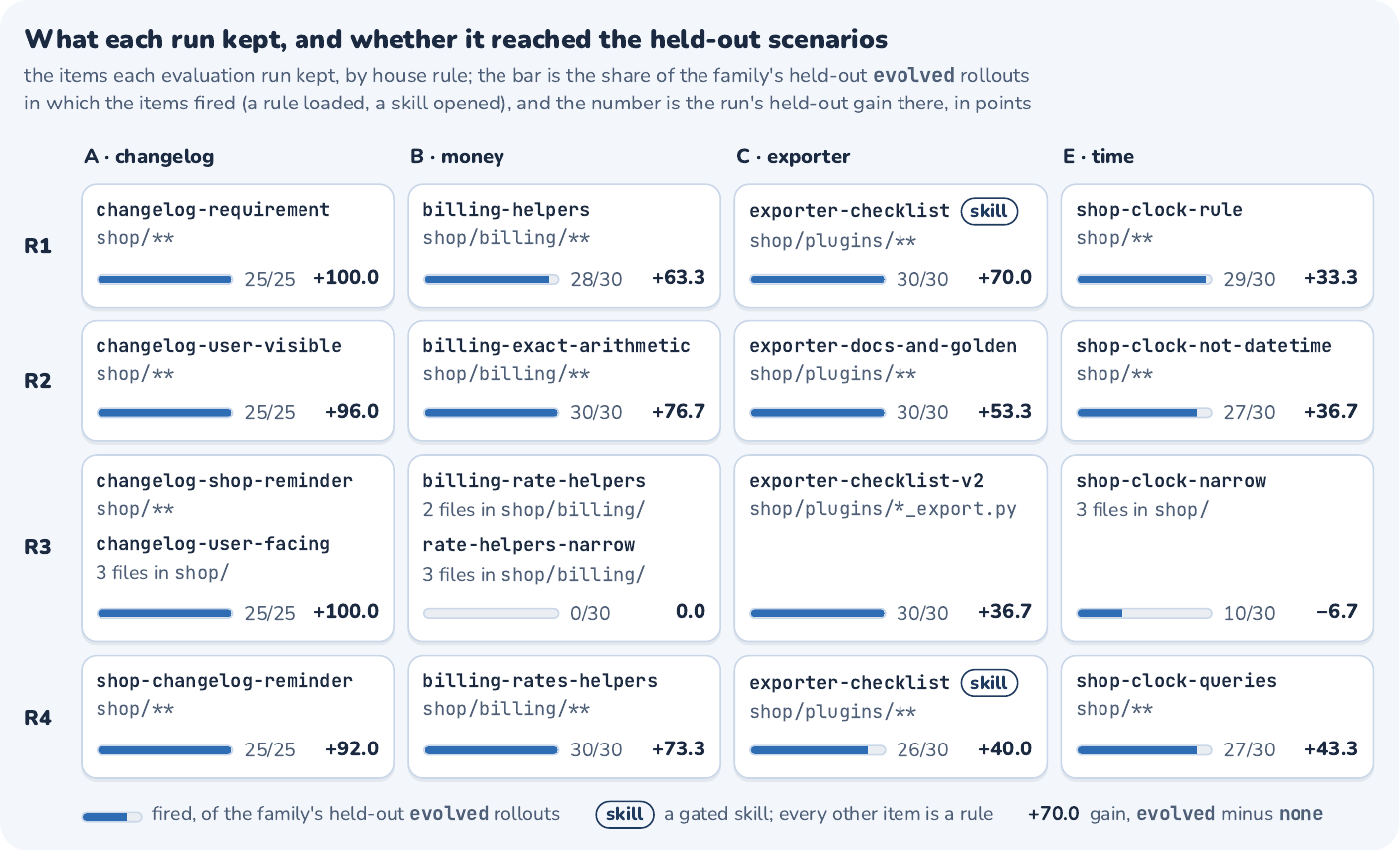}
  \caption{What each evaluation run kept, by house rule, and whether it reached the
  held-out scenarios: the share of the family's held-out \cmd{evolved} rollouts (scenarios
  valid in every run) in which the run's items for that family fired (a rule loaded, a skill
  opened), and the run's gain on the family. \NarrowRun{}'s money rules, which name
  \NarrowMoneyFiles{} files, loaded in none.}
  \label{fig:kept}
\end{figure}

\paragraph{The runs.} Every run gained on the four rule families taken together, by
different amounts (the dots in \cref{fig:headline}), though not on every family:
\NarrowRun{} gained nothing on money and lost on the clock (\cref{fig:kept}). \NarrowRun{}
gained least: its pooled pass rate on the \CompleteTasks{} scenarios valid in every run
rose from
\NarrowPoolNone{} to \NarrowPoolEvolved{}, where the other runs reached between
\OthersPoolLo{} and \OthersPoolHi{}. \Cref{sec:results:loop} traces why. Every held-out
scenario under every arm of R1 to R4 is shown in \cref{fig:heatmap}, in \cref{app:results}.

\paragraph{Training against held-out scenarios.} The benchmark also replayed the training
scenarios, the ones the sessions had played, under \cmd{none} and \cmd{evolved}
(\cref{sec:setup}). Measured on the \SeenTasks{} of them in the four rule families, the
gain would have been \SeenGainEst{}~points [\SeenGainLo, \SeenGainHi] instead of
\HoneEst{}. More of that difference lies in the baseline than in the harness: without a
harness the agent passed \SeenNoneRate{} of the training rollouts in these families and
\HeldNoneRate{} of the held-out ones, so the training scenarios left more to gain, while
the evolved harness's own pass rate went from \SeenEvolvedRate{} to \HeldEvolvedRate{}.
By family, the evolved rate went from training to held-out as follows: A from
\SeenEvoA{} to \HeldEvoA{}, B from \SeenEvoB{} to \HeldEvoB{}, C from \SeenEvoC{} to
\HeldEvoC{}, and E from \SeenEvoE{} to \HeldEvoE{}. It fell most for exporters, and fell
there in every run. For money, the fall is \NarrowRun{}'s, the run that narrowed its
money rule (\cref{sec:results:loop}); without it, the rate rose. For the clock, the rate
held but the gain did not, because the held-out clock scenarios were easier without a
harness: the agent alone passed \HeldNoneE{} of their rollouts, against \SeenNoneE{} of
the training ones. The comparison is descriptive and tests no hypothesis: the two sets are
different scenarios, attempted \TrainK{} times each against \PrimaryK{}; a training
scenario starts from the commit its session started from, not from the run's final code;
and its session's fix is in the repository's history, which a rollout can read
(\cref{sec:limitations}).

\begin{figure}[!htbp]
  \centering
  \includegraphics[width=\linewidth]{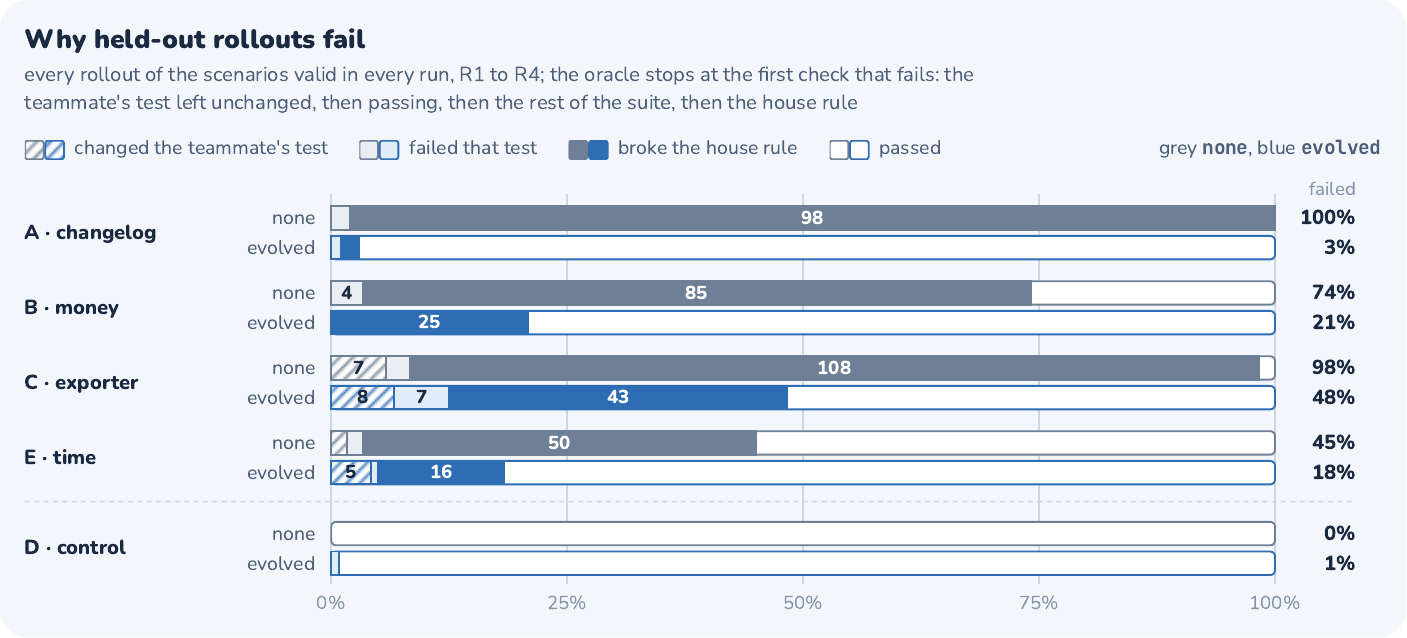}
  \caption{Why held-out rollouts fail, in the \EvalRuns{} evaluation runs, on the scenarios
  valid in every run. Each bar is all of an arm's rollouts on a family, split by the first
  check the oracle found failing: the agent edited the teammate's test (hatched), left it
  unchanged but did not make it pass (light), or passed the tests and broke the house rule
  (solid); white is a pass. The number inside a segment counts its rollouts, and the number
  at the right is the share that failed. Without a harness, nearly every rule-family
  failure passed its tests and broke the house rule.}
  \label{fig:failures}
\end{figure}

\paragraph{Why rollouts fail.} The oracle, the lab's own script that scores a rollout
(\cref{sec:setup}), makes the reviewer's checks in order and stops at the first that
fails, so each failed rollout has exactly one cause in \cref{fig:failures}. First, the teammate's
test, the test file the scenario starts from, must be unchanged: a rollout that edits it
has moved the goalposts. Second, that test must pass: otherwise the task is not solved.
Third, the rest of the test suite must still pass. Last, the house rule must hold: the
linter, which checks the rules of families B, C and E, must be clean, and in family A
the changelog must have gained a line. None of these rollouts failed on the rest of the suite. Without a harness,
\FailRuleShareNone{} of the rule families' failed rollouts (\FailRuleNone{} of
\FailNone{}) passed the tests and broke the house rule: the agent solved the task and
missed the convention, the failure the lab was built around.
The evolved harness turned most of them into passes. Of its \FailEvolved{} failures,
\FailRuleEvolved{} still broke the rule, \FailCRuleEvolved{} of them in the exporter family.
A rollout that changes the teammate's test fails before anything else is checked; that
happened in \TamperedNone{} \cmd{none} rollouts and \TamperedEvolved{} \cmd{evolved} ones,
all in families C and E.

\begin{figure}[!htbp]
  \begin{minipage}[t]{\dimexpr(\linewidth-2em)/2\relax}
    \includegraphics[width=\linewidth]{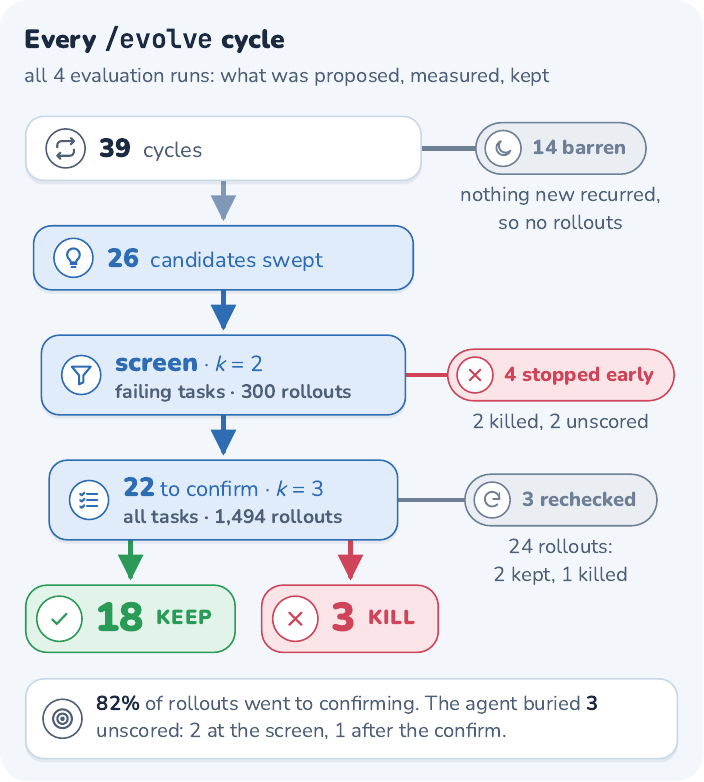}
    \caption{Every recorded \cmd{/evolve} cycle of the \EvalRuns{} evaluation runs, each
    candidate's verdict computed again from its rollouts by the rules of
    \cref{sec:design:gates}. A cycle in which nothing new recurred runs no rollouts; the
    screen, on the failing tasks ($k=\KScreen$), stops weak candidates before the confirm
    sweep over all tasks ($k=\KConfirm$), where most rollouts go. The cycle drawn split
    in \cref{fig:cycles} swept \SplitCycleSwept{} candidates. \cmd{score.sh} returned RERUN
    for \LoopUnscored{} candidates, which the \cmd{/evolve} agent buried without a verdict.}
    \label{fig:funnel}
  \end{minipage}\hfill
  \begin{minipage}[t]{\dimexpr(\linewidth-2em)/2\relax}
    \includegraphics[width=\linewidth]{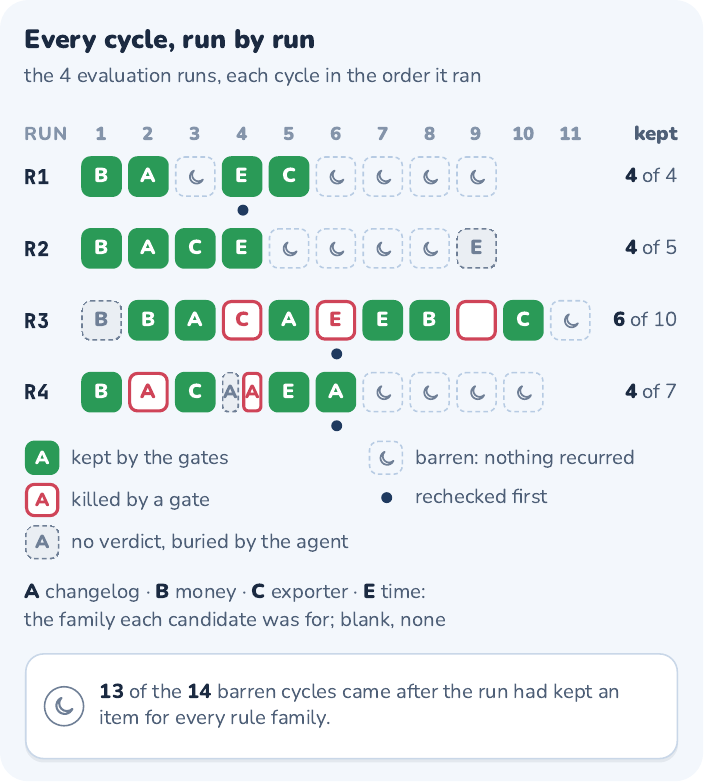}
    \caption{The same cycles, run by run; barren cycles, which sweep nothing, are placed
    where the journal recorded them. Each cell is one cycle, numbered in the order it ran:
    its colour is the verdict the gates give, and its letter the family the candidate was
    written for. Every run kept an item for each rule family, and most barren cycles came
    after that, when nothing new recurred. The cycle of \SplitCycleRun{} drawn split swept
    \SplitCycleSwept{} candidates.}
    \label{fig:cycles}
  \end{minipage}
\end{figure}

\paragraph{Corrections, and the development run.} H4 is that the scripted developer has
to correct the agent less often in the evaluation runs, where Cortex's items go live as
they are kept, than in the control run, where \cmd{/evolve} is off; the count covers
every session, including those before any item was kept. The evaluation runs' sessions needed \CorrEval{} corrections in
\SessEval{} sessions (\CorrEvalRate{} per session), and the control run's \CorrControl{}
in \SessControl{} (\CorrControlRate{}). The margin, fewer corrections per session than in
the control run, is met, so H4 is supported. But with a single control run beside the
\EvalRuns{} evaluation runs, a
permutation test that shuffles which run is the control has too few arrangements to give
$p$ below \HfourMinP{}, so we read the difference as description, not as a tested effect.
H12 is that the result of the development run, D0 (\cref{sec:setup}), holds at
\EvalTag{}. Its harness, measured on the same held-out scenarios
with the evaluation's tooling, gained \DzeroGain{}~points, inside the evaluation runs'
interval over every valid scenario, [\HoneAllLo, \HoneAllHi], so H12 is supported. That
interval is for the mean of the \EvalRuns{} runs, and a single run varies more: on the
same scenarios the runs' own gains ranged from \RunGainAllLo{} to \RunGainAllHi{}, and
the development run's gain lies within that range as well.

\subsection{What the loop did}\label{sec:results:loop}

\paragraph{Cycles and verdicts.} In the \EvalRuns{} evaluation runs \cmd{/evolve} recorded
\LoopCycles{} cycles (\cref{fig:funnel}; run by run in \cref{fig:cycles}). Of these,
\LoopBarren{} were barren, most of them after the run had kept an item for every rule
family; the other \LoopActive{} swept \LoopSwept{} candidates between them, one cycle
sweeping \SplitCycleSwept{}. \cmd{score.sh} kept \KeptItems{} and killed \LoopKilled{};
for the other \LoopUnscored{} it returned RERUN, and the \cmd{/evolve} agent buried them
without a verdict rather than sweep them again. A recheck was needed for \Rechecked{} candidates, and \RecheckKept{} of them were
kept after it. Of the \LoopRollouts{} sweep rollouts, \LoopConfirmRollouts{} were spent on
confirm sweeps, which only candidates that had passed a screen paid for; every confirm
sweep, with the runs it won net and its worst drop on one task, is drawn in
\cref{fig:sweeps}.

\paragraph{What the runs learned.} H9 is that independent runs learn the same families,
in similar forms. It is supported: every run kept an item for each of the four rule
families, and none for the control (\cref{fig:kept}, and every kept item in
\cref{tab:kept}). Every run but \NarrowRun{} chose the same
scopes: \cmd{shop/**} for the changelog and the clock, \cmd{shop/billing/**} for money
and \cmd{shop/plugins/**} for exporters; they differed in form only for exporters, a gated
skill in two runs and a rule in the third. \NarrowRun{} narrowed its money and clock
rules: instead of a folder pattern, their \cmd{paths} name a short list of individual
files, so the rules load only when the agent reads one of those files. Its held-out
result shows what that cost. Its first money rule covered \cmd{shop/billing/**}. The
confirm sweep, over the \BroadMoneyTasks{} tasks then in the suite, won
\BroadMoneyNetRuns{} runs and lost none, but exposure was unknown in \BroadMoneyUnknown{}
of its rollouts, so \cmd{score.sh} returned RERUN (\cref{sec:design:measure}). The
\cmd{/evolve} agent buried the rule and, in the next cycle, proposed a narrower one. It
and a later rule name \NarrowMoneyFiles{} files between them,
and the gates kept both. They loaded in \NarrowMoneyFired{} of the run's \NarrowMoneyN{}
held-out money rollouts, where the run's money pass rate stayed at \NarrowMoneyNone{} in
both arms;
the other runs' money rules loaded in at least \BroadMoneyFiredMin{} of \NarrowMoneyN{}.
Its clock rule on \cmd{shop/**} was killed when a task the live harness passed in all its
runs lost one run in the confirm and another in the recheck (gate~3). The rule that
replaced it names \NarrowClockFiles{} files and loaded in \NarrowClockFired{} of
\NarrowClockN{} held-out clock rollouts, and the run's clock pass rate fell from
\NarrowClockNone{} to \NarrowClockEvolved{}. The gates judge a candidate on the tasks the
loop has harvested, where these rules did load. Nothing in them can see that a scope is
too narrow for tasks the loop has not yet met.

\paragraph{The tier rubric.} H3 is that every kept item lands in a sensible tier and
fires mainly in its own family's work. The lab scored each kept item on three criteria:
its form fits how it loads (a skill fired at least once, a rule has \cmd{paths}); its
scope is limited (it has \cmd{paths}, or is a skill); and it fires more often on its own
family's rollouts than on the others'. The margin was every item scoring three of three.
Of the \RubricScored{} kept items, \RubricGood{} did, so H3 is not supported. Every item
that failed is a rule on \cmd{shop/**} and failed only the third criterion. Rules on
\cmd{shop/**} load on almost every task of every family (in \ShopWideLo{} to
\ShopWideHi{} of rollouts), and the \ShopWidePassed{} that passed it did so narrowly,
firing on their own family at most \ShopWideMargin{}~points more often than on the
others.

\begin{figure}[!htbp]
  \centering
  \includegraphics[width=\linewidth]{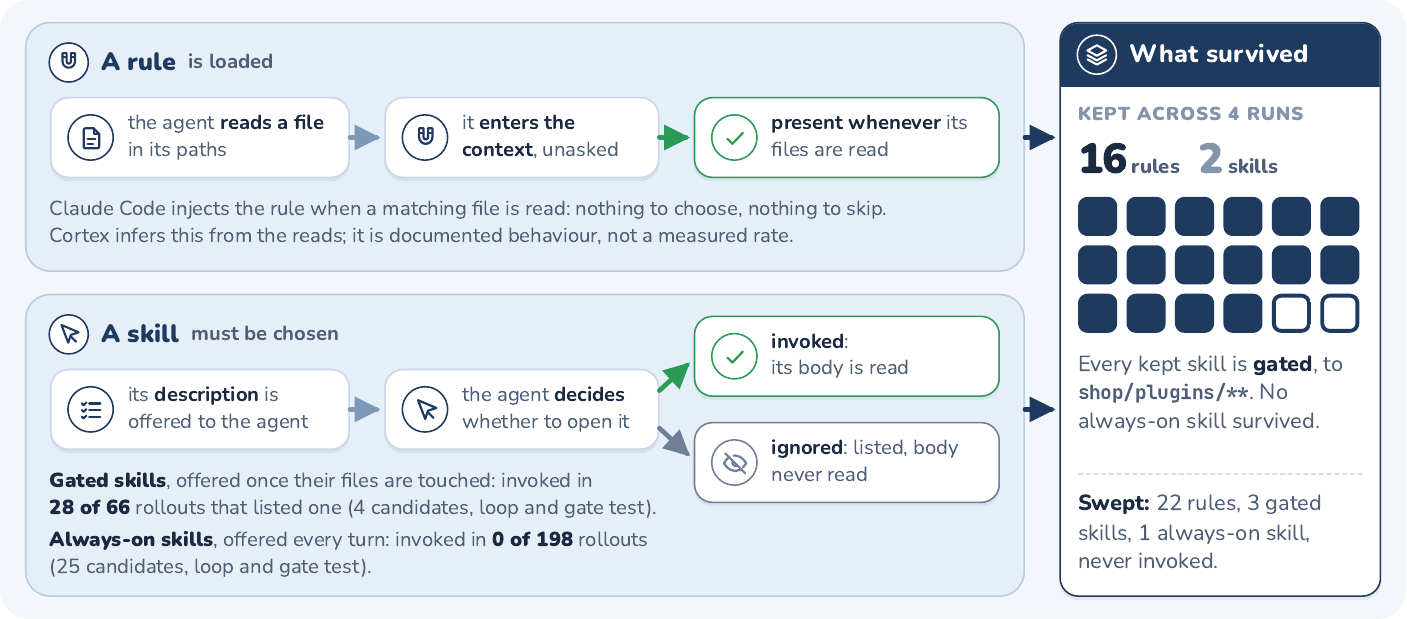}
  \caption{A rule is loaded; a skill must be chosen. Claude Code injects a path-scoped rule
  whenever the agent reads a matching file. A skill is offered by its description, and its
  body enters the context only if the agent opens it: in their own sweeps and in the gate
  test, path-gated candidates were opened in some of the rollouts that listed them, and
  always-on ones in none. Of the items the loop kept, \KeptRules{} are rules
  and \KeptSkills{} are path-gated skills; most candidates were rules too.}
  \label{fig:loaded}
\end{figure}

\paragraph{Loaded or chosen.} Of the \KeptItems{} kept items, \KeptRules{} are rules
(\cref{fig:loaded}). The other \KeptSkills{} are skills, each an exporter checklist gated
on \cmd{shop/plugins/**}. A gated candidate, in its own sweeps and in the gate test, was
listed to the agent in \GatedOffered{} rollouts and opened in \GatedInvoked{}; an
always-on candidate, likewise in its own sweeps and in the gate test, was listed in
\AlwaysOnOffered{} and opened in \AlwaysOnInvoked{}. Elsewhere such a skill was opened, but rarely: the gate test's
placebos that were planted unchanged in the prune test were listed there
\PruneMomentListed{} times and opened \PruneMomentOpened{}. A rule needs no such decision: it is in the context whenever
the agent reads a file it covers. The gate test and the ablations below meet the same
asymmetry.

\subsection{The gates against known answers}\label{sec:results:gates}

\paragraph{The verdicts.} The gate test ran \GateCandidates{} candidates with known
answers through the real pipeline (\cref{sec:setup}). H5 is that the gates are
calibrated: they rarely keep a placebo, kill every harmful candidate, and keep at least
\MarginPositives{} of the \Positives{} known-good ones. None of the \Placebos{} placebos was kept; with so
few candidates, the rate at which the gates keep placebos could still be as high as
\PlaceboUpper{} (the upper end of a two-sided Clopper--Pearson \CI{}). None of the
\Harmful{} harmful candidates was kept either, but \cmd{score.sh} killed only
\GateHarmfulKilled{} of them and reached no verdict on the other \GateHarmfulUnscored{}
(below). And \PositivesKept{} of the \Positives{} known-good candidates were kept, short
of the \MarginPositives{} the margin required. H5 is not supported.
Fisher's exact test, which asks whether the known-good candidates were kept more often
than the placebos, gives $p = \FisherP$.

\paragraph{What decided them.} \Cref{tab:gatetest} gives each kind of candidate's
verdicts. No placebo was opened on its screen: each was listed in all
\ScreenCandRollouts{} candidate rollouts and invoked in none. \cmd{score.sh} killed
\GatePlaceboKilled{} of them for never firing (gate~5) and returned RERUN for the other
\GatePlaceboUnscored{}, whose screen tasks failed in every run of both arms, so that no
gain was possible; the test's procedure, which stops at the screen, recorded all
\Placebos{} as KILLs. In all, \GateNoiseGain{} candidates (\GateNoisePlacebo{} placebos,
\GateNoiseHarmful{} harmful skills and \GateNoisePositive{} known-good skill) showed a
positive gain on their screen although the agent never opened them, \GateNoiseSingle{} of
them by a single run on a single task. Gate~5, which asks that a candidate fired, turned
them away, not their gain.

\paragraph{Harm was not tested.} The \GateHarmfulKilled{} harmful skills were never opened
and fell to gate~5, like the placebos. The \HarmfulRules{} harmful rules loaded in every
candidate rollout of their screens, but a harmful item shows itself only on a task that
passes without it, and their screens ran, as screens do, on failing tasks. One screen
included a task that passed without the rule, and the rule broke it; \cmd{score.sh} still
returned RERUN, because no task in that screen could show a gain. The test therefore shows
that the gates never keep what never fires, not that they catch harm. In the loop itself,
\LoopKilledRegression{} of the \LoopKilled{} KILLs failed a regression gate, and
\LoopKilledRegressionOnly{}, the clock rule of \NarrowRun{} above, failed nothing else.

\paragraph{The positives.} The known-good candidates split along the tier line: the rule
and the gated skill were kept, and the \IdealAlwaysSkills{} always-on skills, never
opened on their screens, were lost.

\paragraph{Do the gates add value?} H6 is that they do: a harness that also keeps what
the loop buried should be no better than \cmd{evolved}, and the margin asks that the
upper end of its difference's interval be below \MarginNotBetter{}~points. \cmd{accept-all}, \AcceptAllRun{}'s harness plus
the one candidate that run buried, scored \AcceptAllEst{}~points against \cmd{evolved}
[\AcceptAllLo, \AcceptAllHi], an interval that resamples scenarios only, since the arm
ran in a single run. That meets H6's margin, though the permutation test, at
$p = \AcceptAllP$, does not by itself rule out chance. That candidate, a second clock rule, was not buried by a gate: both arms
passed every run of its screen, \cmd{score.sh} returned RERUN, and the \cmd{/evolve} agent
buried it as overlapping a live rule. The loss falls on families A (\AcceptAllFamA) and C
(\AcceptAllFamC) more than on E (\AcceptAllFamE), the family the added rule is about. The
rule, on \cmd{shop/**}, was in context in every A and C rollout, so it could have caused
the loss there, but the arm also ran in a later batch, so part of the difference may be
the shift between batches. The margin is met, but it rests on one burial the
\cmd{/evolve} agent made, not on the gates, so H6 is inconclusive.

\subsection{What the harness costs}\label{sec:results:cost}

\begin{figure}[!htbp]
  \begin{minipage}[t]{\dimexpr(\linewidth-2em)/2\relax}
    \includegraphics[width=\linewidth]{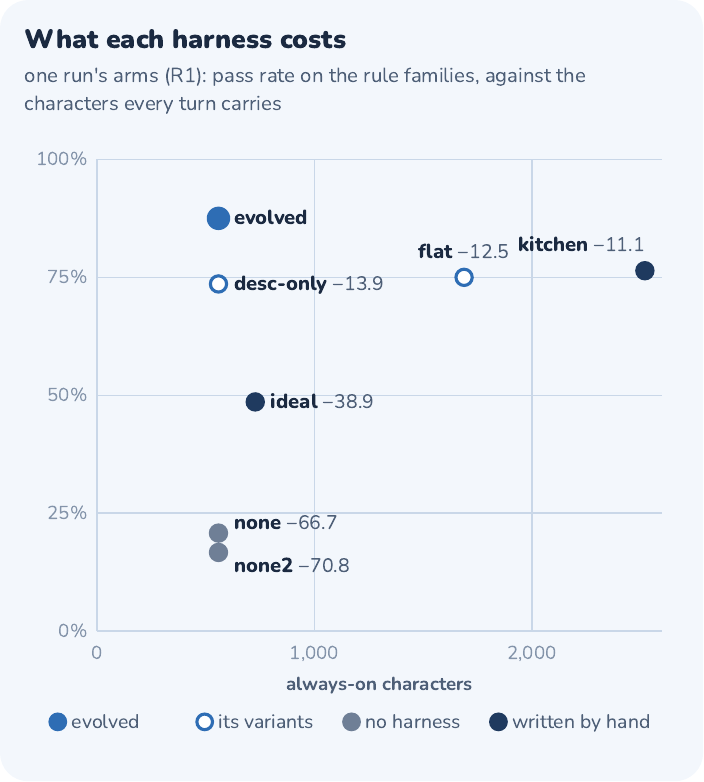}
    \caption{What each harness costs, in \KitchenRun{}: pass rate on the four rule
    families against the characters present in every turn's context. Filled blue is the
    harness the loop evolved; hollow blue, its texts with every scope removed (\cmd{flat})
    or every skill body replaced (\cmd{desc-only}); grey, no harness (\cmd{none}),
    measured twice; navy, the two harnesses that state the rules outright: \cmd{kitchen},
    the contribution guide appended to \cmd{CLAUDE.md}, and \cmd{ideal}, written by hand.
    Beside each arm, its difference from \cmd{evolved}, in points.}
    \label{fig:ablations}
  \end{minipage}\hfill
  \begin{minipage}[t]{\dimexpr(\linewidth-2em)/2\relax}
    \includegraphics[width=\linewidth]{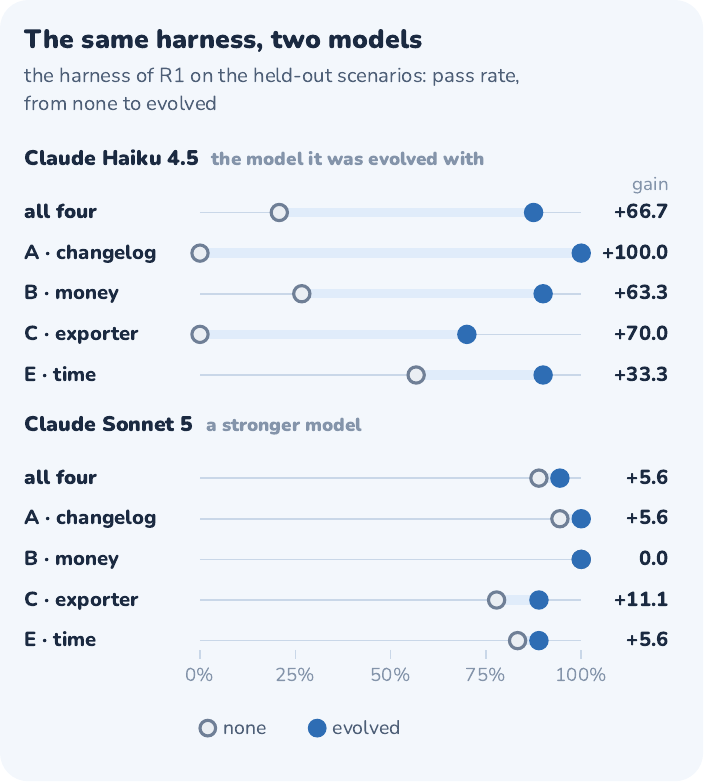}
    \caption{The harness of \ModelChangeSource{} on the held-out scenarios, under the model
    it was evolved with and under a stronger one. Each line runs from \cmd{none} (grey) to
    \cmd{evolved} (blue), and the number beside it is the gain in points. Under the
    stronger model, little is left to gain.}
    \label{fig:model}
  \end{minipage}
\end{figure}

\paragraph{Always-on context.} Two hypotheses concern this cost. H8 is that
\cmd{evolved} comes within \MarginClose{}~points of the better of two harnesses that
know all four rules, \cmd{kitchen} and \cmd{ideal}, while putting fewer characters into
every turn's context. H7 is that the scopes earn their place: \cmd{flat}, the same texts
without their \cmd{paths}, puts more text into every turn and is no better (the upper end
of its difference's interval below \MarginNotBetter{}~points). \Cref{fig:ablations} places
each arm of \KitchenRun{} by its pass rate on the rule families and by the characters it
puts into every turn's context. The rates are over every scenario valid in that run, one
more than the set valid in every run, so they differ slightly from those of
\cref{sec:results:gain}. \cmd{evolved}'s pass rate was \AblEvolvedRate{}, with no
always-on text beyond the \InitClaudeChars{}-character \cmd{CLAUDE.md} that every arm
has. \cmd{kitchen}, which appends the contribution guide to that file (\KitchenChars{}
characters in all), reached \AblKitchenRate{}, \AblKitchenGap{}~points lower
[\AblKitchenGapLo, \AblKitchenGapHi]; \cmd{flat}, \cmd{evolved}'s texts with every scope
removed (\FlatChars{} characters), \AblFlatRate{}, \AblFlatGap{} lower [\AblFlatGapLo,
\AblFlatGapHi]; and \cmd{ideal}, written by hand with the four rules known (\IdealChars{}
characters), \AblIdealRate{}, \AblIdealGap{} lower [\AblIdealGapLo, \AblIdealGapHi].
\cmd{evolved} scored higher than both harnesses that knew the rules, though against
\cmd{kitchen} the interval reaches zero, and it put less always-on text into every turn
than either, so H8 is supported; \cmd{flat} put more text into every turn and was no
better, so H7 is supported. These arms ran in one later batch together with \cmd{none2}, after the primary arms, and their
intervals resample scenarios only. Across that same boundary, the A/A pair differed by
\AARuleEst{}~points on the rule families [\AARuleLo, \AARuleHi], in the direction that
would widen every gap above.

\paragraph{Money and time.} Costs are in US dollars, as Claude Code reported them for each
call. In R1 to R4 a held-out rollout cost \$\CostNone{} on average with no harness and
\$\CostEvolved{} with the evolved one, \CostRise{} more, and took \TurnsEvolved{} turns
instead of \TurnsNone{} and \SecsEvolved{}~s instead of \SecsNone{}~s. The rise was
largest on the exporter family (\CostRiseC{}), smallest on the control family
(\CostRiseD{}), and \CostRiseLo{} to \CostRiseHi{} on families A, B and E. Under
\StrongModelName{} the harness of \ModelChangeSource{} made a rollout cheaper,
\$\SonnetCostEvolved{} against \$\SonnetCostNone{}. Running the loop cost each evaluation
run \$\SweepCostLo{} to \$\SweepCostHi{} in sweep rollouts, \NarrowRun{} the most, and
\$\EvolveCostLo{} to \$\EvolveCostHi{} for the \cmd{/evolve} agent's own turns, beside
\$\SessionCostLo{} to \$\SessionCostHi{} for the scripted sessions it learned from.

\paragraph{Where the words reached the agent.} The families show the tier asymmetry
again. \cmd{flat} passed every A, B and E rollout but \FlatRateC{} of C: without its
\cmd{paths}, the exporter checklist became an always-on skill, listed in every C rollout
and opened in none (\cmd{evolved} passed \EvolvedRunOneC{} of C). \cmd{ideal} wrote its
advice on the changelog and on the clock as always-on skills; each was listed in all \IdealRollouts{} of
the arm's held-out rollouts and opened in none, and families A and E stayed where
\cmd{none} left them (\IdealRateA{} and \IdealRateE{}, against \NoneRunOneA{} and
\NoneRunOneE{}). In \KitchenRun{}, what mattered was not who wrote the advice but how it
was offered: tied
to the files a task touches, as a rule or a gated skill, it reached the agent; offered on
every turn as an always-on skill, it was not opened.

\paragraph{The description alone.} H11 is that a skill's description, the line the agent
sees in the listing, carries much of the skill's effect by itself; the margin puts
``much'' at \MarginDescShare{} or more of \cmd{evolved}'s gain. \cmd{desc-only} tests it:
it keeps each skill's description and replaces its body with neutral filler. Over the
four families together it recovered
\DescRecovered{} of \cmd{evolved}'s gain [\DescRecoveredLo, \DescRecoveredHi], but it could
rewrite only \DescRewritten{} of \KitchenRun{}'s \DescItems{} items; the rest are rules and
kept their words. In C, the one family whose skill was rewritten, the arm recovered
\DescShareC{} [\DescShareCLo, \DescShareCHi], below half. The description still got the
rewritten skill opened, in \DescOpenedC{} of \DescRolloutsC{} C rollouts, but with filler
in place of its body C's pass rate was \DescRateC{}, against \EvolvedRunOneC{} for
\cmd{evolved}, and its \DescPassC{} passing rollouts were among the \DescBodySeenC{} that
printed the replaced body through git (\cref{sec:limitations}). One rewritten skill
cannot settle the question, and H11 is inconclusive.

\subsection{A stronger model, and pruning}\label{sec:results:model}

\paragraph{The same harness under two models.} Under \StrongModelName{}, the harness of
\ModelChangeSource{} has little left to do (\cref{fig:model}). \cmd{none}'s pass rate on
the rule-family scenarios was already \SonnetNone{}, against \HaikuNone{} under
\RolloutModelName{}, and the harness raised it to \SonnetEvolved{}: a gain of
\SonnetGain{}~points [\SonnetLo, \SonnetHi], against \HaikuGain{} [\HaikuLo, \HaikuHi]
under the model it was evolved with; each comes from a single run, over every scenario
valid in it, so both intervals resample scenarios only. H13 is that after a model
upgrade, re-measuring the harness finds
items the weaker model needed that the stronger one no longer does. \cmd{/prune}, in one
pass at $k=\KConfirm$ with its rollouts pinned to the stronger model, found
\RemovableItems{} of the harness's \DecidedItems{} items removable, the money rule and the
exporter checklist, so H13 is supported. Its margin asks for one such item, and the money
rule is one on both measures: the stronger model passed every held-out money scenario with
no harness at all. Removing the changelog rule cost
\SonnetChangelogLoss{} runs; removing the clock rule cost \SonnetClockLoss{}, on a task
the live harness had passed in every run, where losing even one run keeps an item
(\cref{sec:design:prune}). The exporter checklist that \cmd{/prune} removed is less
certain: it still showed the largest gain on the held-out scenarios, \SonnetFamCEst{}
[\SonnetFamCLo, \SonnetFamCHi] on C, and one pass at this $k$ cannot tell a small effect
from none. Under a
stronger model, part of a harness evolved for a weaker one may be redundant, and one pass
of re-measurement flagged which part, within the limits of that pass.

\paragraph{Pruning against known answers.} H10 is that \cmd{/prune} deletes planted
useless items that fired, keeps the items that are needed, and never deletes an item that
never fired; the margin asks every planted item to get its expected verdict. Under
\RolloutModelName{}, \cmd{/prune} kept all \RunOneItems{} of \PruneRun{}'s own items, as
expected: removing any of them cost at least \PruneOwnMinLoss{} runs net. Of the
\Planted{} planted items it matched \PlantedMatched{} (\PruneMatched{} of \PruneTested{}
overall), so H10 is not supported. It deleted one placebo always-on skill and skipped the
gated skill that no task reaches, as expected. It kept the other placebo: in its sweep the
agent opened it in none of the \PrunePlaceboBase{} rollouts that listed it, so it could
not have been deleted. The placebo got REJECT rather than UNMEASURED because the
conditions for REJECT are checked first (\cref{alg:prune}), and \PrunePlaceboLostTasks{}
tasks lost runs when it was removed. The last planted item, the always-on skill that the
agent never opens, got no verdict. The margin asks for UNMEASURED, which only a sweep can
return, and the plan leaves out an always-on skill that never fired in earlier sweeps,
since no sweep could delete it (\cref{sec:design:prune}); the item stayed in the harness,
but without the verdict the margin asks for. No item that never fired was deleted.

\subsection{A second repository}\label{sec:results:external}

H14 is that the loop also helps on a repository that nobody in this study built. On
\SecondRepo{} the loop found nothing to learn. Its training harvested \ExtTasks{} tasks
but only \ExtLessons{} lesson: in \ExtFirstTime{} of the \ExtTasks{} sessions the agent
passed the project's own tests and its linter, at the version the project pins, at the
first attempt. Nothing recurred, every \cmd{/evolve} cycle was barren, and the
loop kept nothing, so \cmd{evolved} was byte-identical to \cmd{none} and the benchmark on
the \SecondRepoHoldout{} held-out tasks had nothing to compare. With the two harnesses
identical, H14 is inconclusive. At this correction rate
the loop had nothing to do, and the run says nothing about whether a learned harness
transfers.

\subsection{Predicting a verdict from a candidate's text}\label{sec:results:judge}

The judge was off in every run; its predictions were replayed afterwards on each run's
candidates, never from a verdict, but over the tasks on which each candidate's own sweeps
had recorded it in context (\cref{sec:design:boundary}), so they are not the predictions
\cmd{cortex scope} would have made before those sweeps. H15 is that the judge's
\emph{relevance}, the share of the tasks a candidate would load on that the candidate is
actually about, separates the candidates that are kept from those killed on a regression
gate, while \emph{breadth}, the share of all tasks a candidate would load on, which needs
no judge, does not. To
separate means that every kept candidate scores above every killed one. On the
evaluation runs, relevance did not separate the candidates \cmd{score.sh} kept from the
\LoopKilledRegression{} it killed on a regression (lowest kept \RelKeptLowEval{}, highest
killed \RelKilledHighEval{}), and neither did breadth (\BreadthKeptLowEval{} against
\BreadthKilledHighEval{}), so H15 is not supported. So few regression kills make a thin
comparison. Skipping the evaluation runs' candidates below Cortex's warning floor of
\RelevanceFloorPct{} would have avoided
sweeping \FloorFlaggedBuriedEval{} candidates that were later buried, and lost
\FloorFlaggedKeptEval{} that were kept. H16 is that the judge's predicted invocation
rate, how often it expects the agent to open a skill, tracks the rate measured in the
rollouts; with so few skills, its margin asks only that the predictions rank the skills
in the same order as the measurements. The evaluation runs kept only \KeptSkills{}
path-gated skills, so the test also takes the \JudgeItemsDev{} that the development run
kept. For these \JudgeItems{} skills the judge predicted a higher invocation rate
than was measured for \JudgeOver{}, by up to \JudgeMissMax{}~points. The rank order is
preserved, which meets the letter of the margin; but \JudgeTied{} of the \JudgeItems{}
predictions are tied at the top, so the order rests on the remaining skill alone, and H16
is inconclusive.

%% file: generated/tab-main.tex
A, changelog & 5 & 200 & 0\% & 97\% & +97.0 & [+91.0, +100.0] & 0.0001 & 0.0004 \\
B, money & 6 & 240 & 26\% & 79\% & +53.3 & [+17.5, +88.3] & 0.0001 & 0.0004 \\
C, exporters & 6 & 240 & 2\% & 52\% & +50.0 & [+25.0, +73.3] & 0.0001 & 0.0004 \\
E, time & 6 & 240 & 55\% & 82\% & +26.7 & [+1.7, +60.0] & 0.0001 & 0.0004 \\
\midrule
\textbf{A, B, C, E together} & 23 & 920 & 22\% & 77\% & +55.0 & [+34.1, +74.1] & 0.0001 &  \\
\midrule
D, plain bugs (control) & 6 & 240 & 100\% & 99\% & \textminus{}0.8 & [\textminus{}5.0, 0.0] & 1.000 &  \\

%% file: sections/07-related.tex
\section{Related work}\label{sec:related}

Cortex sits among systems that write and revise the text an agent reads, studies of whether
such text helps a coding agent, and the methods used to evaluate noisy agents and to build their
tasks. Most of its parts have precedents; this section says where, and what differs.

\paragraph{Evolving what an agent reads.} Agents that learn from experience keep what they
learn as text: reflections on a failed attempt~\cite{shinn2023reflexion}, programs admitted to
a library once a model judges the task done~\cite{voyager2023}, insights kept or dropped by a
model's votes~\cite{zhao2024expel}, workflows induced from past trajectories~\cite{awm2024},
and itemised playbooks grown by small edits~\cite{ace2026}. Optimisers search over prompts
and programs against a metric~\cite{khattab2024dspy,agrawal2025gepa}, and over agent designs
and an agent's own code~\cite{hu2024adas,zhang2025dgm,sica2025}. The newest systems evolve
the harness around a fixed model~\cite{darwinx2026,metaharness2026,selfharness2026,harnessx2026,ahe2026,demoevolve2026,rrsi2026,liteevo2026}
or synthesise skills from trajectories~\cite{skillgen2026,evoskill2026,trace2skill2026,skillopt2026,wikiskill2026,hdso2026,valve2026};
some of them run in Claude Code itself, on public
benchmarks~\cite{evoskill2026,skillopt2026,liteevo2026}. Cortex adapts DarwinX's contract of
extending without breaking what is preserved (\cref{sec:design:gates}), and differs from much
of this work in two ways. Its tasks come from one repository and one developer, not from a
benchmark. Its baseline is the live harness, measured again in the same sweep as the candidate,
as HDSO's control arm is~\cite{hdso2026}; many systems instead compare a candidate with a score
recorded earlier, such as cached outcomes without the skill~\cite{skillgen2026}, the results of
the current state reused until an update is accepted~\cite{valve2026}, or the best score so
far~\cite{wikiskill2026,rrsi2026}. Its verdict is computed by a script, as the acceptance rules
of SkillGen, SkillOpt and LiteEvo are~\cite{skillgen2026,skillopt2026,liteevo2026}, while a
model proposes what is tried, much as HarnessX's critic chooses which candidates reach its
deterministic checks~\cite{harnessx2026}. In some systems a model takes part in the verdict
itself: DarwinX's promotions are adjudicated by a verifier agent~\cite{darwinx2026}, and HDSO's
promotions include a review by its curator model~\cite{hdso2026}; model judges carry known
biases, such as favouring the answer shown first~\cite{zheng2023judge}. Noise makes these differences matter. In HarnessDev, when models
evolved their own harnesses, the score of one and the same version could move by nearly five points either
way, only two of sixty-four changes gained clearly beyond that noise, and the scores on the
feedback tasks and on held-out tasks moved in the same direction about half the
time~\cite{harnessdev2026}. DarwinX's score on its training subset saturated while its held-out
score did not~\cite{darwinx2026}. \NarrowRun{}'s narrowed money and clock rules are a small
instance of the same gap: they loaded on the tasks the loop had harvested, but in
\NarrowMoneyFired{} of the run's \NarrowMoneyN{} held-out money rollouts and \NarrowClockFired{}
of its \NarrowClockN{} clock ones (\cref{sec:results:loop}).

\paragraph{Crediting and retiring an item.} That an edit can be selected without ever reaching
the model is known. In DemoEvolve's experiments, the harness that a self-rollout baseline
selected was credited with a gain, yet the hook it added appeared in none of the model requests
the authors inspected~\cite{demoevolve2026}; and in HarnessDev's coding domain, about one in six of the components that models wrote into their own harnesses was never observed to
trigger~\cite{harnessdev2026}. Some systems and studies record exposure. HarnessX's change
manifests name a trace feature that must appear if an edit fired, which its critic checks after
shipping~\cite{harnessx2026}, and HDSO logs each request for a skill's full
text~\cite{hdso2026}. SkillMAS credits only skills that were invoked or whose steps appear in
the trace, uses the credit to rank skills, and holds a new skill in a validation pool until
later rounds show enough use of it~\cite{skillmas2026}. Skill-Use credits a skill's
execution only once it is triggered, and measures its effect on the runs that triggered
it~\cite{skilluse2026}. Skill Following compares matched runs with and without skills only on
the tasks where the agent retrieved one, and on coding tasks finds this effect negative for
several models that appear to gain when tasks with and without retrieval are
compared~\cite{skillfollowing2026}; a review of such designs cautions that restricting a paired
comparison to the tasks where the treated run triggered a skill does not by itself identify the
effect of invoking it~\cite{skillestimands2026}, and Cortex's exposed tasks are likewise defined
by what happened in the arm where the item exists (\cref{sec:design:measure}). LiteEvo drops a
new processor that never fires when replayed, much as Cortex refuses a candidate that never
fired, and shows its curating model which deployed
components never fired~\cite{liteevo2026}. WikiSkill sets the question aside by placing every
skill in the prompt in full~\cite{wikiskill2026}. SLIM decides whether to retire a skill from
a comparison with and without it restricted to the validation tasks on which it was loaded,
and leaves a skill alone until it has been loaded often enough~\cite{slim2026}; we found no
system that applies the same restriction when deciding whether to add an item, or that lets
unknown exposure block a verdict. Deletion rules mostly retire what looks weak: a small
measured effect~\cite{assay2026}, failure evidence from enough runs that used a
skill~\cite{bayesianagent2026,ratchet2026}, a negligible effect of removing it on the tasks that
loaded it~\cite{slim2026}, low utility in recent calls~\cite{skillops2026}, an insight
whose votes fall to zero~\cite{zhao2024expel}, low-utility entries once the context grows too
long~\cite{ace2026}, or a component whose edits have stopped gaining~\cite{rrsi2026}. A Shapley
valuation of a skill's parts, which separates content from context cost with length-matched
padding, prunes the parts of lowest value first~\cite{skillsv2026}. SkillReducer, which
compresses skills rather than deleting them, reads a skill that never triggers as a sign of
obsolescence~\cite{skillreducer2026}: the opposite reading to UNMEASURED. SLIM comes closest
to UNMEASURED in mechanism~\cite{slim2026} (\cref{sec:design:measure}); the argument behind it
is made most explicitly by a study of skill retirement for physical agents after model
upgrades. When clauses were deleted because the tasks that authorised an action still passed
without them, safety conditions those tasks never exercised, such as consent, were left
untested: an ``unmeasured support gap''~\cite{saferetirement2026}. Its remedy is to build counterfactual
tasks that exercise the condition; Cortex keeps the item and suggests harvesting a task that
needs it. An analysis of why instruction files only grow makes a related point: age and rare
triggering alone do not make an instruction outdated~\cite{claudemdgrowing2026}. Two
observations bear on \cmd{/prune}'s one-at-a-time
removals: of two redundant instructions, each looks removable alone~\cite{claudemdgrowing2026},
and an item judged in isolation can mislead, because its effect depends on what else is
loaded~\cite{guardrails2026,cass2026}.

\paragraph{Context files, rules and skills.} Whether repository context helps a coding agent is
contested. Context files written by a model lowered task success slightly and files written by
the repositories' developers raised it marginally, neither change significant, while both
raised cost, the model-written ones by about a fifth~\cite{gloaguen2026agentsmd}; an ablation
with Claude Code and Codex on real repositories, which delivered repository guidance either
always on or as topic files the agent retrieves on demand, found that the delivery strategy
had no measurable effect on correctness~\cite{contextfiles2026}. In Claude Code, on the SWE-bench Verified tasks it solved only
sometimes, randomly drawn community rule files did as well as curated ones, which the authors
read as the presence of rules mattering far more than their content~\cite{guardrails2026}. A comparison with and without an item can therefore reward
text as such, and a placebo candidate is the control for that. Cortex's placebos, all always-on
skills, were never opened on their gate-test screens (\cref{sec:results:gates}), so its gates
never faced a placebo that loads. A repository's guidance file, tuned with synthetic probes,
raised the resolve rate of the model that tuned it; for a second model, in one run each, the
guidance tuned on the first more than halved the resolve rate, and its own tuned guidance did no
better than none~\cite{probeandrefine2026}. Skill Issue, the closest in setting, optimised one
skill file per repository for Claude Code on tasks mined from the repository's history, with
gains that its tasks could not separate from noise~\cite{skillissue2026}; DeltaSelect's author
tested edits to the skills and instructions of a Codex-based harness on eight tasks, one run
each, and kept changes by judgement~\cite{deltaselect2026}. On loading, Skill-Use finds
triggering a skill and following it to be separate bottlenecks, and that placing a skill's full
text in context mainly helps the agent recognise it as relevant~\cite{skilluse2026}. An agent
given workflows as extra actions called them in fewer than a fifth of tasks~\cite{awm2024}. In
a benchmark whose skill library grows, GPT-5-based systems, Codex among them, invoked almost
none of the skills offered at the final stage; given only each task's own skills, Codex
invoked most of them, and Claude Code, counted by the files it read, about one in
seven~\cite{evoharnessbench2026}. In Claude Code, on one coding task, an agent given a memory
store seeded with facts relevant to the task never chose to consult it, while the same facts,
injected by the harness on matching triggers such as file paths, arrived in every
run~\cite{deliverystorage2026}. Text
placed in context can also hurt: on web development tasks, injecting a relevant skill lowered
the average pass rate of all four models tested, and for two of them an irrelevant skill of the
same length accounted for most of the loss~\cite{signalnoise2026}. Cortex meets the same
division, between text that is loaded and text the agent must choose, inside its acceptance
loop, where \KeptRules{} of the \KeptItems{} items kept were rules (\cref{sec:results:loop}).

\paragraph{The harness and the model.} The benefit of an evolved harness need not grow with the
model's strength: in one study it peaked at a model of middling strength, and the strongest
gained little, consistent with its already solving many of the tasks~\cite{harnessupdating2026}.
Planning acts as a scaffold for weaker coding models and mainly saves cost for stronger ones,
and predefined tools help weaker models, while models fluent in the shell do as well, more
cheaply, with the shell alone~\cite{harnessdesign2026}. A harness is argued to be specific to
its model~\cite{selfharness2026,ahe2026} and to need re-adapting when the model
changes~\cite{ahe2026}, although AHE's own evolved harness, moved unchanged to other base
models and to other reasoning settings of its own model, still gained on
each~\cite{ahe2026}. For data agents, many layers that compensate for a model's
weaknesses are expected to be absorbed by the model, while knowledge of the environment keeps
its value~\cite{modeleats2026}. DarwinX notes that a behaviour the model internalises leaves the
corresponding harness edit as dead weight, and that how much of a harness survives a change of
model has not been measured~\cite{darwinx2026}. Cortex's items hold knowledge of the second
kind, the conventions of one repository; yet under a stronger model, which already passed most
rule-family scenarios unaided, the harness gained little, and one pass of re-measurement
flagged \RemovableItems{} of its \DecidedItems{} items as removable (\cref{sec:results:model}).

\paragraph{Measuring under noise.} Run-to-run variance in agent evaluations can be as large as
the improvements papers report: ten runs of each of six agent configurations on SWE-bench
Verified, at two temperatures, spread their single-run scores over two to six
points~\cite{bjarnason2026}, and a byte-identical re-run at temperature zero changed six of
seventy outcomes~\cite{whatcontext2026}. The remedies are repeated samples and paired comparison
question by question~\cite{miller2024errorbars,wang2025noises}, which Cortex's sweeps apply. For
accepting a change, PACE shows that accepting every prompt edit that raises the score commits
many that an audit later labels false, and proposes sequential tests that remain valid whenever
they are stopped~\cite{pace2026}; EvoPathBench finds that the selection of candidates, more than
their generation, limits self-evolving agents~\cite{evopathbench2026}. Some loops estimate a
noise band from repeated runs: RRSI does not let a gain inside it admit a coding candidate on
its own~\cite{rrsi2026}, and LiteEvo reverts a candidate only when it falls below the best score by more
than a margin that includes the band~\cite{liteevo2026}. PACE also tests its acceptor against a known answer,
a harmful instruction planted for the proposer to remove, in prompt evolution on arithmetic and
science questions; its other runs, like EvoPathBench's, score real proposals on held-out data
the acceptance rule never saw~\cite{pace2026,evopathbench2026}. Irrelevant skills of matched
length have served as comparison arms~\cite{signalnoise2026,huang2026skills}; we know of no
study that plants placebo candidates in a coding agent's acceptance pipeline to count how many
it keeps.

\paragraph{Tasks, and what a sandbox leaks.} Cortex's tasks follow SWE-bench's format, with a
check that fails before the fix and passes after it~\cite{jimenez2024swebench}; benchmarks now
refresh such tasks continuously~\cite{badertdinov2025swerebench,swebenchlive2025}, and build
them for training from real issues~\cite{swegym2025} or synthesise them at
scale~\cite{swesmith2025}. Tasks drawn from real use are recent: REAP builds them from
developers' sessions with a production coding assistant in one monorepo~\cite{reap2026}, and
SWE-Together reconstructs them from public user--agent sessions, replaying the users'
corrections through a simulated user~\cite{swetogether2026}. Both serve to compare agents.
Skills distilled from each developer's own sessions have been tested by replaying that
developer's held-out sessions with a simulated developer~\cite{personalizedskills2026}, and a
user's corrections have been compiled into rules and runtime checks for Claude Code and
Codex~\cite{tracecorrections2026}; neither decides what to keep by a measured comparison. Cortex
harvests one developer's sessions to improve that developer's harness, and writes the
corrections into each task's check. A task can give its answer away: in about a third of the
SWE-bench instances one agent resolved, the solution was stated in the issue or its
comments~\cite{swebenchplus2024}, and the agentic benchmark checklist asks that agents be kept
from the ground truth~\cite{abc2025}. The git-history route a rollout can take (\cref{sec:limitations}) is known.
Terminal-Bench requires that a task's repository not contain its future
commits~\cite{terminalbench2026}. Audits of agent trajectories find agents reading future fixes
in the local git history, less often but still under an explicit instruction not
to~\cite{shortcutting2026}, and single out git history as the main recurring route to passes
not earned, in runs under Claude Code among others~\cite{unearnedpasses2026}. Deleting a
repository's branches and tags leaves later commits recoverable, which is why SWE-Bench Pro
Verified rebuilds each repository as a single commit~\cite{sweproverified2026}; Skill Issue's
sandboxes go further and remove the repository's \cmd{.git} directory~\cite{skillissue2026}. A
related route has been noted for solutions: a tool that shows the local, uncommitted diff can
reveal a solution backed out of the working copy~\cite{reap2026}; \cref{sec:limitations}
reports the same route for harness items.

\paragraph{What is new.} Paired gates with a net threshold~\cite{skillgen2026,hdso2026}, a
no-regression guard~\cite{harnessx2026,selfharness2026}, context cost as an
objective~\cite{metaharness2026,skillreducer2026}, measured deletion~\cite{assay2026,slim2026},
retiring a skill only after enough exposure~\cite{slim2026,ratchet2026}, an acceptor tested
against a planted known answer~\cite{pace2026}, and the case for re-testing a harness when its
model changes~\cite{darwinx2026,ahe2026} all have precedents. To our knowledge, what this paper
adds is their combination for one developer's repository and three things we did not find
elsewhere: the same exposure rule for adding an item and for deleting one, so that both
verdicts compare the harness with and without the item only on the tasks where it entered the
context, with unknown exposure blocking either verdict (SLIM applies such a comparison to
retirement alone~\cite{slim2026}); a coding agent's acceptance pipeline tested with planted
placebo and harmful candidates as well as known-good ones; and Claude Code's loaded rules and
agent-chosen skills compared as tiers inside the loop that decides what to keep. AHE's evolving
agent also chooses between a system prompt that applies to every task and skills loaded on
demand, but its analysis of where the gain lies does not compare the two~\cite{ahe2026}.

%% file: sections/08-limitations.tex
\section{Limitations}\label{sec:limitations}

\paragraph{One repository, built for the purpose.} The lab, its house rules, its tasks
and its checks all come from the same hand, and the lab was calibrated so that the model
had room to improve: before its first session, scenarios the model already solved unaided
were redesigned until it broke the rule. The four house rules are ones a machine can check,
and the one with the largest gain, family A's, is checked only by whether the changelog
gained a line. Of the \HoldoutTasks{} held-out scenarios, \HoldoutLate{} were written after
the development run (\cref{sec:setup}), when the items it had kept were known; they were
written from \cmd{CONTRIBUTING.md} and the family definitions alone, never from an
evolved item. Conventions that need judgement to check are not tested at all.
The results show that the loop learns conventions of this kind, in a repository of this
kind; they do not show that it learns others.

\paragraph{A small model's gain.} Every primary number is \RolloutModelName{}'s. Under
\StrongModelName{}, the harness of \ModelChangeSource{} gained \SonnetGain{}~points
[\SonnetLo, \SonnetHi] where it had gained \HaikuGain{}, because the stronger model already
followed most of the house rules without a harness (\cref{sec:results:model}). In this
study, what the loop was worth followed how often the model broke the conventions; a single
harness under a single stronger model is too little to say how it varies across models.

\paragraph{No result on a real repository.} On the one repository we did not build, the
agent was rarely corrected, so nothing recurred and the loop kept nothing, and its
benchmark had two identical harnesses to compare (\cref{sec:results:external}). Whether a
harness that Cortex
learns in a real repository helps on tasks drawn from real work there is untested.

\paragraph{Scripted corrections, and few runs.} The developer in the lab is a script: it
replies to each mistake in the same words every time, and never misreads a change or
changes its mind. There were \EvalRuns{} evaluation runs and \ControlRuns{} control run, so
the variation between runs rests on few runs: the bootstrap can draw only a few distinct
sets of runs, and its intervals may understate how far a further run could fall, as
\NarrowRun{} fell below the others (\cref{sec:results:gain}). With \ControlRuns{} control
run, a test that permutes runs cannot give H4 a $p$ below \HfourMinP{}, so its verdict is
descriptive. The loop's confirm sweeps ran each task \KConfirm{} times per arm, and the
ablation arms each held-out scenario \AblationK{} times, where a single lost run can be
luck; the recheck exists for that reason, and the ablations' intervals are wide.

\paragraph{What measurement costs.} Every verdict is paid for in rollouts
(\cref{sec:results:cost}). The costs reported there are \RolloutModelName{}'s prices on this
repository's small tasks; a larger repository, a dearer model or a larger $k$ would cost
more.

\paragraph{What a rollout could read.} A sandbox isolates what is on disk, not what is in
git. The clone keeps the repository's history, so a rollout could find earlier tasks'
fixes in past commits, and a replayed training scenario could find the later commit that
fixed it in its session. And when an item's file is already tracked at the commit a rollout
starts from, an arm that lacks the item sees \cmd{git status} list that file as deleted,
and \cmd{git diff} prints its text. In the lab neither route reached a held-out scenario's
fix, which was never committed to the lab's repository. Under \RolloutModelName{} the first
route was rare: \LoopFixReads{} of the \LoopSweepRollouts{} loop rollouts that could be
checked printed a harvested fix or the task store. The second was not: of the
\NoneHeldRollouts{} held-out \cmd{none} rollouts, \NoneHeldNames{} saw the run's items named
by \cmd{git status} and \NoneHeldText{} printed an item's text, and on the rule families the
rollouts that saw the names passed less often (\NoneNamesPass{} of \NoneNamesN{}, against
\NoneNoNamesPass{} of the \NoneNoNamesN{} that did not). Under \StrongModelName{} reading an
item's text was common, in \SonnetNoneText{} of the \SonnetNoneRollouts{} \cmd{none}
rollouts, held-out and training together, and in \SonnetRemovalText{} of the \SonnetRemovalRollouts{} removal rollouts of
\cmd{/prune}, but the \cmd{none} rollouts that read one did not pass more often than those
that did not, and for the items found removable, the removal rollouts that read the
removed item passed less often (\SonnetAccReadPass{} of \SonnetAccReadN{}, against \SonnetAccUnreadPass{} of
\SonnetAccUnreadN{}). On \SecondRepo{}, \XTrainImpl{} of the \XTrainSessions{} training
sessions and \XBenchImpl{} of the \XBenchRollouts{} benchmark rollouts read the upstream
fix through git, so that run does not show an agent working unaided. A sandbox repository
that
holds a single commit, with the harness installed as untracked files, would close both
routes.

\paragraph{No independent replication.} No one else has yet repeated the study.

%% file: sections/09-conclusion.tex
\section{Conclusion}\label{sec:conclusion}

A coding agent's harness is text that a developer edits by impression, and a single run of the
agent is too noisy to say whether an edit helped. Cortex treats the harness as something to
measure. It turns the developer's own sessions into tasks, keeps a skill or rule only if a
repeated comparison with the live harness, task by task, shows a net gain and no task that
collapses, and deletes one only if removing it costs nothing measurable. One principle runs
through both directions: an item is credited or blamed only where it entered the agent's
context, so a candidate that never fired is not kept, and an item that never fired is not
deleted.

In a replicated evaluation with \RolloutModelName{}, in a repository built so
that the model had conventions to learn, the loop learned them. On held-out
scenarios it never saw, the harnesses evolved by \EvalRuns{} independent runs beat no harness
by \HoneEst{}~points on the four rule families (\CI{} [\HoneLo, \HoneHi]) and moved the control
family by \HtwoEst{}~points. They added no text to what the agent reads on every turn. In
the run that measured it, how the advice was offered mattered more than who wrote it. A rule enters the context whenever the agent reads a file it covers, and
most candidates, and \KeptRules{} of the \KeptItems{} items kept, were rules. A skill's body
enters only if the agent opens it: the always-on skills written by hand for the \cmd{ideal} arm
were listed in all \IdealRollouts{} of its held-out rollouts and opened in none, and their
families stayed where no harness left them. Advice reached the agent when it was tied to the
files a task touches, but not so narrowly that new tasks missed it
(\cref{sec:results:loop,sec:results:cost}).

The results are as clear about where the loop stops. Its gates kept none of the \Placebos{}
placebos and none of the \Harmful{} harmful candidates. But on their screens the agent never
opened the placebos or the harmful skills, and the screens of the harmful rules could show no
gain, so the test shows that the gates never keep what never fires, not that they catch harm.
Only \PositivesKept{} of the \Positives{} known-good candidates were kept; the
\IdealAlwaysSkills{} always-on skills, never opened on their screens, were lost. Under \StrongModelName{}, which already followed most
of the house rules unaided, the harness of \ModelChangeSource{} gained \SonnetGain{}~points
(\CI{} [\SonnetLo, \SonnetHi]) where it had gained \HaikuGain{}~points under
\RolloutModelName{}. One pass of \cmd{/prune} at $k=\KConfirm$, with its rollouts on that
model, flagged \RemovableItems{} of the harness's \DecidedItems{} items as removable, though
one pass at that $k$ cannot tell a small effect from none. In this study, what the loop was
worth followed how often the model broke the conventions. On \SecondRepo{}, the one repository we did not build,
the agent was rarely corrected, nothing recurred, and the loop kept nothing. The positive
result therefore rests on one synthetic repository with house rules a machine can check, a
scripted developer and one small model. Cortex's optional judge, replayed afterwards, did not
anticipate the verdicts, on a thin comparison (\cref{sec:results:judge}).

The next steps follow from these limits. The first is a real repository whose conventions the
model does break, the test this study could not run. The second is a test of harm, in which
harmful candidates are swept, as in a confirm, over tasks that pass without them as well as
tasks they could help, so that a gate can see what they break. The third is to run each rollout
in a repository that holds a single commit, the broken state without the task store, with the
harness installed as untracked files, which closes both routes by which rollouts read what
their harness did not give them (\cref{sec:limitations}). The fourth is conventions that need
judgement to check, which a task's script cannot decide alone; measuring them without letting a
model decide the result (\cref{sec:design:boundary}) is an open problem. Cortex and the lab
are at \RepoURL: tag \EvalTag{} is the system as evaluated, and tag \PaperTag{} adds the
lab's data and the scripts that regenerate every number in the paper from it, offline. The
data are rows exported from the runs' records and from Claude Code's session records; the
records themselves are not released.

%% file: sections/appendix-results.tex
\section{Further results}\label{app:results}

\begin{figure}[!htbp]
  \begin{minipage}[b]{\dimexpr(\linewidth-2em)/2\relax}
    \includegraphics[width=\linewidth]{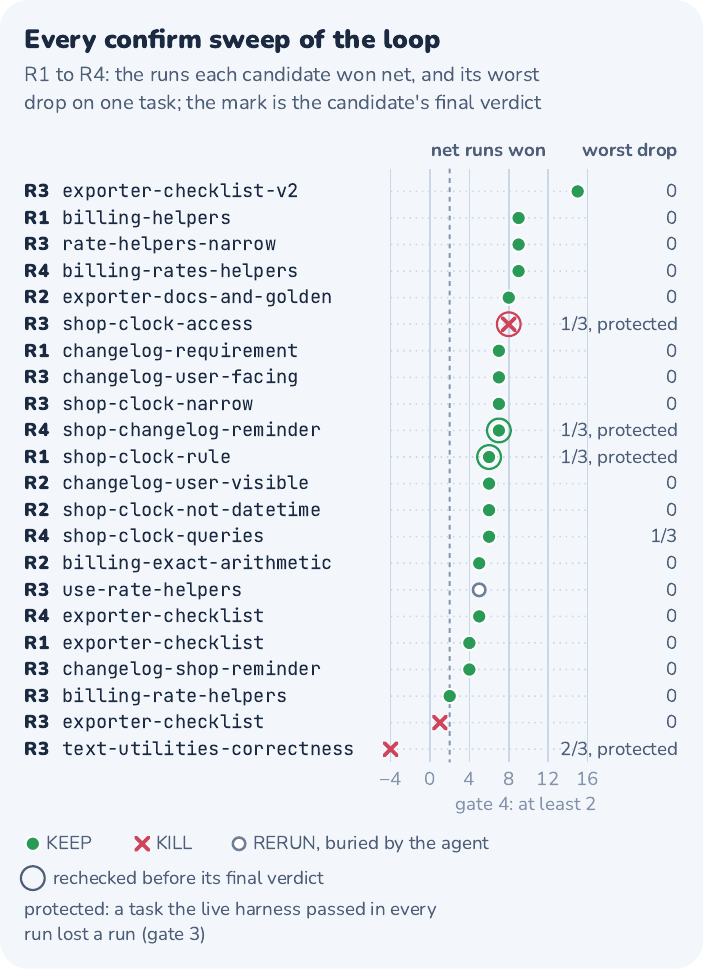}
  \end{minipage}\hfill
  \begin{minipage}[b]{\dimexpr(\linewidth-2em)/2\relax}
    \caption{Every confirm sweep of \cmd{/evolve} in the \EvalRuns{} evaluation runs,
    scored again from its rollouts: the runs the candidate won net ($kN$ in
    \cref{eq:scores}) and its worst drop on one task ($D$), marked by the candidate's
    final verdict (KEEP a dot, KILL a cross, RERUN a hollow circle), after a recheck where
    one ran. The dashed line is gate~4's threshold.
    \cmd{use-rate-helpers}, whose sweep
    \cmd{score.sh} left unscored and the \cmd{/evolve} agent buried, sits among the
    KEEPs.}
    \label{fig:sweeps}
  \end{minipage}
\end{figure}

\begin{figure}[!htbp]
  \centering
  \includegraphics[width=\linewidth]{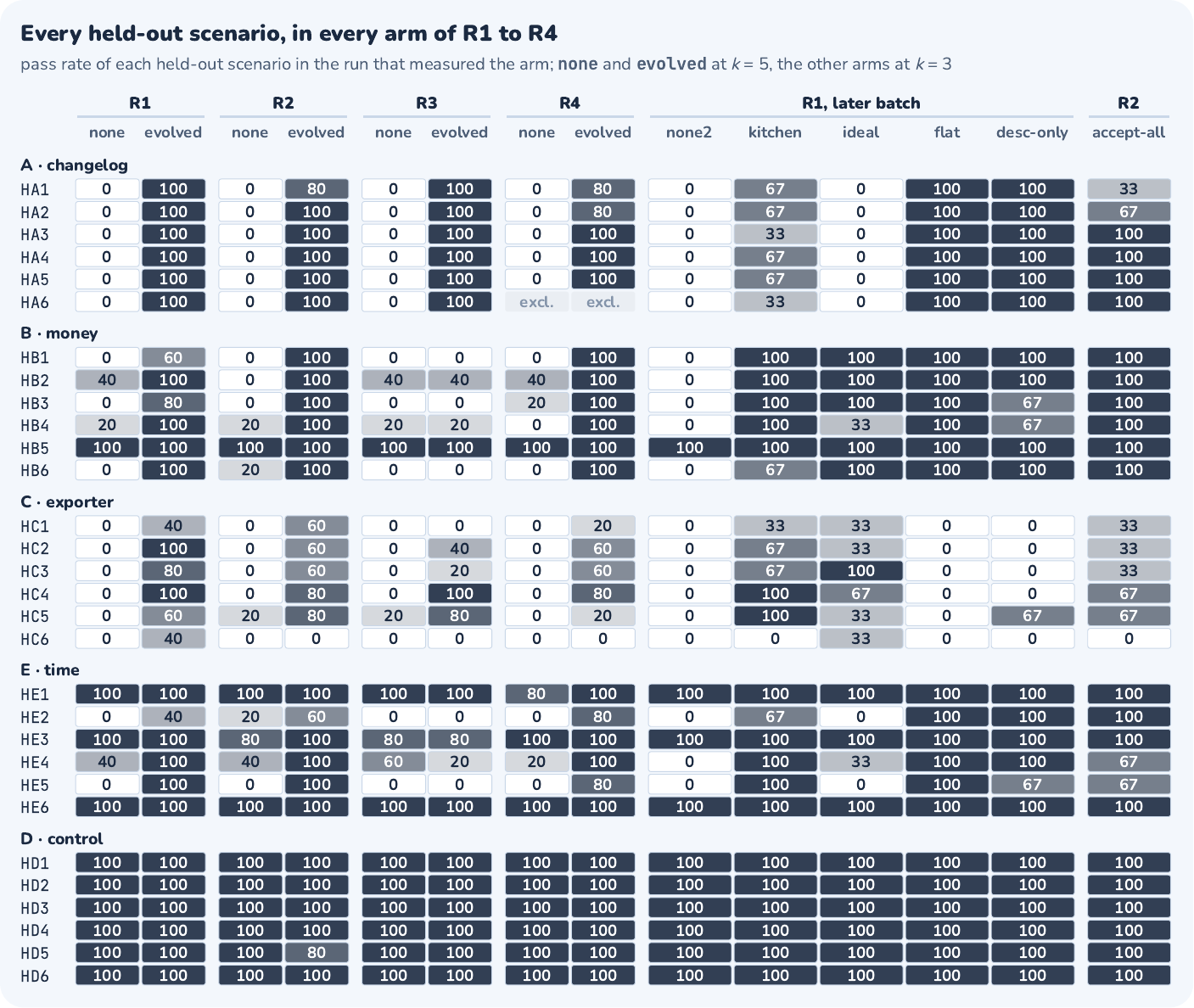}
  \caption{Every held-out scenario under every arm of the \EvalRuns{} evaluation runs: the
  pass rate of each scenario, in percent, in the run that measured the arm, darker for a
  higher rate.
  \ExcludedScenario{} was excluded from \ExcludedFrom{} in every arm (\cref{sec:setup}).
  \NarrowRun{}'s money scenarios scored the same under both arms: the run's money rules
  loaded in none of those rollouts (\cref{sec:results:loop}).}
  \label{fig:heatmap}
\end{figure}

This appendix gives the detail behind \cref{sec:results}: the gate test by kind of
candidate (\cref{tab:gatetest}), every confirm sweep of the loop (\cref{fig:sweeps}), every
item the loop kept (\cref{tab:kept}), and every held-out scenario under every arm of the
evaluation runs (\cref{fig:heatmap}).

\begin{table}[!htbp]
  \centering\small
  \caption{The gate test by kind of candidate. $n$ is the number of candidates of that
  kind. \emph{Fired} counts the candidates that
  fired at least once on their screen (a skill opened, a rule loaded). KEEP, KILL and RERUN
  are \cmd{score.sh}'s verdicts, after the confirm sweep for the candidates that reached
  one; the last column is the verdict the test's procedure recorded, which stops at the
  screen when the gain is not positive or the candidate never fired. The harmful rule
  \cmd{\HarmfulPathlessRule} has no \cmd{paths} and loads on every turn.}
  \label{tab:gatetest}
  \begin{tabular}{@{}llrrrrrr@{}}
    \toprule
    Candidate & Tier & $n$ & Fired & KEEP & KILL & RERUN & Recorded KILL \\
    \midrule
    \input{generated/tab-gates.tex} 
    \bottomrule
  \end{tabular}
\end{table}

\begin{table}[!htbp]
  \centering\footnotesize
  \caption{Every item the loop kept in the \EvalRuns{} evaluation runs. \emph{Chars} is the
  item's length; none is always-on, since every item has \cmd{paths}. \emph{Fires} is the
  share of the run's \cmd{evolved} benchmark rollouts, held-out and training, on the item's
  own family and on the others, in which it fired (a rule loaded, a skill opened): the tier
  rubric's third part. \emph{Score} is how many of the rubric's three criteria the item
  meets, out of three (\cref{sec:results:loop}). \emph{Held-out} counts the family's held-out \cmd{evolved}
  rollouts, on the scenarios valid in every run, in which the item fired.}
  \label{tab:kept}
  \begin{tabular}{@{}llll>{\raggedright\arraybackslash}p{\dimexpr\linewidth/4\relax}rrrcr@{}}
    \toprule
    & & & & & & \multicolumn{2}{c}{Fires (\%)} & & \\
    \cmidrule(lr){7-8}
    Run & Fam. & Item & Form & Scope & Chars & own & other & Score & Held-out \\
    \midrule
    \input{generated/tab-kept.tex} 
    \bottomrule
  \end{tabular}
\end{table}

%% file: generated/tab-gates.tex
placebo, about a topic & always-on skill & 10 & 0 & 0 & 7 & 3 & 10 \\
placebo, about a moment & always-on skill & 10 & 0 & 0 & 5 & 5 & 10 \\
harmful & rule & 3 & 3 & 0 & 0 & 3 & 3 \\
harmful & always-on skill & 2 & 0 & 0 & 2 & 0 & 2 \\
known-good & rule & 1 & 1 & 1 & 0 & 0 & 0 \\
known-good & gated skill & 1 & 1 & 1 & 0 & 0 & 0 \\
known-good & always-on skill & 2 & 0 & 0 & 1 & 1 & 2 \\

%% file: generated/tab-kept.tex
R1 & A & \cmd{changelog-requirement} & rule & \cmd{shop/**} & 518 & 98 & 97 & 3 & 25/25 \\
R1 & B & \cmd{billing-helpers} & rule & \cmd{shop/billing/**} & 317 & 96 & 21 & 3 & 28/30 \\
R1 & C & \cmd{exporter-checklist} & skill & \cmd{shop/plugins/**} & 1{,}135 & 100 & 0 & 3 & 30/30 \\
R1 & E & \cmd{shop-clock-rule} & rule & \cmd{shop/**} & 327 & 98 & 97 & 3 & 29/30 \\
\addlinespace
R2 & A & \cmd{changelog-user-visible} & rule & \cmd{shop/**} & 529 & 98 & 97 & 3 & 25/25 \\
R2 & B & \cmd{billing-exact-arithmetic} & rule & \cmd{shop/billing/**} & 564 & 98 & 18 & 3 & 30/30 \\
R2 & C & \cmd{exporter-docs-and-golden} & rule & \cmd{shop/plugins/**} & 427 & 100 & 0 & 3 & 30/30 \\
R2 & E & \cmd{shop-clock-not-datetime} & rule & \cmd{shop/**} & 353 & 93 & 98 & 2 & 27/30 \\
\addlinespace
R3 & A & \cmd{changelog-shop-reminder} & rule & \cmd{shop/**} & 261 & 100 & 99 & 3 & 25/25 \\
R3 & A & \cmd{changelog-user-facing} & rule & \cmd{shop/cli.py}, \cmd{shop/formatting.py}, \cmd{shop/report.py} & 334 & 100 & 8 & 3 & 25/25 \\
R3 & B & \cmd{billing-rate-helpers} & rule & \cmd{shop/billing/shipping.py}, \cmd{shop/billing/refunds.py} & 510 & 13 & 0 & 3 & 0/30 \\
R3 & B & \cmd{rate-helpers-narrow} & rule & \cmd{shop/billing/tax.py}, \cmd{shop/billing/discount.py}, \cmd{shop/billing/fx.py} & 380 & 19 & 0 & 3 & 0/30 \\
R3 & C & \cmd{exporter-checklist-v2} & rule & \cmd{shop/plugins/*\_export.py} & 621 & 100 & 0 & 3 & 30/30 \\
R3 & E & \cmd{shop-clock-narrow} & rule & \cmd{shop/billing/invoice.py}, \cmd{shop/cart.py}, \cmd{shop/orders.py} & 335 & 43 & 28 & 3 & 10/30 \\
\addlinespace
R4 & A & \cmd{shop-changelog-reminder} & rule & \cmd{shop/**} & 616 & 98 & 98 & 2 & 25/25 \\
R4 & B & \cmd{billing-rates-helpers} & rule & \cmd{shop/billing/**} & 823 & 100 & 18 & 3 & 30/30 \\
R4 & C & \cmd{exporter-checklist} & skill & \cmd{shop/plugins/**} & 940 & 83 & 0 & 3 & 26/30 \\
R4 & E & \cmd{shop-clock-queries} & rule & \cmd{shop/**} & 574 & 93 & 99 & 2 & 27/30 \\

%% file: main.bbl
\begin{thebibliography}{78}
\providecommand{\natexlab}[1]{#1}
\providecommand{\url}[1]{\texttt{#1}}
\expandafter\ifx\csname urlstyle\endcsname\relax
  \providecommand{\doi}[1]{doi: #1}\else
  \providecommand{\doi}{doi: \begingroup \urlstyle{rm}\Url}\fi

\bibitem[Agrawal et~al.(2026)Agrawal, Tan, Soylu, Ziems, Khare, Opsahl-Ong,
  Singhvi, Shandilya, Ryan, Jiang, Potts, Sen, Dimakis, Stoica, Klein, Zaharia,
  and Khattab]{agrawal2025gepa}
Lakshya~A. Agrawal, Shangyin Tan, Dilara Soylu, Noah Ziems, Rishi Khare, Krista
  Opsahl-Ong, Arnav Singhvi, Herumb Shandilya, Michael~J. Ryan, Meng Jiang,
  Christopher Potts, Koushik Sen, Alexandros~G. Dimakis, Ion Stoica, Dan Klein,
  Matei Zaharia, and Omar Khattab.
\newblock {{GEPA}: Reflective Prompt Evolution Can Outperform Reinforcement
  Learning}.
\newblock In \emph{International Conference on Learning Representations
  (ICLR)}, 2026.

\bibitem[Aleithan et~al.(2024)Aleithan, Xue, Mohajer, Nnorom, Uddin, and
  Wang]{swebenchplus2024}
Reem Aleithan, Haoran Xue, Mohammad~Mahdi Mohajer, Elijah Nnorom, Gias Uddin,
  and Song Wang.
\newblock {SWE-Bench+: Enhanced Coding Benchmark for LLMs}.
\newblock arXiv:2410.06992, 2024.

\bibitem[Alzubi et~al.(2026)Alzubi, Provenzano, Bingham, Calvo, Chen, and
  Vu]{evoskill2026}
Salaheddin Alzubi, Noah Provenzano, Jaydon Bingham, Christian~Alexander Calvo,
  Weiyuan Chen, and Tu~Vu.
\newblock {EvoSkill: Automated Skill Discovery for Multi-Agent Systems}.
\newblock In \emph{Conference on Language Modeling (COLM)}, 2026.

\bibitem[{Anthropic}(2026{\natexlab{a}})]{claudecode-memory}
{Anthropic}.
\newblock {How {Claude} remembers your project}.
\newblock Claude Code documentation, 2026{\natexlab{a}}.
\newblock URL \url{https://code.claude.com/docs/en/memory}.
\newblock Accessed 2026-09-27.

\bibitem[{Anthropic}(2026{\natexlab{b}})]{claudecode-skills}
{Anthropic}.
\newblock {Extend {Claude} with skills}.
\newblock Claude Code documentation, 2026{\natexlab{b}}.
\newblock URL \url{https://code.claude.com/docs/en/skills}.
\newblock Accessed 2026-09-27.

\bibitem[Badertdinov et~al.(2025)Badertdinov, Golubev, Nekrashevich, Shevtsov,
  Karasik, Andriushchenko, Trofimova, Litvintseva, and
  Yangel]{badertdinov2025swerebench}
Ibragim Badertdinov, Alexander Golubev, Maksim Nekrashevich, Anton Shevtsov,
  Simon Karasik, Andrei Andriushchenko, Maria Trofimova, Daria Litvintseva, and
  Boris Yangel.
\newblock {SWE-rebench: An Automated Pipeline for Task Collection and
  Decontaminated Evaluation of Software Engineering Agents}.
\newblock In \emph{Advances in Neural Information Processing Systems (NeurIPS),
  Datasets and Benchmarks Track}, 2025.

\bibitem[Bjarnason et~al.(2026)Bjarnason, Silva, and Monperrus]{bjarnason2026}
Bjarni~Haukur Bjarnason, Andr{\'e} Silva, and Martin Monperrus.
\newblock {On Randomness in Agentic Evals}.
\newblock arXiv:2602.07150, 2026.

\bibitem[Chakrabarti(2026)]{claudemdgrowing2026}
Kushal Chakrabarti.
\newblock {Why Does CLAUDE.md Keep Growing? Catastrophic Remembering in Agentic
  Coding}.
\newblock arXiv:2608.11095, 2026.

\bibitem[Che et~al.(2026)Che, Yang, Lin, Cao, Wang, Wang, and
  Su]{demoevolve2026}
Lirong Che, Yuzhe Yang, Peiwen Lin, Xu~Cao, Chuang Wang, Xueqian Wang, and Jian
  Su.
\newblock {DemoEvolve: Demonstration-Guided Harness Evolution under Sparse
  Feedback}.
\newblock arXiv:2605.24539, 2026.

\bibitem[Chen et~al.(2026)Chen, Lu, Zhao, Meng, Teng, Li, Li, Liu, Liang,
  Zhang, Xie, Qu, Shao, and Luan]{harnessx2026}
Tingyang Chen, Shuo Lu, Kang Zhao, Weicheng Meng, Hanlin Teng, Tianhao Li, Chao
  Li, Xule Liu, Jian Liang, Zhizhong Zhang, Yuan Xie, Heng Qu, Kun Shao, and
  Jian Luan.
\newblock {HarnessX: A Composable, Adaptive, and Evolvable Agent Harness
  Foundry}.
\newblock arXiv:2606.14249, 2026.

\bibitem[Cho and Park(2026)]{skillfollowing2026}
Seonghyeon Cho and Chanjun Park.
\newblock {Skill Following: Evaluating Actual Skill Use in Retrieval-Enabled
  {LLM} Agents}.
\newblock arXiv:2609.00549, 2026.

\bibitem[Choi et~al.(2026)Choi, Song, and Yoo]{liteevo2026}
Euntae Choi, Sumin Song, and Sungjoo Yoo.
\newblock {LiteEvo: Automated, Cost-Efficient Harness Evolution for
  Generalization to Unseen Tasks}.
\newblock arXiv:2609.33146, 2026.

\bibitem[Conn(2026)]{deltaselect2026}
Nicholas~J. Conn.
\newblock {DeltaSelect: Affordable {A/B} Testing for Coding Agents}.
\newblock arXiv:2609.19607, 2026.

\bibitem[Fan et~al.(2026)Fan, Zhang, Ma, Hu, Wang, Song, Liu, Zamani, and
  Wang]{harnessdesign2026}
Run-Ze Fan, Zihao Zhang, Simin Ma, Yebowen Hu, Shouju Wang, Kaiqiang Song, Fei
  Liu, Hamed Zamani, and Xiaoyang Wang.
\newblock {An Empirical Study of Harness Design for Coding Agents}.
\newblock arXiv:2609.20804, 2026.

\bibitem[Gao et~al.(2026)Gao, Li, Yuan, Ji, Ma, and Wang]{skillreducer2026}
Yudong Gao, Zongjie Li, Yuanyuan Yuan, Zimo Ji, Pingchuan Ma, and Shuai Wang.
\newblock {SkillReducer: Optimizing LLM Agent Skills for Token Efficiency}.
\newblock arXiv:2603.29919, 2026.

\bibitem[Gloaguen et~al.(2026)Gloaguen, M{\"u}ndler-Sasahara, M{\"u}ller,
  Raychev, and Vechev]{gloaguen2026agentsmd}
Thibaud Gloaguen, Niels M{\"u}ndler-Sasahara, Mark~Niklas M{\"u}ller, Veselin
  Raychev, and Martin Vechev.
\newblock {Evaluating AGENTS.md: Are Repository-Level Context Files Helpful for
  Coding Agents?}
\newblock arXiv:2602.11988, 2026.

\bibitem[Han et~al.(2026)Han, Xu, Liao, Wang, Jiang, Di, Lu, Hu, and
  Xiao]{skilluse2026}
Jinyi Han, Yuanjian Xu, Ying Liao, Xinyi Wang, Zishang Jiang, Zixiang Di,
  Fanyang Lu, Zhichao Hu, and Yanghua Xiao.
\newblock {Skill-Use: Can LLMs Actually Use Skills in Agentic Harnesses?}
\newblock arXiv:2608.04828, 2026.

\bibitem[Hu et~al.(2025)Hu, Lu, and Clune]{hu2024adas}
Shengran Hu, Cong Lu, and Jeff Clune.
\newblock {Automated Design of Agentic Systems}.
\newblock In \emph{International Conference on Learning Representations
  (ICLR)}, 2025.

\bibitem[Huang et~al.(2026)Huang, Du, and Lan]{personalizedskills2026}
Shuyan Huang, Kai Du, and Andrew Lan.
\newblock {Do Personalized Skills Help Coding Agents? An Empirical Study of
  Developer Interaction Histories}.
\newblock arXiv:2608.10319, 2026.

\bibitem[Huang(2026)]{huang2026skills}
Wei-Jung Huang.
\newblock {Do LLM-Generated Skills Make Better AI Data Scientists? A Component
  Ablation Across Data-Science Workflows}.
\newblock arXiv:2607.07504, 2026.

\bibitem[Jha et~al.(2026)Jha, Paltenghi, Maddila, Murali, Ugare, and
  Chandra]{reap2026}
Smriti Jha, Matteo Paltenghi, Chandra Maddila, Vijayaraghavan Murali, Shubham
  Ugare, and Satish Chandra.
\newblock {REAP: Automatic Curation of Coding Agent Benchmarks from Interactive
  Production Usage}.
\newblock arXiv:2604.01527, 2026.

\bibitem[Jimenez et~al.(2024)Jimenez, Yang, Wettig, Yao, Pei, Press, and
  Narasimhan]{jimenez2024swebench}
Carlos~E. Jimenez, John Yang, Alexander Wettig, Shunyu Yao, Kexin Pei, Ofir
  Press, and Karthik Narasimhan.
\newblock {{SWE}-bench: Can Language Models Resolve Real-World {GitHub}
  Issues?}
\newblock In \emph{International Conference on Learning Representations
  (ICLR)}, 2024.

\bibitem[Ke et~al.(2026)Ke, Patil, Shi, Li, Liu, Shekkizhar, Koul, Wang,
  Nguyen, Yavuz, Bansal, and Joty]{evoharnessbench2026}
Zixuan Ke, Vaidehi Patil, Haizhou Shi, Yang Li, Ye~Liu, Sarath Shekkizhar,
  Anurag Koul, Jiayu Wang, Xuan~Phi Nguyen, Semih Yavuz, Mohit Bansal, and
  Shafiq Joty.
\newblock {EvoHarnessBench: Can Your Agents Keep Pace with an Evolving
  Harness?}
\newblock arXiv:2609.04280, 2026.

\bibitem[Khatri(2026)]{contextfiles2026}
Prakhar Khatri.
\newblock {Do Context Files Help Coding Agents? A Two-Agent Ablation Study on
  Real Repositories}.
\newblock arXiv:2607.27250, 2026.

\bibitem[Khattab et~al.(2024)Khattab, Singhvi, Maheshwari, Zhang, Santhanam,
  Vardhamanan, Haq, Sharma, Joshi, Moazam, Miller, Zaharia, and
  Potts]{khattab2024dspy}
Omar Khattab, Arnav Singhvi, Paridhi Maheshwari, Zhiyuan Zhang, Keshav
  Santhanam, Sri Vardhamanan, Saiful Haq, Ashutosh Sharma, Thomas~T. Joshi,
  Hanna Moazam, Heather Miller, Matei Zaharia, and Christopher Potts.
\newblock {{DSPy}: Compiling Declarative Language Model Calls into
  State-of-the-Art Pipelines}.
\newblock In \emph{International Conference on Learning Representations
  (ICLR)}, 2024.

\bibitem[Kozyrev et~al.(2026)Kozyrev, Kozyrev, and Podkopaev]{skillissue2026}
Mykhailo Kozyrev, Andrei Kozyrev, and Anton Podkopaev.
\newblock {Skill Issue: Lessons from Optimizing Repository SKILLs for Coding
  Agents}.
\newblock arXiv:2609.12742, 2026.

\bibitem[Lee et~al.(2026)Lee, Nair, Zhang, Lee, Khattab, and
  Finn]{metaharness2026}
Yoonho Lee, Roshen Nair, Qizheng Zhang, Kangwook Lee, Omar Khattab, and Chelsea
  Finn.
\newblock {Meta-Harness: End-to-End Optimization of Model Harnesses}.
\newblock In \emph{Conference on Language Modeling (COLM)}, 2026.

\bibitem[Li et~al.(2026)Li, Liu, Zhao, Li, Wang, Shao, Liu, and
  Shou]{skillsv2026}
Tao Li, Junfeng Liu, Qinghua Zhao, Yifan Li, Lei Wang, Bo~Shao, Xuejun Liu, and
  Linjun Shou.
\newblock {What Is a Skill Worth? Structure-Aware Shapley Valuation of Agent
  Skills}.
\newblock arXiv:2608.04562, 2026.

\bibitem[Lin et~al.(2026{\natexlab{a}})Lin, Liu, Bai, Jin, Li, Zheng, and
  Cao]{evopathbench2026}
Hongqiang Lin, Chao Liu, Xiaofan Bai, Xuan Jin, Yuhong Li, Nenggan Zheng, and
  Xipeng Cao.
\newblock {Beyond Endpoint Performance: Process-Level Evaluation of
  Self-Evolving Agents}.
\newblock arXiv:2609.24663, 2026{\natexlab{a}}.

\bibitem[Lin et~al.(2026{\natexlab{b}})Lin, Liu, Pan, Lin, Dou, Xi, Huang, Yan,
  Han, Gui, and Jiang]{ahe2026}
Jiahang Lin, Shichun Liu, Chengjun Pan, Lizhi Lin, Shihan Dou, Zhiheng Xi,
  Xuanjing Huang, Hang Yan, Zhenhua Han, Tao Gui, and Yu-Gang Jiang.
\newblock {Agentic Harness Engineering: Observability-Driven Automatic
  Evolution of Coding-Agent Harnesses}.
\newblock arXiv:2604.25850, 2026{\natexlab{b}}.

\bibitem[Lin et~al.(2026{\natexlab{c}})Lin, Wu, Wang, Shi, Sang, He, Liu, Wei,
  Wu, Zhang, Wang, Zhang, Dumoulin, Xie, Zhou, Wang, and
  Lu]{harnessupdating2026}
Minhua Lin, Juncheng Wu, Zijun Wang, Zhan Shi, Yisi Sang, Bing He, Zewen Liu,
  Tianxin Wei, Zongyu Wu, Zhiwei Zhang, Dakuo Wang, Xiang Zhang, Benoit
  Dumoulin, Cihang Xie, Yuyin Zhou, Suhang Wang, and Hanqing Lu.
\newblock {Harness Updating Is Not Harness Benefit: Disentangling Evolution
  Capabilities in Self-Evolving LLM Agents}.
\newblock arXiv:2605.30621, 2026{\natexlab{c}}.

\bibitem[Ludwig et~al.(2026)Ludwig, Ahmad, Majumdar, and
  Ginsburg]{shortcutting2026}
Nikolai Ludwig, Wasi~Uddin Ahmad, Somshubra Majumdar, and Boris Ginsburg.
\newblock {Shortcutting the Fix: Identifying and Categorizing Agentic Exploits
  in Software Engineering Benchmarks}.
\newblock arXiv:2609.06780, 2026.

\bibitem[Luo et~al.(2026)Luo, Luu, Liu, Luo, Calvo, Dan, Zhang, Liu, and
  Elfeki]{unearnedpasses2026}
Weijun Luo, Kelvin Luu, Xinyi Liu, Guangze Luo, Miguel~Romero Calvo, Soham Dan,
  Daniel~Yue Zhang, Ying Liu, and Mohamed Elfeki.
\newblock {Maintaining Benchmarks Against Increasingly Capable Agents:
  Detection and Remediation of Unearned Passes}.
\newblock arXiv:2609.34262, 2026.

\bibitem[Ma et~al.(2026)Ma, Huang, Bao, Zhuang, Shukla, Galley, Zhang, and
  Feuerriegel]{skillgen2026}
Yuchen Ma, Yue Huang, Han Bao, Haomin Zhuang, Swadheen Shukla, Michel Galley,
  Xiangliang Zhang, and Stefan Feuerriegel.
\newblock {SkillGen: Verified Inference-Time Agent Skill Synthesis}.
\newblock arXiv:2605.10999, 2026.

\bibitem[Merrill et~al.(2026)Merrill, Shaw, Carlini, Li, Raj, Bercovich, Shi,
  Shin, Walshe, Buchanan, et~al.]{terminalbench2026}
Mike~A. Merrill, Alexander~G. Shaw, Nicholas Carlini, Boxuan Li, Harsh Raj,
  Ivan Bercovich, Lin Shi, Jeong~Yeon Shin, Thomas Walshe, E.~Kelly Buchanan,
  et~al.
\newblock {Terminal-Bench: Benchmarking Agents on Hard, Realistic Tasks in
  Command Line Interfaces}.
\newblock In \emph{International Conference on Learning Representations
  (ICLR)}, 2026.

\bibitem[Miller(2024)]{miller2024errorbars}
Evan Miller.
\newblock {Adding Error Bars to Evals: A Statistical Approach to Language Model
  Evaluations}.
\newblock arXiv:2411.00640, 2024.

\bibitem[Ni et~al.(2026)Ni, Liu, Liu, Sun, Zhou, Cheng, Wang, Zhao, Jiang, and
  Jiang]{trace2skill2026}
Jingwei Ni, Yihao Liu, Xinpeng Liu, Yutao Sun, Mengyu Zhou, Pengyu Cheng, Dexin
  Wang, Erchao Zhao, Xiaoxi Jiang, and Guanjun Jiang.
\newblock {Trace2Skill: Distill Trajectory-Local Lessons into Transferable
  Agent Skills}.
\newblock arXiv:2603.25158, 2026.

\bibitem[Pan et~al.(2025)Pan, Wang, Neubig, Jaitly, Ji, Suhr, and
  Zhang]{swegym2025}
Jiayi Pan, Xingyao Wang, Graham Neubig, Navdeep Jaitly, Heng Ji, Alane Suhr,
  and Yizhe Zhang.
\newblock {Training Software Engineering Agents and Verifiers with SWE-Gym}.
\newblock In \emph{International Conference on Machine Learning (ICML)}, 2025.

\bibitem[Pan et~al.(2026)Pan, Liu, Gao, Gao, Liu, Lin, Fu, Wang, Zhang, and
  Yu]{skillmas2026}
Shuai Pan, Yixiang Liu, Jiaye Gao, Te~Gao, Weiwen Liu, Jianghao Lin, Zhihui Fu,
  Jun Wang, Weinan Zhang, and Yong Yu.
\newblock {SkillMAS: Skill Co-Evolution with LLM-based Multi-Agent System}.
\newblock arXiv:2605.09341, 2026.

\bibitem[Patel et~al.(2026)Patel, Jha, Arabzadeh, Guestrin, Stoica, and
  Zaharia]{modeleats2026}
Liana Patel, Siddharth Jha, Negar Arabzadeh, Carlos Guestrin, Ion Stoica, and
  Matei Zaharia.
\newblock {What Happens When the Model Eats the Stack? Rethinking the Research
  Agenda for Data Agents to Withstand the Bitter Lesson}.
\newblock arXiv:2609.03141, 2026.

\bibitem[Pu et~al.(2026)Pu, Song, and Zhao]{skillops2026}
Hongji Pu, Xinyuan Song, and Liang Zhao.
\newblock {SkillOps: Managing LLM Agent Skill Libraries as Self-Maintaining
  Software Ecosystems}.
\newblock arXiv:2605.13716, 2026.

\bibitem[Robeyns et~al.(2025)Robeyns, Szummer, and Aitchison]{sica2025}
Maxime Robeyns, Martin Szummer, and Laurence Aitchison.
\newblock {A Self-Improving Coding Agent}.
\newblock arXiv:2504.15228, 2025.

\bibitem[Saha(2026)]{deliverystorage2026}
Swapnanil Saha.
\newblock {Delivery, Not Storage: Cue-Anchored Working Memory as a Harness
  Property for Coding Agents}.
\newblock arXiv:2607.20972, 2026.

\bibitem[Sam-Bodden(2026)]{whatcontext2026}
Brian Sam-Bodden.
\newblock {What Context Does a Coding Agent Actually Need to Act?}
\newblock arXiv:2607.09691, 2026.

\bibitem[Shang and Yang(2026)]{hdso2026}
Fangxin Shang and Yehui Yang.
\newblock {Hypothesis-Driven Skill Optimization for LLM Agents}.
\newblock arXiv:2606.22330, 2026.

\bibitem[Shawn(2026)]{pace2026}
Zayx Shawn.
\newblock {PACE: Anytime-Valid Acceptance Tests for Self-Evolving Agents}.
\newblock arXiv:2606.08106, 2026.

\bibitem[Shen et~al.(2026)Shen, Zhang, Zhao, and Cheng]{slim2026}
Junhao Shen, Teng Zhang, Xiaoyan Zhao, and Hong Cheng.
\newblock {Dynamic Skill Lifecycle Management for Agentic Reinforcement
  Learning}.
\newblock arXiv:2605.10923, 2026.

\bibitem[Shepard and Albrecht(2026)]{probeandrefine2026}
Asa Shepard and Jeannie Albrecht.
\newblock {Probe-and-Refine Tuning of Repository Guidance for Coding Agents}.
\newblock arXiv:2606.20512, 2026.

\bibitem[Shinn et~al.(2023)Shinn, Cassano, Gopinath, Narasimhan, and
  Yao]{shinn2023reflexion}
Noah Shinn, Federico Cassano, Ashwin Gopinath, Karthik Narasimhan, and Shunyu
  Yao.
\newblock {Reflexion: Language Agents with Verbal Reinforcement Learning}.
\newblock In \emph{Advances in Neural Information Processing Systems
  (NeurIPS)}, 2023.

\bibitem[Tang et~al.(2026)Tang, Rashtchian, Ferng, Tomkins, Juan, and
  Vu]{wikiskill2026}
Liyan Tang, Cyrus Rashtchian, Chun-Sung Ferng, Andrew Tomkins, Da-Cheng Juan,
  and Tu~Vu.
\newblock {WikiSkill: Compiling Agent Experience into Persistent Knowledge for
  Skill Evolution}.
\newblock arXiv:2608.27454, 2026.

\bibitem[Wang et~al.(2024)Wang, Xie, Jiang, Mandlekar, Xiao, Zhu, Fan, and
  Anandkumar]{voyager2023}
Guanzhi Wang, Yuqi Xie, Yunfan Jiang, Ajay Mandlekar, Chaowei Xiao, Yuke Zhu,
  Linxi Fan, and Anima Anandkumar.
\newblock {Voyager: An Open-Ended Embodied Agent with Large Language Models}.
\newblock \emph{Transactions on Machine Learning Research (TMLR)}, 2024.

\bibitem[Wang(2025)]{wang2025noises}
Sida Wang.
\newblock {Measuring all the noises of LLM Evals}.
\newblock arXiv:2512.21326, 2025.

\bibitem[Wang et~al.(2026{\natexlab{a}})Wang, Cheng, Li, Hao, Ramesh, Jung,
  Tang, Wang, Zhou, Wu, Yao, Grama, Banburski-Fahey, Peng, Lanier, and
  Gao]{valve2026}
Yifan Wang, Hao Cheng, Xiaomin Li, Yuexing Hao, Hemanth~Neelgund Ramesh,
  Dongwon Jung, Hao Tang, Keru Wang, Chenliang Zhou, Qianhui Wu, Wenlin Yao,
  Ananth Grama, Andrzej Banburski-Fahey, Baolin Peng, Jaron Lanier, and
  Jianfeng Gao.
\newblock {Certified Long-Horizon Code Agent Evolution via Validation-Gated
  Skill Optimization}.
\newblock arXiv:2609.32990, 2026{\natexlab{a}}.

\bibitem[Wang et~al.(2026{\natexlab{b}})Wang, Zhou, Liang, Zhang, Liu, Zhou,
  and Yao]{assay2026}
Yixuan Wang, Yiyang Zhou, Yiming Liang, Congyu Zhang, Fuxiao Liu, Jiawei Zhou,
  and Huaxiu Yao.
\newblock {Not All Skills Help: Measuring and Repairing Agent Knowledge}.
\newblock arXiv:2606.15390, 2026{\natexlab{b}}.

\bibitem[Wang et~al.(2025)Wang, Mao, Fried, and Neubig]{awm2024}
Zora~Zhiruo Wang, Jiayuan Mao, Daniel Fried, and Graham Neubig.
\newblock {Agent Workflow Memory}.
\newblock In \emph{International Conference on Machine Learning (ICML)}, 2025.

\bibitem[Wu et~al.(2026{\natexlab{a}})Wu, Yang, Liu, Lin, Zhang, Shi, Jiang,
  Xu, Li, and Guo]{bayesianagent2026}
Xiaojun Wu, Cehao Yang, Honghao Liu, Xueyuan Lin, Wenjie Zhang, Zhichao Shi,
  Xuhui Jiang, Chengjin Xu, Jia Li, and Jian Guo.
\newblock {Bayesian-Agent: Posterior-Guided Skill Evolution Across {LLM} Agent
  Harnesses}.
\newblock arXiv:2606.08348, 2026{\natexlab{a}}.

\bibitem[Wu et~al.(2026{\natexlab{b}})Wu, Zhao, Li, Lee, Zhu, Wu, Yu, Li,
  Zhang, Fan, and Li]{swetogether2026}
Yifan Wu, Zhuokai Zhao, Songlin Li, Ho~Hin Lee, Jiacheng Zhu, Shirley Wu,
  Tianhe Yu, Serena Li, Lizhu Zhang, Xiangjun Fan, and Shengzhi Li.
\newblock {SWE-Together: Evaluating Coding Agents in Interactive User
  Sessions}.
\newblock arXiv:2606.29957, 2026{\natexlab{b}}.

\bibitem[Wu et~al.(2026{\natexlab{c}})Wu, Zhang, Shi, Lei, Gu, Zhang, Wang, He,
  Huang, Song, Zeng, Wang, Liu, Shi, Liu, Yan, Huang, Zhang, and
  Zhang]{harnessdev2026}
Yuhao Wu, Jingyuan Zhang, Jiajun Shi, Xinping Lei, Qingshui Gu, Yuxuan Zhang,
  Zexuan Wang, Chen He, Chen Huang, Maojia Song, Zhiyuan Zeng, Shaowen Wang,
  Jinkai Liu, Yunfeng Shi, Jiaheng Liu, Shen Yan, Wenhao Huang, Ge~Zhang, and
  Wenxuan Zhang.
\newblock {HarnessDev: Can LLMs Create and Evolve Their Own Agent Harness?}
\newblock arXiv:2609.01437, 2026{\natexlab{c}}.

\bibitem[Xia et~al.(2026)Xia, Han, Wang, Chen, Zhuang, Lee, Huang, Yu, CuiZhu,
  Ming, Yao, Gokturk, Pfister, and Lee]{rrsi2026}
Peng Xia, Rujun Han, Zifeng Wang, Yanfei Chen, Yufan Zhuang, Yoonho Lee,
  Chengsong Huang, Han Yu, Zhongying CuiZhu, Yifei Ming, Huaxiu Yao, Burak
  Gokturk, Tomas Pfister, and Chen-Yu Lee.
\newblock {RRSI: Regularized Recursive Self-Improvement of Agent Harnesses}.
\newblock arXiv:2609.24972, 2026.

\bibitem[Yang et~al.(2025)Yang, Lieret, Jimenez, Wettig, Khandpur, Zhang, Hui,
  Press, Schmidt, and Yang]{swesmith2025}
John Yang, Kilian Lieret, Carlos~E. Jimenez, Alexander Wettig, Kabir Khandpur,
  Yanzhe Zhang, Binyuan Hui, Ofir Press, Ludwig Schmidt, and Diyi Yang.
\newblock {SWE-smith: Scaling Data for Software Engineering Agents}.
\newblock In \emph{Advances in Neural Information Processing Systems (NeurIPS),
  Datasets and Benchmarks Track}, 2025.

\bibitem[Yang et~al.(2026)Yang, Gong, Huang, Yang, Zhou, Huang, Li, Gao, Dai,
  Liu, Qiu, Yang, Chen, Yang, and Luo]{skillopt2026}
Yifan Yang, Ziyang Gong, Weiquan Huang, Qihao Yang, Ziwei Zhou, Zisu Huang, Yan
  Li, Xuemei Gao, Qi~Dai, Bei Liu, Kai Qiu, Yuqing Yang, Dongdong Chen, Xue
  Yang, and Chong Luo.
\newblock {SkillOpt: Executive Strategy for Self-Evolving Agent Skills}.
\newblock arXiv:2605.23904, 2026.

\bibitem[Yang and Ding(2026)]{signalnoise2026}
Ziyue Yang and Fan Ding.
\newblock {Signal or Noise? A Benchmark Study of Agent Skills in Web
  Development}.
\newblock arXiv:2608.23067, 2026.

\bibitem[Yao et~al.(2025)Yao, Shinn, Razavi, and Narasimhan]{yao2025taubench}
Shunyu Yao, Noah Shinn, Pedram Razavi, and Karthik Narasimhan.
\newblock {$\tau$-bench: A Benchmark for Tool-Agent-User Interaction in
  Real-World Domains}.
\newblock In \emph{International Conference on Learning Representations
  (ICLR)}, 2025.

\bibitem[Zhan et~al.(2026)Zhan, Ma, and Haddadi]{saferetirement2026}
Zhonghao Zhan, Xiao Ma, and Hamed Haddadi.
\newblock {Safe Skill Retirement for Physical Agents}.
\newblock arXiv:2609.29543, 2026.

\bibitem[Zhang et~al.(2026{\natexlab{a}})Zhang, Zhang, Li, Zhang, Chen, Zhang,
  Bai, and Hu]{selfharness2026}
Hangfan Zhang, Shao Zhang, Kangcong Li, Chen Zhang, Yang Chen, Yiqun Zhang, Lei
  Bai, and Shuyue Hu.
\newblock {Self-Harness: Harnesses That Improve Themselves}.
\newblock arXiv:2606.09498, 2026{\natexlab{a}}.

\bibitem[Zhang et~al.(2026{\natexlab{b}})Zhang, Hu, Lu, Lange, and
  Clune]{zhang2025dgm}
Jenny Zhang, Shengran Hu, Cong Lu, Robert Lange, and Jeff Clune.
\newblock {Darwin {G}{\"o}del Machine: Open-Ended Evolution of Self-Improving
  Agents}.
\newblock In \emph{International Conference on Learning Representations
  (ICLR)}, 2026{\natexlab{b}}.

\bibitem[Zhang et~al.(2025)Zhang, He, Zhang, Kang, Li, Xie, Wang, Wang, Huang,
  Fu, Nallipogu, Lin, Dang, Rajmohan, and Zhang]{swebenchlive2025}
Linghao Zhang, Shilin He, Chaoyun Zhang, Yu~Kang, Bowen Li, Chengxing Xie,
  Junhao Wang, Maoquan Wang, Yufan Huang, Shengyu Fu, Elsie Nallipogu, Qingwei
  Lin, Yingnong Dang, Saravan Rajmohan, and Dongmei Zhang.
\newblock {SWE-bench Goes Live!}
\newblock In \emph{Advances in Neural Information Processing Systems (NeurIPS),
  Datasets and Benchmarks Track}, 2025.

\bibitem[Zhang et~al.(2026{\natexlab{c}})Zhang, Hu, Upasani, Ma, Hong,
  Kamanuru, Rainton, Wu, Ji, Li, Thakker, Zou, and Olukotun]{ace2026}
Qizheng Zhang, Changran Hu, Shubhangi Upasani, Boyuan Ma, Fenglu Hong,
  Vamsidhar Kamanuru, Jay Rainton, Chen Wu, Mengmeng Ji, Hanchen Li, Urmish
  Thakker, James Zou, and Kunle Olukotun.
\newblock {Agentic Context Engineering: Evolving Contexts for Self-Improving
  Language Models}.
\newblock In \emph{International Conference on Learning Representations
  (ICLR)}, 2026{\natexlab{c}}.

\bibitem[Zhang(2026)]{skillestimands2026}
Shuyang Zhang.
\newblock {What Does a Skill Actually Do? Estimands and Evaluation Validity for
  Tool and Skill Use in {LLM} Agents: A Critical Review}.
\newblock arXiv:2609.33153, 2026.

\bibitem[Zhang et~al.(2026{\natexlab{d}})Zhang, Cui, Wang, Li, Qiu, Zhu, and
  He]{ratchet2026}
Xing Zhang, Yanwei Cui, Guanghui Wang, Ziyuan Li, Wei Qiu, Bing Zhu, and
  Peiyang He.
\newblock {Ratchet: How Reliable Must an {LLM} Judge Be to Retire a Skill?}
\newblock arXiv:2605.22148, 2026{\natexlab{d}}.

\bibitem[Zhang et~al.(2026{\natexlab{e}})Zhang, Wang, Cui, Qiu, Li, Zhu, and
  He]{guardrails2026}
Xing Zhang, Guanghui Wang, Yanwei Cui, Wei Qiu, Ziyuan Li, Bing Zhu, and
  Peiyang He.
\newblock {Guardrails Beat Guidance: A Large-Scale Study of Rules, Skills, and
  Persistent Configuration for Coding Agents}.
\newblock arXiv:2604.11088, 2026{\natexlab{e}}.

\bibitem[Zhang et~al.(2026{\natexlab{f}})Zhang, Dai, Tan, Yang, Mullur, Hoang,
  Hu, Zhu, Mui, Savarese, Xu, and Chen]{darwinx2026}
Yifan Zhang, Yutong Dai, Juntao Tan, Luyu Yang, Rishi Mullur, Thai Hoang,
  Zhiyuan Hu, James Zhu, Phil Mui, Silvio Savarese, Ran Xu, and Zeyuan Chen.
\newblock {DarwinX: Evolving Agent Harnesses Through Natural Selection}.
\newblock arXiv:2608.07545, 2026{\natexlab{f}}.

\bibitem[Zhao et~al.(2024)Zhao, Huang, Xu, Lin, Liu, and Huang]{zhao2024expel}
Andrew Zhao, Daniel Huang, Quentin Xu, Matthieu Lin, Yong-Jin Liu, and Gao
  Huang.
\newblock {ExpeL: LLM Agents Are Experiential Learners}.
\newblock In \emph{AAAI Conference on Artificial Intelligence (AAAI)}, 2024.

\bibitem[Zhao et~al.(2026)Zhao, Zhang, Liu, Chen, Xiong, Chen, Ye, Yu, Yuan,
  Han, Wang, Ji, Huang, and Zhang]{cass2026}
Qiyan Zhao, Xiaofeng Zhang, Bo~Liu, Minda Chen, Wei Xiong, Jingyang Chen,
  Guanting Ye, Wenhao Yu, Xiaosong Yuan, Shijie Han, Da-Han Wang, Jianmin Ji,
  Fei Huang, and Xu-Yao Zhang.
\newblock {Coalition-Aware Skill Reliability for Self-Evolving Agents}.
\newblock arXiv:2608.22610, 2026.

\bibitem[Zheng et~al.(2023)Zheng, Chiang, Sheng, Zhuang, Wu, Zhuang, Lin, Li,
  Li, Xing, Zhang, Gonzalez, and Stoica]{zheng2023judge}
Lianmin Zheng, Wei-Lin Chiang, Ying Sheng, Siyuan Zhuang, Zhanghao Wu, Yonghao
  Zhuang, Zi~Lin, Zhuohan Li, Dacheng Li, Eric~P. Xing, Hao Zhang, Joseph~E.
  Gonzalez, and Ion Stoica.
\newblock {Judging LLM-as-a-Judge with MT-Bench and Chatbot Arena}.
\newblock In \emph{Advances in Neural Information Processing Systems (NeurIPS),
  Datasets and Benchmarks Track}, 2023.

\bibitem[Zheng et~al.(2026)Zheng, Shang, Jiang, Tian, Zhu, Ma, Yuan, and
  Zhang]{sweproverified2026}
Pujun Zheng, Zixin Shang, Shufan Jiang, Wenhui Tian, Dongsheng Zhu, Zerun Ma,
  Dingbo Yuan, and Qi~Zhang.
\newblock {SWE-Bench Pro Verified: A Reliable Benchmark for Software
  Engineering Agents}.
\newblock arXiv:2609.08149, 2026.

\bibitem[Zhou et~al.(2026)Zhou, Guo, Zhuang, Wang, Huang, Liang, Chen, Gao,
  Moniz, Chawla, and Zhang]{tracecorrections2026}
Yujun Zhou, Kehan Guo, Haomin Zhuang, Xiangqi Wang, Yue Huang, Zhenwen Liang,
  Pin-Yu Chen, Tian Gao, Nuno Moniz, Nitesh~V. Chawla, and Xiangliang Zhang.
\newblock {Getting Better at Working With You: Compiling User Corrections into
  Runtime Enforcement for Coding Agents}.
\newblock arXiv:2606.13174, 2026.

\bibitem[Zhu et~al.(2025)Zhu, Jin, Pruksachatkun, Zhang, Liu, Cui, Kapoor,
  Longpre, Meng, Weiss, et~al.]{abc2025}
Yuxuan Zhu, Tengjun Jin, Yada Pruksachatkun, Andy Zhang, Shu Liu, Sasha Cui,
  Sayash Kapoor, Shayne Longpre, Kevin Meng, Rebecca Weiss, et~al.
\newblock {Establishing Best Practices for Building Rigorous Agentic
  Benchmarks}.
\newblock In \emph{Advances in Neural Information Processing Systems (NeurIPS),
  Datasets and Benchmarks Track}, 2025.

\end{thebibliography}
